\documentclass[12pt]{article}
\usepackage{a4wide,amsmath,amssymb,bm,graphicx}
\newcommand{\1}[1]{\mathbf{1}^{#1}}
\newcommand{\2}[1]{\mathbf{2}^{#1}}
\newcommand{\3}[1]{\mathbf{3}^{#1}}
\newcommand{\4}[1]{\mathbf{4}^{#1}}
\newcommand{\5}[1]{\mathbf{5}^{#1}}
\newcommand{\D}{\rlap{\hspace{0.2em}/}D}
\newcommand{\ke}{\bm{k}}
\newcommand{\pe}{\bm{p}}
\newcommand{\MS}{\ensuremath{\overline{\text{MS}}}}
\DeclareMathOperator{\Tr}{Tr}
\numberwithin{equation}{section}
\newlength{\prep}

\begin{document}

\begin{flushright}
\Large
\settowidth{\prep}{SFB/CPP-05-43}
\begin{minipage}{\prep}
\begin{flushleft}
TTP05-15\\
SFB/CPP-05-43\\
hep-ph/0508242
\end{flushleft}
\end{minipage}
\end{flushright}
\vspace{2em}
\begin{center}
\LARGE Lectures on QED and QCD${}^1$\\[1.5em]
\large Andrey Grozin\\[0.5em]
\large Institut f\"ur Theoretische Teilchenphysik, Universit\"at Karlsruhe
\end{center}
\addtocounter{footnote}{1}
\footnotetext{Given at Dubna International Advanced School
on Theoretical Physics, 29 Jan.\ -- 6 Feb.\ 2005}
\vspace{2em}

\begin{abstract}
The lectures are a practical introduction
to perturbative calculations in QED and QCD.
I discuss methods of calculation of one- and two-loop diagrams
in dimensional regularization,
\MS{} and on-shell renormalization schemes,
decoupling of heavy-particle loops.
\end{abstract}

\tableofcontents

\section{Introduction}
\label{S:Intro}

These lectures are an addendum to a standard textbook
on quantum field theory, such as~\cite{PS:95}.
We shall discuss practical ways to calculate one- and two-loop diagrams
in QED and QCD and to obtain renormalized physical results.
More details about calculation of Feynman integrals
can be found in my lectures at another Dubna school~\cite{G:04},
or in the textbook~\cite{S:05}.
Calculation of colour factors of diagrams in QCD (Appendix~\ref{S:Colour}),
and group theory in general, is discussed in an excellent book
by P.~Cvitanovi\'c~\cite{Cv} available on the Web.
In Sect.~\ref{S:Dec} about decoupling of heavy-particle loops,
we follow~\cite{CKS:98}, but in a much simpler case
(QED at two loops).

It is well-known that Feynman diagrams with loops can diverge.
The first thing we have to do is to introduce some regularization.
Regularized diagrams must converge,
so that working with them has sense.
When the regularization parameter tends to some value,
the original theory must be restored.
There exist many regularization methods.
A good choice of regularization is crucial for any
non-trivial computation:
the regularization should preserve as much of the symmetry
of the theory as possible,
manipulating regularized Feynman integrals should be simple, etc.
Cut-off regularization is difficult to define unambiguously;
in gauge theories, it breaks gauge invariance;
it is difficult to integrate by parts because of boundary terms.
Pauli-Villars regularization also breaks gauge invariance.
Lattice regularization preserves it, but breaks Lorentz invariance.
The method used in practically all multiloop calculations
is dimensional regularization.
It preserves both Lorentz and gauge invariance,
and rules for manipulating integrals are simple
(no boundary terms, etc.).
We shall use it throughout these lectures.
We'll discuss the most widely used method of multiloop calculations,
integration by parts~\cite{CT:81}, in Sect.~\ref{S:q2}.

Another important concept is renormalization.
Physical results should be expressed via physical parameters
(measurable, or expressible via measurable quantities),
not via parameters in the bare Lagrangian.
This re-expressing in physically necessary,
independently of presence or absence of divergences.
After it, we can switch off regularization;
in a sensible theory, results must be finite.

Renormalization procedure is not unique.
There are many renormalization schemes;
some of them depend on parameters.
The most widely used scheme is \MS{}.
It depends on a single parameter --- renormalization scale.
Physical results should not depend on this artificial quantity.
This requirement leads to renormalization group equations ---
a very powerful instrument in quantum field theory.
We shall use this scheme in Sects.~\ref{S:QED}--\ref{S:2L}.
On-shell renormalization scheme (Sect.~\ref{S:OS})
is also widely used, especially in QED at low energies.
In QCD, it is often useful to use on-shell renormalization
for heavy quarks, and \MS{} for all the rest.

\section{One-loop diagrams}
\label{S:1l}

\subsection{Massive vacuum diagram}
\label{S:V1}

During these lectures,
we are going to live in $d$-dimensional space--time:
1 time and $d-1$ space dimensions.
The dimensionality $d$ must appear in all formulas as a symbol,
it is not enough to obtain separate results
for a few integer values of $d$.
After calculating a physical result in terms of physical parameters,
we take the limit $d\to4$.
Therefore, $d$ is often written as $4-2\varepsilon$,
where the physical limit is $\varepsilon\to0$.
It is not possible to take this limit in intermediate formulas,
because they often contain poles $1/\varepsilon$.
Such divergences must disappear in physical (renormalized) results.

\begin{figure}[ht]
\begin{center}
\begin{picture}(22,22)
\put(11,11){\makebox(0,0){\includegraphics{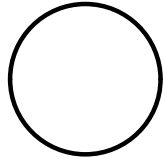}}}
\end{picture}
\end{center}
\caption{One-loop massive vacuum diagram}
\label{F:V1}
\end{figure}

Let's consider the simplest diagram shown in Fig.~\ref{F:V1}:
\begin{equation}
\int\frac{d^d k}{D^n} = i \pi^{d/2} m^{d-2n} V(n)\,,\quad
D = m^2 - k^2 - i0\,.
\label{V1:def}
\end{equation}
The power of $m$ is evident from the dimensional counting,
and our aim is to find the dimensionless function $V(n)$;
we can put $m=1$ to simplify the calculation.
Due to the position of the poles in the complex $k_0$ plane
(determined by $-i0$),
we may rotate the integration contour counterclockwise by $\pi/2$
without crossing the poles.
After this Wick rotation,
we integrate along the imaginary axis in the $k_0$ plane
in the positive direction: $k_0=i\ke_0$.
Here $\ke_0$ is the 0-th component of the vector $\ke$
in $d$-dimensional euclidean space
(euclidean vectors will be denoted by the bold font),
and $k^2=-\ke^2$.
Then our definition~(\ref{V1:def}) of $V(n)$ becomes
\begin{equation}
\int \frac{d^d\ke}{(\ke^2+1)^n} = \pi^{d/2} V(n)\,.
\label{V1:eucl}
\end{equation}

It is often useful to turn denominators into exponentials
using the $\alpha$-parametrization
\begin{equation}
\frac{1}{a^n} = \frac{1}{\Gamma(n)}
\int_0^\infty e^{-a\alpha} \alpha^{n-1} d\alpha\,.
\label{V1:alpha}
\end{equation}
For our integral, this gives
\begin{equation}
V(n) = \frac{\pi^{-d/2}}{\Gamma(n)}
\int e^{-\alpha(\ke^2+1)} \alpha^{n-1} d\alpha\,d^d\ke\,.
\label{V1:Valpha}
\end{equation}
The $d$-dimensional integral of the exponent of the quadratic form
is the product of $d$ one-dimensional integrals:
\begin{equation}
\int e^{-\alpha\ke^2} d^d\ke
= \left[ \int_{-\infty}^{+\infty} e^{-\alpha\ke_x^2} d\ke_x \right]^d
= \left(\frac{\pi}{\alpha}\right)^{d/2}\,.
\label{V1:gauss}
\end{equation}
This is the definition of $d$-dimensional integration;
note that the result contains $d$ as a symbol.
Now it is easy to calculate
\begin{equation}
V(n) = \frac{1}{\Gamma(n)} \int_0^\infty e^{-\alpha} \alpha^{n-d/2-1} d\alpha\,;
\end{equation}
the result is
\begin{equation}
V(n) = \frac{\Gamma(-d/2+n)}{\Gamma(n)}\,.
\label{V1:res}
\end{equation}
For all integer $n$, the results are proportional to
\begin{equation}
V_1 = V(1) = \frac{4}{(d-2)(d-4)} \Gamma(1+\varepsilon)\,,
\label{V1:V1}
\end{equation}
where the coefficients are rational functions of $d$.

The denominator in~(\ref{V1:eucl}) behaves as $(\ke^2)^n$ at $\ke\to\infty$.
Therefore, the integral diverges if $d\ge2n$.
At $d\to4$ this means $n\le2$.
This ultraviolet divergence shows itself as a $1/\varepsilon$ pole
in~(\ref{V1:res}) for $n=1$, $2$.

Let's calculate the full solid angle in $d$-dimensional space
(we shall need it later).
To this end, we'll calculate one and the same integral
in Cartesian coordinates
\begin{equation}
\int e^{-\ke^2} d^d\ke
= \left[ \int_{-\infty}^{+\infty} e^{-\ke_x^2} d\ke_x \right]^d
= \pi^{d/2}
\label{V1:Cartesian}
\end{equation}
and in spherical coordinates
\begin{equation}
\Omega_d \int_0^\infty e^{-\ke^2} \ke^{d-1} d\ke
= \frac{\Omega_d}{2} \int_0^\infty e^{-\ke^2} (\ke^2)^{d/2-1} d\ke^2
= \frac{\Omega_d \Gamma(d/2)}{2}\,.
\label{V1:Spherical}
\end{equation}
Therefore, the full solid angle is
\begin{equation}
\Omega_d = \frac{2 \pi^{d/2}}{\Gamma(d/2)}\,.
\label{V1:Omega}
\end{equation}
For example,
\begin{equation}
\Omega_1 = 2\,,\quad
\Omega_2 = 2 \pi\,,\quad
\Omega_3 = 4 \pi\,,\quad
\Omega_4 = 2 \pi^2\,,\quad
\ldots
\label{V1:Omega1}
\end{equation}
In 1-dimensional space, a sphere consists of 2 points,
hence $\Omega_1=2$;
the values for $d=2$ and $3$ are also well known.

\begin{figure}[ht]
\begin{center}
\begin{picture}(22,22)
\put(11,11){\makebox(0,0){\includegraphics{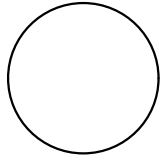}}}
\end{picture}
\end{center}
\caption{One-loop massless vacuum diagram}
\label{F:V0}
\end{figure}

What is the value of the massless vacuum diagram
\begin{equation*}
\int \frac{d^d k}{(-k^2-i0)^n}
\end{equation*}
(Fig.~\ref{F:V0})?
Its dimensionality is $d-2n$.
But it contains no dimensionfull parameters
from which such a value could be constructed.
The only result we can write for this diagram is
\begin{equation}
\int \frac{d^d k}{(-k^2-i0)^n} = 0\,.
\label{V1:V0}
\end{equation}
This argument fails at $n=d/2$;
surprises can be expected at this point.

\begin{figure}[ht]
\begin{center}
\begin{picture}(38,32)
\put(16,16){\makebox(0,0){\includegraphics{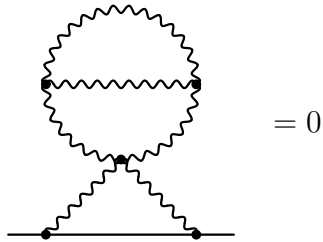}}}
\put(35,16){\makebox(0,0)[l]{${}=0$}}
\end{picture}
\end{center}
\caption{A diagram for the quark propagator}
\label{F:00}
\end{figure}

This dimensions-counting argument is much more general.
For example, the diagram in Fig.~\ref{F:00} contains a sub-diagram
which is attached to the rest of the diagram at a single vertex,
and which contains no scale.
This subdiagram is given by an integral (maybe, a tensor one)
with some $d$-dependent dimensionality.
It has no dimensionfull parameters;
the only value we can construct for such an integral is 0.

Now we shall discuss a trick (invented by Feynman)
to combine several denominators into a single one
at the price of introducing some integrations.
We'll not use it during these lectures;
however, it is widely used, and it is important to know it.
Let's multiply two $\alpha$-parametrizations~(\ref{V1:alpha}):
\begin{equation}
\frac{1}{a_1^{n_1}a_2^{n_2}} = \frac{1}{\Gamma(n_1)\Gamma(n_2)}
\int e^{-a_1\alpha_1 - a_2\alpha_2} \alpha_1^{n_1-1} \alpha_2^{n_2-1}
d\alpha_1\,d\alpha_2\,.
\label{V1:alpha2}
\end{equation}
It is always possible to calculate one integral in such a product,
namely, to integrate in a common scale $\eta$ of all $\alpha_i$.
If we denote $\eta=\alpha_1+\alpha_2$,
i.e., make the substitution $\alpha_1=\eta x$, $\alpha_2=\eta(1-x)$,
then we arrive at the Feynman parametrization
\begin{equation}
\frac{1}{a_1^{n_1}a_2^{n_2}} = \frac{\Gamma(n_1+n_2)}{\Gamma(n_1)\Gamma(n_2)}
\int_0^1 \frac{x^{n_1-1} (1-x)^{n_2-1} d x}%
{\left[a_1 x + a_2 (1-x)\right]^{n_1+n_2}}\,.
\label{V1:Feynman2}
\end{equation}
This is not the only possibility.
We can also take, say, $\eta=\alpha_2$,
i.e., make the substitution $\alpha_1=\eta x$, $\alpha_2=\eta$.
This results in a variant of the Feynman parametrization,
which can be more useful in some cases:
\begin{equation}
\frac{1}{a_1^{n_1}a_2^{n_2}} = \frac{\Gamma(n_1+n_2)}{\Gamma(n_1)\Gamma(n_2)}
\int_0^\infty \frac{x^{n_1-1} d x}{\left[a_1 x + a_2\right]^{n_1+n_2}}\,.
\label{V1:HQET}
\end{equation}

Let's consider the general case of $k$ denominators:
\begin{equation}
\begin{split}
\frac{1}{a_1^{n_1}a_2^{n_2}\cdots a_k^{n_k}} ={}&
\frac{1}{\Gamma(n_1)\Gamma(n_2)\cdots\Gamma(n_k)}\\
&{}\times \int e^{-a_1\alpha_1-a_2\alpha_2\cdots-a_k\alpha_k}
\alpha_1^{n_1-1}\alpha_2^{n_2-1}\cdots\alpha_k^{n_k-1}
d\alpha_1\,d\alpha_2\cdots d\alpha_k\,.
\end{split}
\label{V1:alphan}
\end{equation}
We can choose the common scale $\eta$ to be the sum
of any subset of $\alpha_i$;
the numbering of the denominators is not fixed,
and we can always re-number them in such a way
that $\eta=\alpha_1+\alpha_2\cdots+\alpha_l$,
where $1\le l\le k$.
We insert
\begin{equation*}
\delta(\alpha_1+\alpha_2\cdots+\alpha_l-\eta) d\eta
\end{equation*}
into the integrand,
and make the substitution $\alpha_i=\eta x_i$.
Then the integral in $\eta$ can be easily calculated,
and we obtain the general Feynman parametrization
\begin{equation}
\begin{split}
\frac{1}{a_1^{n_1}a_2^{n_2}\cdots a_k^{n_k}} ={}&
\frac{\Gamma(n_1+n_2\cdots+n_k)}{\Gamma(n_1)\Gamma(n_2)\cdots\Gamma(n_k)}\\
&{}\times \int \frac{\delta(x_1+x_2\cdots+x_l-1)
x_1^{n_1-1} x_2^{n_2-1} \cdots x_k^{n_k-1}
d x_1\,d x_2\cdots d x_k}%
{\left[a_1 x_1 + a_2 x_2 \cdots + a_k x_k\right]^{n_1+n_2\cdots+n_k}}\,.
\end{split}
\label{V1:Feynman}
\end{equation}
Let's stress once more that the $\delta$ function here
can contain the sum of any subset of the variables $x_i$,
from a single variable up to all of them.

\subsection{Massless propagator diagram}
\label{S:q1}

\begin{figure}[ht]
\begin{center}
\begin{picture}(32,24)
\put(16,12){\makebox(0,0){\includegraphics{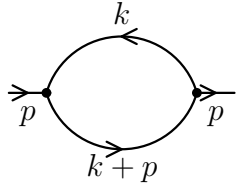}}}
\put(16,0){\makebox(0,0)[b]{$k+p$}}
\put(16,24){\makebox(0,0)[t]{$k$}}
\put(3.5,7.5){\makebox(0,0)[b]{$p$}}
\put(28.5,7.5){\makebox(0,0)[b]{$p$}}
\end{picture}
\end{center}
\caption{One-loop massless propagator diagram}
\label{F:q1}
\end{figure}

\begin{figure}[ht]
\begin{center}
\begin{picture}(26,17)
\put(13,8.5){\makebox(0,0){\includegraphics{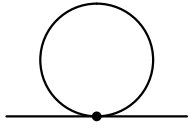}}}
\end{picture}
\end{center}
\caption{Massless vacuum diagram}
\label{F:q0}
\end{figure}

Now we shall consider the massless propagator diagram (Fig.~\ref{F:q1}),
\begin{equation}
\int \frac{d^d k}{D_1^{n_1}D_2^{n_2}}
= i \pi^{d/2} (-p^2)^{d/2-n_1-n_2} G(n_1,n_2)\,,\quad
D_1 = - (k+p)^2\,,\quad
D_2 = - k^2
\label{q1:G1def}
\end{equation}
(from now on, we'll not write $-i0$ in denominators explicitly,
but they are implied).
We shall consider this diagram at $p^2<0$,
i.e., below the threshold of production of a real pair,
where the result is an analytic function.
The power of $-p^2$ is evident from dimensionality.
Our aim is to calculate the dimensionless function $G(n_1,n_2)$;
we can put $-p^2=1$ to simplify things.
This diagram is symmetric with respect to $1\leftrightarrow2$.
It vanishes for integer $n_1\le0$ or $n_2\le0$,
because then it becomes the massless vacuum diagram (Fig.~\ref{F:q0}),
possibly, with some polynomial numerator.

Using Wick rotation and $\alpha$ parametrization~(\ref{V1:alpha}),
we rewrite the definition~(\ref{q1:G1def}) of $G(n_1,n_2)$ as
\begin{equation}
G(n_1,n_2) = \frac{\pi^{-d/2}}{\Gamma(n_1)\Gamma(n_2)}
\int e^{-\alpha_1(\ke+\pe)^2-\alpha_2\ke^2}
\alpha_1^{n_1-1} \alpha_2^{n_2-1} d\alpha_1\,d\alpha_2\,d^d\ke\,.
\label{q1:G1a}
\end{equation}
We want to separate a full square in the exponent;
to this end, we shift the integration momentum:
\begin{equation*}
\ke' = \ke + \frac{\alpha_1}{\alpha_1+\alpha_2} \pe\,,
\end{equation*}
and obtain
\begin{equation}
\begin{split}
G(n_1,n_2) &{}= \frac{\pi^{-d/2}}{\Gamma(n_1)\Gamma(n_2)}
\int \exp\left[-\frac{\alpha_1\alpha_2}{\alpha_1+\alpha_2}\right]
\alpha_1^{n_1-1} \alpha_2^{n_2-1} d\alpha_1\,d\alpha_2
\int e^{-(\alpha_1+\alpha_2)\ke^2} d^d\ke\\
&{}= \frac{1}{\Gamma(n_1)\Gamma(n_2)}
\int \exp\left[-\frac{\alpha_1\alpha_2}{\alpha_1+\alpha_2}\right]
(\alpha_1+\alpha_2)^{-d/2}
\alpha_1^{n_1-1} \alpha_2^{n_2-1} d\alpha_1\,d\alpha_2\,.
\end{split}
\label{q1:G1b}
\end{equation}
Now we make the usual substitution $\alpha_1=\eta x$, $\alpha_2=\eta(1-x)$,
and obtain
\begin{equation}
\begin{split}
G(n_1,n_2) &{}= \frac{1}{\Gamma(n_1)\Gamma(n_2)}
\int_0^1 x^{n_1-1} (1-x)^{n_2-1} dx
\int_0^\infty e^{-\eta x(1-x)} \eta^{-d/2+n_1+n_2-1} d\eta\\
&{}= \frac{\Gamma(-d/2+n_1+n_2)}{\Gamma(n_1)\Gamma(n_2)}
\int_0^1 x^{d/2-n_2-1} (1-x)^{d/2-n_1-1} dx\,.
\end{split}
\label{q1:G1c}
\end{equation}
This integral is the Euler $B$ function,
and we arrive at the final result
\begin{equation}
G(n_1,n_2) = \frac{\Gamma(-d/2+n_1+n_2)\Gamma(d/2-n_1)\Gamma(d/2-n_2)}%
{\Gamma(n_1)\Gamma(n_2)\Gamma(d-n_1-n_2)}\,.
\label{q1:G1}
\end{equation}
For all integer $n_{1,2}$, these integrals are proportional to
\begin{equation}
G_1 = G(1,1) = - \frac{2 g_1}{(d-3)(d-4)}\,,\quad
g_1 = \frac{\Gamma(1+\varepsilon)\Gamma^2(1-\varepsilon)}%
{\Gamma(1-2\varepsilon)}\,,
\label{q1:g1}
\end{equation}
with coefficients which are rational functions of $d$.

The denominator in~(\ref{q1:G1def}) behaves as $(k^2)^{n_1+n_2}$
at $k\to\infty$.
Therefore, the integral diverges if $d\ge2(n_1+n_2)$.
At $d\to4$ this means $n_1+n_2\le2$.
This ultraviolet divergence shows itself as a $1/\varepsilon$ pole
of the first $\Gamma$ function in the numerator of~(\ref{q1:G1})
for $n_1=n_2=1$ (this $\Gamma$ function depends on $n_1+n_2$,
i.e., on the behaviour of the integrand at $k\to\infty$).
The integral~(\ref{q1:G1def}) can also have infrared divergences.
Its denominator behaves as $(k^2)^{n_2}$ at $k\to0$,
and the integral diverges in this region if $d\le2n_2$.
At $d\to4$ this means $n_2\ge2$.
This infrared divergence shows itself as a $1/\varepsilon$ pole
of the third $\Gamma$ function in the numerator of~(\ref{q1:G1})
for $n_2\ge2$ (this $\Gamma$ function depends on $n_2$,
i.e., on the behaviour of the integrand at $k\to0$).
Similarly, the infrared divergence at $k+p\to0$ appears,
at $d\to4$, as a pole of the second $\Gamma$ function,
if $n_1\ge2$.

\begin{figure}[ht]
\begin{center}
\begin{picture}(32,32)
\put(16,16){\makebox(0,0){\includegraphics{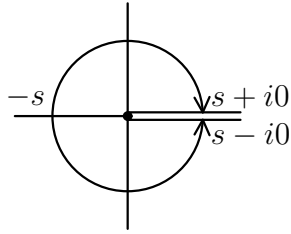}}}
\put(27,17){\makebox(0,0)[lb]{$s+i0$}}
\put(27,15){\makebox(0,0)[lt]{$s-i0$}}
\put(5,17){\makebox(0,0)[rb]{$-s$}}
\end{picture}
\end{center}
\caption{Complex $p^2$ plane}
\label{F:analyt}
\end{figure}

Let's consider the integral (Fig.~\ref{F:q1})
\begin{equation}
I(p^2) = - \frac{i}{\pi^{d/2}} \int
\frac{d^d k}{(-k^2-i0)(-(k+p)^2-i0)}
= G_1 (-p^2)^{-\varepsilon}
\label{q1:Ip2}
\end{equation}
as a function of the complex variable $p^2$.
It has a cut along the positive half-axis $p^2>0$,
starting at the branching point $p^2=0$
at the threshold of the real pair production (Fig.~\ref{F:analyt}).
Analytically continuing it from a negative $p^2=-s$
(where the function is regular) along the upper contour in Fig.~\ref{F:analyt},
$p^2=-s \exp(-i\alpha)$ with $\alpha$ varying from 0 to $\pi$,
we obtain
\begin{equation}
I(s+i0) = G_1 s^{-\varepsilon} e^{i\pi\varepsilon}\,.
\label{q1:analyt}
\end{equation}
Similarly, continuing along the lower contour,
we get $I(s-i0)$ having the opposite sign of the imaginary part,
and the discontinuity across the cut is
\begin{equation}
I(s+i0) - I(s-i0) = G_1 s^{-\varepsilon} 2 i \sin(\pi\varepsilon)\,.
\label{q1:disc}
\end{equation}
At $\varepsilon\to0$, $G_1$~(\ref{q1:g1}) has a $1/\varepsilon$ pole,
and the discontinuity is finite:
\begin{equation}
I(s+i0) - I(s-i0) = 2 \pi i\,.
\label{q1:disc0}
\end{equation}

\section{QED at one loop}
\label{S:QED}

\subsection{Lagrangian and Feynman rules}
\label{S:QEDl}

First we shall discuss quantum electrodynamics with massless electron.
This theory has many similarities with QCD.
Its Lagrangian is
\begin{equation}
L = \bar{\psi}_0 i\D \psi_0 - \frac{1}{4} F_{0\mu\nu} F_0^{\mu\nu}\,,
\label{QEDl:L}
\end{equation}
where $\psi_0$ is the electron field,
\begin{equation}
D_\mu \psi_0 = \left(\partial_\mu - i e_0 A_{0\mu}\right) \psi_0
\label{QEDl:D}
\end{equation}
is its covariant derivative,
$A_{0\mu}$ is the photon field, and
\begin{equation}
F_{0\mu\nu} = \partial_\mu A_{0\nu} - \partial_\nu A_{0\mu}
\label{QEDl:F}
\end{equation}
is the field strength tensor.

As explained in the textbooks,
it is not possible to obtain the photon propagator from this Lagrangian:
the matrix which should be inverted has zero determinant.
This is due to the gauge invariance;
in order to have a photon propagator,
we have to fix the gauge in some way.
The most popular way (called the covariant gauge)
is to add the term
\begin{equation}
\Delta L = - \frac{1}{2a_0} \left(\partial_\mu A_0^\mu\right)^2
\label{QEDl:DL}
\end{equation}
to the Lagrangian~(\ref{QEDl:L})
($a_0$ is the gauge-fixing parameter).
This additional term does not change physics:
physical results are gauge-invariant,
and do not depend on the choice of the gauge
(in particular, on $a_0$).

\begin{figure}[ht]
\begin{center}
\begin{picture}(50,34)
\put(14,30){\makebox(0,0){\includegraphics{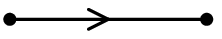}}}
\put(14,28){\makebox(0,0)[t]{$p$}}
\put(30,30){\makebox(0,0)[l]{${}=i S_0(p)$}}
\put(14,22){\makebox(0,0){\includegraphics{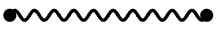}}}
\put(14,20){\makebox(0,0)[t]{$p$}}
\put(1.5,22){\makebox(0,0){$\mu$}}
\put(26,22){\makebox(0,0){$\nu$}}
\put(30,22){\makebox(0,0)[l]{${}=-i D^0_{\mu\nu}(p)$}}
\put(14,7){\makebox(0,0){\includegraphics{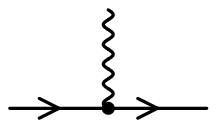}}}
\put(14,14){\makebox(0,0){$\mu$}}
\put(30,2){\makebox(0,0)[l]{${}=i e_0 \gamma^\mu$}}
\end{picture}
\end{center}
\caption{QED Feynman rules}
\label{F:QEDFr}
\end{figure}

Now we can derive the Feynman rules (Fig.~\ref{F:QEDFr}).
The free electron propagator is
\begin{equation}
S_0(p) = \frac{1}{\rlap/p} = \frac{\rlap/p}{p^2}\,,
\label{QEDl:S0}
\end{equation}
and the free photon propagator is
\begin{equation}
D^0_{\mu\nu}(p) = \frac{1}{p^2}
\left[g_{\mu\nu} - (1-a_0) \frac{p_\mu p_\nu}{p^2}\right]\,.
\label{QEDl:D0}
\end{equation}

The Lagrangian~(\ref{QEDl:L}), (\ref{QEDl:DL})
contains the bare fields $\psi_0$, $A_0$
and the bare parameters $e_0$, $a_0$.
The renormalized quantities are related to them by
\begin{equation}
\psi_0 = Z_\psi^{1/2} \psi\,,\quad
A_0 = Z_A^{1/2} A\,,\quad
a_0 = Z_A a\,,\quad
e_0 = Z_\alpha^{1/2} e
\label{QEDl:ren}
\end{equation}
(we shall see why the same renormalization constant $Z_A$
describes renormalization of both the photon field $A$
and the gauge-fixing parameter $a$ in Sect.~\ref{S:Photon}).
We shall mostly use the \textit{minimal renormalization scheme} (\MS).
In this scheme, renormalization constants have the structure
\begin{equation}
Z_i(\alpha) = 1 + \frac{z_1}{\varepsilon} \frac{\alpha}{4\pi}
+ \left(\frac{z_{22}}{\varepsilon^2} + \frac{z_{21}}{\varepsilon}\right)
\left(\frac{\alpha}{4\pi}\right)^2 + \cdots
\label{QEDl:min}
\end{equation}
They contain neither finite parts ($\varepsilon^0$)
nor positive powers of $\varepsilon$,
just $1/\varepsilon^n$ terms
necessary to remove $1/\varepsilon^n$ divergences
in renormalized (physical) results.
Sometimes, other renormalization schemes are useful, see Sect.~\ref{S:OS}.

In order to write renormalization constants~(\ref{QEDl:min}),
we must define the renormalized coupling $\alpha$
in such a way that it is exactly dimensionless at any $d$.
The action is dimensionless, because it appears in the exponent
in the Feynman path integral.
The action is an integral of $L$ over $d$-dimensional space--time;
therefore, the dimensionality of the Lagrangian $L$ is $[L]=d$
(in mass units).
From the kinetic terms in the Lagrangian~(\ref{QEDl:L})
we obtain the dimensionalities of the fields:
$[A_0]=1-\varepsilon$, $[\psi_0]=3/2-\varepsilon$.
Both terms in the covariant derivative~(\ref{QEDl:D})
must have the same dimensionality 1, therefore $[e_0]=\varepsilon$.
In order to define a dimensionless coupling $\alpha$,
we have to introduce a parameter $\mu$ with the dimensionality of mass
(called the renormalization scale):
\begin{equation}
\frac{\alpha(\mu)}{4\pi} = \mu^{-2\varepsilon}
\frac{e^2}{(4\pi)^{d/2}} e^{-\gamma\varepsilon}\,,
\label{QEDl:alpha}
\end{equation}
where $\gamma$ is the Euler constant.
In practice, this equation is more often used in the opposite direction:
\begin{equation}
\frac{e_0^2}{(4\pi)^{d/2}} = \mu^{2\varepsilon}
\frac{\alpha(\mu)}{4\pi} Z_\alpha(\alpha(\mu)) e^{\gamma\varepsilon}\,.
\label{QEDl:e2}
\end{equation}
We first calculate some physical quantity in terms of the bare charge $e_0$,
and then re-express it via the renormalized $\alpha(\mu)$.

\subsection{Photon propagator}
\label{S:Photon}

\begin{figure}[ht]
\begin{center}
\begin{picture}(111,9)
\put(55.5,4.5){\makebox(0,0){\includegraphics{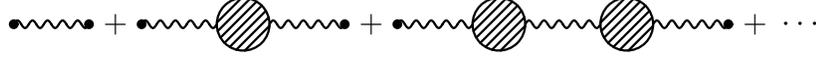}}}
\put(14.5,4.5){\makebox(0,0){+}}
\put(48.5,4.5){\makebox(0,0){+}}
\put(99.5,4.5){\makebox(0,0){+}}
\put(103,4.5){\makebox(0,0)[l]{$\cdots$}}
\end{picture}
\end{center}
\caption{Photon propagator}
\label{F:photon}
\end{figure}

The photon propagator has the structure (Fig.~\ref{F:photon})
\begin{equation}
\begin{split}
-i D_{\mu\nu}(p) ={}& -i D^0_{\mu\nu}(p)
+ (-i)D^0_{\mu\alpha}(p) i\Pi^{\alpha\beta}(p) (-i)D^0_{\beta\nu}(p)\\
&{} + (-i)D^0_{\mu\alpha}(p) i\Pi^{\alpha\beta}(p) (-i) D^0_{\beta\gamma}(p)
i\Pi^{\gamma\delta}(p) (-i)D^0_{\gamma\nu}(p) + \cdots
\end{split}
\label{Photon:Dyson1}
\end{equation}
where the photon self-energy $i\Pi_{\mu\nu}(p)$
(denoted by a shaded blob in Fig.~\ref{F:photon})
is the sum of all one-particle-irreducible diagrams
(diagrams which cannot be cut into two disconnected pieces
by cutting a single photon line),
not including the external photon propagators.

The series~(\ref{Photon:Dyson1}) can also be rewritten as an equation
\begin{equation}
D_{\mu\nu}(p) = D^0_{\mu\nu}(p)
+ D^0_{\mu\alpha}(p) \Pi^{\alpha\beta}(p) D_{\beta\nu}(p)\,.
\label{Photon:Dyson2}
\end{equation}
In order to solve this equation,
let's introduce, for any tensor of the form
\begin{equation*}
A_{\mu\nu} = A_\bot \left[g_{\mu\nu}-\frac{p_\mu p_\nu}{p^2}\right]
+ A_{||} \frac{p_\mu p_\nu}{p^2}\,,
\end{equation*}
the inverse tensor
\begin{equation*}
A^{-1}_{\mu\nu} = A^{-1}_\bot \left[g_{\mu\nu}-\frac{p_\mu p_\nu}{p^2}\right]
+ A^{-1}_{||} \frac{p_\mu p_\nu}{p^2}
\end{equation*}
satisfying
\begin{equation*}
A^{-1}_{\mu\lambda} A^{\lambda\nu} = \delta_\mu^\nu\,.
\end{equation*}
Then
\begin{equation}
D^{-1}_{\mu\nu}(p) = (D^0)^{-1}_{\mu\nu}(p) - \Pi_{\mu\nu}(p)\,.
\label{Photon:Dyson3}
\end{equation}

As we shall see in Sect.~\ref{S:Ward},
\begin{equation}
\Pi_{\mu\nu}(p) p^\nu = 0\,,\quad
\Pi_{\mu\nu}(p) p^\mu = 0\,.
\label{Photon:Ward1}
\end{equation}
Therefore, the photon self-energy has the form
\begin{equation}
\Pi_{\mu\nu}(p) = (p^2 g_{\mu\nu} - p_\mu p_\nu) \Pi(p^2)\,,
\label{Photon:Ward2}
\end{equation}
and the full photon propagator is
\begin{equation}
D_{\mu\nu}(p) =
\frac{1}{p^2(1-\Pi(p^2))} \left[g_{\mu\nu}-\frac{p_\mu p_\nu}{p^2}\right]
+ a_0 \frac{p_\mu p_\nu}{(p^2)^2}\,.
\label{Photon:D}
\end{equation}
Its longitudinal part gets no corrections,
to all orders of perturbation theory.
The full bare propagator is related to the renormalized one by
\begin{equation*}
D_{\mu\nu}(p) = Z_A(\alpha(\mu)) D^r_{\mu\nu}(p;\mu)\,.
\end{equation*}
Therefore,
\begin{equation}
D^r_{\mu\nu}(p;\mu) =
D^r_\bot(p^2;\mu) \left[g_{\mu\nu}-\frac{p_\mu p_\nu}{p^2}\right]
+ a(\mu) \frac{p_\mu p_\nu}{(p^2)^2}\,.
\label{Photon:Dr}
\end{equation}
The minimal~(\ref{QEDl:min}) renormalization constant $Z_A(\alpha)$
is constructed to make
\begin{equation*}
D^r_\bot(p^2;\mu) = Z_A^{-1}(\alpha(\mu)) \frac{1}{p^2(1-\Pi(p^2))}
\end{equation*}
finite at $\varepsilon\to0$.
But the longitudinal part of~(\ref{Photon:Dr}) containing
\begin{equation*}
a(\mu) = Z_A^{-1}(\alpha(\mu)) a_0
\end{equation*}
must be finite too.
This explains why $Z_A$ appears twice in~(\ref{QEDl:ren}),
in renormalization of $A_0$ and of $a_0$.

\subsection{Ward identity}
\label{S:Ward}

\begin{figure}[ht]
\begin{center}
\begin{picture}(86,12.5)
\put(43,7.75){\makebox(0,0){\includegraphics{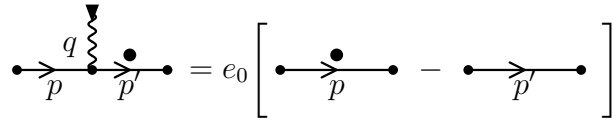}}}
\put(28.5,4){\makebox(0,0){${}=e_0\Biggl[\Biggr.$}}
\put(56,4){\makebox(0,0){$-$}}
\put(80,4){\makebox(0,0){$\Biggl.\Biggr]$}}
\put(6,0){\makebox(0,0)[b]{$p$}}
\put(16,0){\makebox(0,0)[b]{$p'$}}
\put(43.5,0){\makebox(0,0)[b]{$p$}}
\put(68.5,0){\makebox(0,0)[b]{$p'$}}
\put(9,7){\makebox(0,0)[r]{$q$}}
\end{picture}
\end{center}
\caption{Ward identity}
\label{F:Ward0}
\end{figure}

It is easy to check by a direct calculation that (Fig.~\ref{F:Ward0})
\begin{equation}
i S_0(p')\;i e_0 \rlap/q\;i S_0(p) = e_0 \left[ i S_0(p') - i S_0(p) \right]\,.
\label{Ward:Ward0}
\end{equation}
Here $q=p'-p$; substituting $\rlap/q$ and using $S_0(p)=1/\rlap/p$,
we immediately obtain~(\ref{Ward:Ward0}).
We shall use graphical notation (Fig.~\ref{F:Ward0}):
a photon line with a black triangle at the end
means an external gluon leg (no propagator!)
contracted in its polarization index
with the incoming photon momentum $q$.
A dot near an electron line means that its momentum is shifted by $q$,
as compared to the diagram without the longitudinal photon insertion.
For an infinitesimal $q$, we obtain from~(\ref{Ward:Ward0})
a useful identity
\begin{equation}
\frac{\partial S_0(p)}{\partial p^\mu} = - S_0(p) \gamma_\mu S_0(p)\,.
\label{Ward:Ward00}
\end{equation}

Let's calculate $\Pi_{\mu\nu}(p) p^\nu$~(\ref{Photon:Ward1}) at one loop:
\begin{equation}
\raisebox{-5.25mm}{\includegraphics{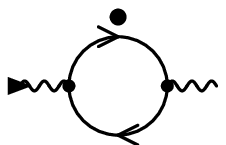}}
= e_0 \left[ \raisebox{-5.25mm}{\includegraphics{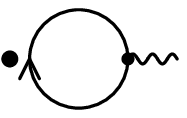}}
- \raisebox{-5.25mm}{\includegraphics{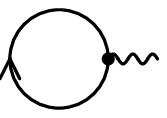}} \right]
= 0\,.
\label{Ward:Ward1}
\end{equation}
Two integrals here differ by a shift of the integration momentum,
and hence are equal.

Now we'll do the same at two loops.
All two-loop diagrams for $\Pi_{\mu\nu}(p) p^\nu$
can be obtained from one two-loop diagram with a single photon leg,
and using the Ward identity (Fig.~\ref{F:Ward0}) we obtain
\begin{equation}
\begin{split}
&\raisebox{-16.25mm}{\includegraphics{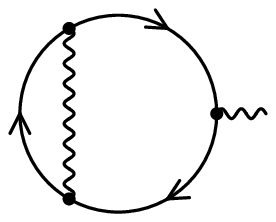}} \Rightarrow{}\\
&\raisebox{-16.25mm}{\includegraphics{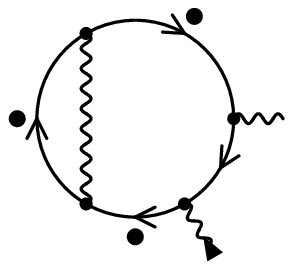}}
+ \raisebox{-16.25mm}{\includegraphics{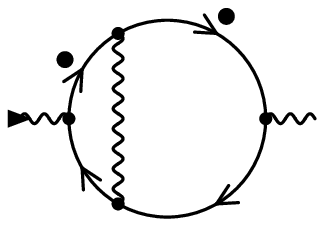}}
+ \raisebox{-16.25mm}{\includegraphics{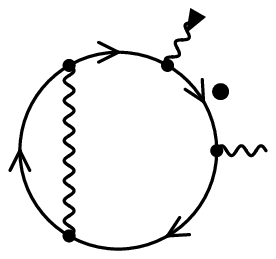}}\\
&{} = e_0 \left[ \raisebox{-16.25mm}{\includegraphics{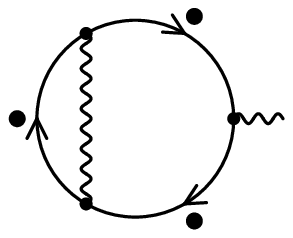}}
- \raisebox{-16.25mm}{\includegraphics{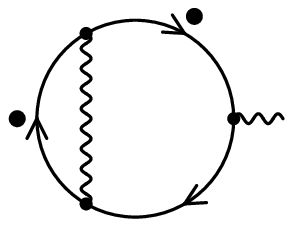}} \right.\\
&\hphantom{{}=e_0\Biggl[\Biggr.} + \raisebox{-16.25mm}{\includegraphics{wpf.eps}}
- \raisebox{-16.25mm}{\includegraphics{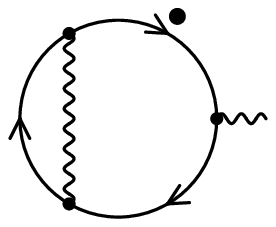}}\\
&\hphantom{{}=e_0\Biggl[\Biggr.} \left.{}
+ \raisebox{-16.25mm}{\includegraphics{wpg.eps}}
- \raisebox{-16.25mm}{\includegraphics{wpa.eps}} \right]
= 0\,.
\end{split}
\label{Ward:Ward2}
\end{equation}
All diagrams cancel pairwise, except the first one and the last one;
these two diagrams differ only by a shift in an integration momentum.

It is clear that this proof works at any order of perturbation theory.

\subsection{Photon self-energy}
\label{S:Photon1}

\begin{figure}[ht]
\begin{center}
\begin{picture}(64,38)
\put(32,19){\makebox(0,0){\includegraphics{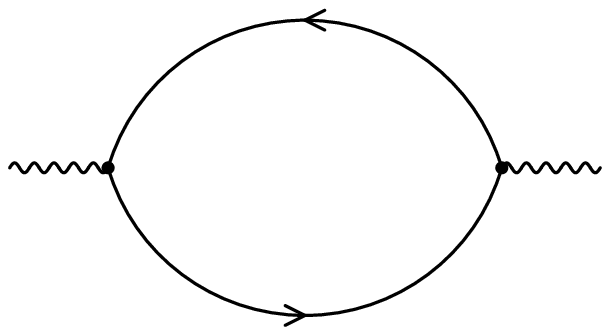}}}
\put(32,0){\makebox(0,0)[b]{$k+p$}}
\put(32,38){\makebox(0,0)[t]{$k$}}
\end{picture}
\end{center}
\caption{One-loop photon self-energy}
\label{F:Photon1}
\end{figure}

Now we shall explicitly calculate photon self-energy
at one loop (Fig.~\ref{F:Photon1}).
The fermion loop gives the factor $-1$, and
\begin{equation}
i (p^2 g_{\mu\nu} - p_\mu p_\nu) \Pi(p^2)
= - \int \frac{d^d k}{(2\pi)^d} \Tr i e_0 \gamma_\mu i \frac{\rlap/k+\rlap/p}{(k+p)^2}
i e_0 \gamma_\nu \frac{\rlap/k}{k^2}\,.
\label{Photon1:Pi0}
\end{equation}
To simplify finding the scalar function $\Pi(p^2)$,
we contract in $\mu$ and $\nu$.
In $d$-dimensional space--time
\begin{equation}
\delta_\mu^\mu = d\,,
\label{Photon1:dmm}
\end{equation}
and we obtain
\begin{equation}
\Pi(p^2) = \frac{-i e_0^2}{(d-1) (-p^2)}
\int \frac{d^d k}{(2\pi)^d}
\frac{\Tr \gamma_\mu (\rlap/k+\rlap/p) \gamma^\mu \rlap/k}%
{\left[-(k+p)^2\right] (-k^2)}\,.
\label{Photon1:Pi1}
\end{equation}

Now we make a short digression and discuss $\gamma$ matrices in $d$ dimensions.
Their defining property is
\begin{equation}
\gamma^\mu \gamma^\nu + \gamma^\nu \gamma^\mu = 2 g^{\mu\nu}\,.
\label{Photon1:gamma}
\end{equation}
Therefore,
\begin{equation}
\gamma_\mu \gamma^\mu = d\,.
\label{Photon1:gamma0}
\end{equation}
How to find $\gamma_\mu\rlap/a\gamma^\mu$?
We anticommute $\gamma^\mu$ to the left:
\begin{equation}
\gamma_\mu \rlap/a \gamma^\mu
= \gamma_\mu (-\gamma^\mu\rlap/a + 2a^\mu)
= - (d-2) \rlap/a\,.
\label{Photon1:gamma1}
\end{equation}
Similarly,
\begin{equation}
\gamma_\mu \rlap/a \rlap/b \gamma^\mu
= \gamma_\mu \rlap/a (-\gamma^\mu\rlap/b + 2b^\mu)
= (d-2) \rlap/a \rlap/b + 2 \rlap/b \rlap/a
= 4 a \cdot b + (d-4) \rlap/a \rlap/b\,,
\label{Photon1:gamma2}
\end{equation}
and
\begin{equation}
\gamma_\mu \rlap/a \rlap/b \rlap/c \gamma^\mu
= \gamma_\mu \rlap/a \rlap/b (-\gamma^\mu\rlap/c + 2c^\mu)
= - 4 a\cdot b - (d-4) \rlap/a \rlap/b \rlap/c + 2 \rlap/c \rlap/a \rlap/b
= - 2 \rlap/c \rlap/b \rlap/a - (d-4) \rlap/a \rlap/b \rlap/c\,.
\label{Photon1:gamma3}
\end{equation}
The usual convention is\footnote{It would be more natural to use $2^{d/2}$,
because $\gamma$ matrices can be defined in any even-dimensional space,
and they are $2^{d/2}\times2^{d/2}$ matrices;
moreover, with this definition Fierz identities can be generalized
to $d$ dimensions.}
\begin{equation}
\Tr 1 = 4\,,
\label{Photon1:Tr1}
\end{equation}
then the usual formulas for traces of 2, 4,\dots{} $\gamma$ matrices hold.

Now we can find the trace in the numerator of~(\ref{Photon1:Pi1}):
\begin{equation}
\Pi(p^2) = \frac{d-2}{d-1} \frac{i e_0^2}{-p^2}
\int \frac{d^d k}{(2\pi)^d}
\frac{4 (k+p)\cdot k}{\left[-(k+p)^2\right] (-k^2)}\,.
\label{Photon1:Pi2}
\end{equation}
We can set $-p^2=1$; it is easy to restore the power of $-p^2$
by dimensionality at the end of calculation.
The denominators~(\ref{q1:G1def}) are
\begin{equation*}
D_1 = - (k+p)^2\,,\quad
D_2 = - k^2\,.
\end{equation*}
All scalar products in the numerator
can be expressed via the denominators:
\begin{equation}
p^2 = -1\,,\quad
k^2 = - D_2\,,\quad
p\cdot k = \frac{1}{2} \left( 1 + D_2 - D_1 \right)\,.
\label{Photon1:Mult}
\end{equation}
We obtain
\begin{equation}
\Pi(p^2) = 2  \frac{d-2}{d-1} i e_0^2
\int \frac{d^d k}{(2\pi)^d} \frac{-2D_2+1+D_2-D_1}{D_1 D_2}\,.
\label{Photon1:Pi3}
\end{equation}
Terms with $D_1$ or $D_2$ in the numerator can be omitted,
because integrals with a single massless denominator (Fig.~\ref{F:q0})
vanish.

Restoring the power of $-p^2$,
we arrive at the final result
\begin{equation}
\Pi(p^2) = - \frac{e_0^2 (-p^2)^{-\varepsilon}}{(4\pi)^{d/2}}
2 \frac{d-2}{d-1} G_1\,,
\label{Photon1:Res1}
\end{equation}
or, recalling~(\ref{q1:g1}),
\begin{equation}
\Pi(p^2) = \frac{e_0^2 (-p^2)^{-\varepsilon}}{(4\pi)^{d/2}}
4 \frac{d-2}{(d-1)(d-3)(d-4)} g_1\,.
\label{Photon1:Res2}
\end{equation}

\subsection{Photon field renormalization}
\label{S:PhotonZ}

The transverse part of the photon propagator~(\ref{Photon:D}) is,
with the one-loop accuracy,
\begin{equation}
p^2 D_\bot(p^2) = \frac{1}{1-\Pi(p^2)}
= 1 + \frac{e_0^2 (-p^2)^{-\varepsilon}}{(4\pi)^{d/2}}
4 \frac{d-2}{(d-1)(d-3)(d-4)} g_1\,.
\label{PhotonZ:D}
\end{equation}
Re-expressing it via the renormalized $\alpha(\mu)$~(\ref{QEDl:e2}),
we obtain
\begin{equation}
\begin{split}
&p^2  D_\bot(p^2) = 1
+ \frac{\alpha(\mu)}{4\pi} e^{-L\varepsilon} e^{\gamma\varepsilon} g_1\,
4 \frac{d-2}{(d-1)(d-3)(d-4)}\,,\\
&L = \log \frac{-p^2}{\mu^2}\,.
\end{split}
\label{PhotonZ:Da}
\end{equation}

We want to expand this at $\varepsilon\to0$.
Using
\begin{equation}
\Gamma(1+\varepsilon) = \exp \left[ - \gamma \varepsilon
+ \sum_{n=2}^\infty \frac{(-1)^n \zeta_n}{n} \varepsilon^n \right]\,,
\label{PhotonZ:Gamma}
\end{equation}
where $\gamma$ is the Euler constant, and
\begin{equation}
\zeta_n = \sum_{k=1}^\infty \frac{1}{k^n}
\label{PhotonZ:zeta}
\end{equation}
is the Riemann $\zeta$ function,
\begin{equation}
\zeta_2 = \frac{\pi^2}{6}\,,\quad
\zeta_3 \approx 1.202\,,\quad
\zeta_4 = \frac{\pi^4}{90}\,,\quad
\cdots
\end{equation}
we see that
\begin{equation*}
e^{\gamma\varepsilon}g_1 = 1 + \mathcal{O}(\varepsilon^2)\,.
\end{equation*}
This is exactly the reason of including $\exp(-\gamma\varepsilon)$
into the definition~(\ref{QEDl:alpha}).

We obtain
\begin{equation}
p^2  D_\bot(p^2) = 1 - \frac{4}{3} \frac{\alpha(\mu)}{4\pi\varepsilon}
\left[1 - \left(L-\frac{5}{3}\right) \varepsilon + \cdots\right]\,.
\label{PhotonZ:De}
\end{equation}
This should be equal to $Z_A(\alpha(\mu)) p^2 D^r_\bot(p^2;\mu)$,
where $Z_A(\alpha)$ is a minimal~(\ref{QEDl:min}) renormalization constant,
and $D^r_\bot(p^2;\mu)$ is finite at $\varepsilon\to0$.
It is easy to see that
\begin{equation}
Z_A(\alpha) = 1 - \frac{4}{3} \frac{\alpha}{4\pi\varepsilon}\,,
\label{PhotonZ:ZA}
\end{equation}
and
\begin{equation}
p^2 D^r_\bot(p^2;\mu) = 1 + \frac{4}{3} \frac{\alpha(\mu)}{4\pi}
\left(L - \frac{5}{3}\right)\,.
\label{PhotonZ:Dr}
\end{equation}

The bare propagator $D_\bot(p^2)=Z_A(\alpha(\mu))D^r_\bot(p^2;\mu)$
does not depend on $\mu$.
Differentiating it in $\log\mu$
we obtain the renormalization group (RG) equation
\begin{equation}
\frac{\partial D^r_\bot(p^2;\mu)}{\partial\log\mu}
+ \gamma_A(\alpha(\mu)) D^r_\bot(p^2;\mu) = 0\,,
\label{PhotonZ:RG}
\end{equation}
where the anomalous dimension is defined by
\begin{equation}
\gamma_A(\alpha(\mu)) = \frac{d\log Z_A(\alpha(\mu))}{d\log\mu}\,.
\label{PhotonZ:gamma}
\end{equation}

For any minimal renormalization constant
\begin{equation*}
Z_i(\alpha) = 1 + z_1 \frac{\alpha}{4\pi\varepsilon} + \cdots
\end{equation*}
using~(\ref{QEDl:alpha})
\begin{equation*}
\frac{d\log\alpha(\mu)}{d\log\mu} = - 2 \varepsilon + \cdots
\end{equation*}
we find the corresponding anomalous dimension
\begin{equation*}
\gamma_i(\alpha(\mu)) = \frac{d\log Z_i(\alpha(\mu))}{d\log\mu}
= \gamma_0 \frac{\alpha(\mu)}{4\pi} + \cdots
\end{equation*}
to be
\begin{equation*}
\gamma_i(\alpha) = - 2 z_1 \frac{\alpha}{4\pi} + \cdots
\end{equation*}
In other words, the renormalization constant with the one-loop accuracy is
\begin{equation*}
Z_i(\alpha) = 1 - \frac{\gamma_0}{2} \frac{\alpha}{4\pi\varepsilon} + \cdots
\end{equation*}
The anomalous dimension of the photon field is, from~(\ref{PhotonZ:ZA}),
\begin{equation}
\gamma_A(\alpha) = \frac{8}{3} \frac{\alpha}{4\pi} + \cdots
\label{PhotonZ:gammaA}
\end{equation}

The dependence of the renormalized propagator~(\ref{PhotonZ:Dr}) on $L$
can be completely determined from the RG equation.
We can rewrite it as
\begin{equation}
\frac{\partial p^2 D^r_\bot}{\partial L} = \frac{\gamma_A}{2} p^2 D^r_\bot\,.
\label{PhotonZ:RGL}
\end{equation}
Solving this equation with the initial condition
\begin{equation*}
p^2 D^r_\bot(p^2;\mu^2=-p^2) = 1 - \frac{20}{9} \frac{\alpha(\mu)}{4\pi}
\end{equation*}
at $L=0$, we can reconstruct~(\ref{PhotonZ:Dr}).

We can also write the RG equation for $a(\mu)$.
The bare gauge-fixing parameter $a_0=Z_A(\alpha(\mu))a(\mu)$
does not depend on $\mu$.
Differentiating it in $\log\mu$, we obtain
\begin{equation}
\frac{d a(\mu)}{d\log\mu} + \gamma_A(\alpha(\mu)) a(\mu) = 0\,.
\label{PhotonZ:RGa}
\end{equation}

\subsection{Electron propagator}
\label{S:Electron}

\begin{figure}[ht]
\begin{center}
\begin{picture}(111,9)
\put(55.5,4.5){\makebox(0,0){\includegraphics{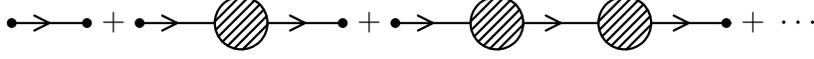}}}
\put(14.5,4.5){\makebox(0,0){${}+{}$}}
\put(48.5,4.5){\makebox(0,0){${}+{}$}}
\put(99.5,4.5){\makebox(0,0){+}}
\put(103,4.5){\makebox(0,0)[l]{$\cdots$}}
\end{picture}
\end{center}
\caption{Electron propagator}
\label{F:electron}
\end{figure}

The electron propagator has the structure (Fig.~\ref{F:electron})
\begin{equation}
i S(p) = i S_0(p) + i S_0(p) (-i) \Sigma(p) i S_0(p)
+ i S_0(p) (-i) \Sigma(p) i S_0(p) (-i) \Sigma(p) i S_0(p)
+ \cdots
\label{Electron:Dyson1}
\end{equation}
where the electron self-energy $-i\Sigma(p)$
is the sum of all one-particle-irreducible diagrams
(diagrams which cannot be cut into two disconnected pieces
by cutting a single electron line),
not including the external electron propagators.

The series~(\ref{Electron:Dyson1}) can also be rewritten as an equation
\begin{equation}
S(p) = S_0(p) + S_0(p) \Sigma(p) S(p)\,.
\label{Electron:Dyson2}
\end{equation}
Its solution is
\begin{equation}
S(p) = \frac{1}{S_0^{-1}(p)-\Sigma(p)}\,.
\label{Electron:Dyson3}
\end{equation}
Electron self-energy $\Sigma(p)$ depends on a single vector $p$,
and can have two $\gamma$-matrix structures: 1 and $\rlap/p$.
When electron is massless, any diagram for $\Sigma$
contains an odd number of $\gamma$ matrices,
and the structure 1 cannot appear:
\begin{equation}
\Sigma(p) = \rlap/p \Sigma_V(p^2)\,.
\label{Electron:SigmaV}
\end{equation}
This is due to helicity conservation.
In massless QED, the electrons with helicity $\lambda=\mp\frac{1}{2}$,
\begin{equation*}
\psi_{L,R} = \frac{1\pm\gamma_5}{2} \psi\,,
\end{equation*}
cannot transform into each other.
Operators with an odd number of $\gamma$ matrices,
like~(\ref{Electron:SigmaV}), conserve helicity,
and those with an even number of $\gamma$ matrices flip helicity.
Therefore, the massless electron propagator has the form
\begin{equation}
S(p) = \frac{1}{1-\Sigma_V(p^2)} \frac{1}{\rlap/p}\,.
\label{Electron:Dyson0}
\end{equation}

\begin{figure}[ht]
\begin{center}
\begin{picture}(64,25.5)
\put(32,12.75){\makebox(0,0){\includegraphics{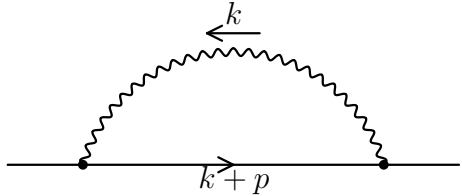}}}
\put(32,0){\makebox(0,0)[b]{$k+p$}}
\put(32,25.5){\makebox(0,0)[t]{$k$}}
\end{picture}
\end{center}
\caption{One-loop electron self-energy}
\label{F:Electron1}
\end{figure}

Let's calculate electron self-energy at one loop (Fig.~\ref{F:Electron1}):
\begin{equation}
-i \rlap/p \Sigma_V(p^2) = \int \frac{d^d k}{(2\pi)^d}
i e_0 \gamma^\mu i \frac{\rlap/k+\rlap/p}{(k+p)^2} i e_0 \gamma^\nu
\frac{-i}{k^2} \left( g_{\mu\nu} - \xi \frac{k_\mu k_\nu}{k^2} \right)\,,
\label{Electron:Sigma0}
\end{equation}
where we introduced the notation
\begin{equation}
\xi = 1 - a_0\,.
\label{Electron:xi}
\end{equation}
To find the scalar function $\Sigma_V(p^2)$,
we take $\frac{1}{4}\Tr\rlap/p$ of both sides:
\begin{equation}
\begin{split}
&\Sigma_V(p^2) = \frac{i e_0^2}{-p^2} \int \frac{d^d k}{(2\pi)^d}
\frac{N}{D_1 D_2}\,,\\
&N = \frac{1}{4} \Tr \rlap/p \gamma^\mu (\rlap/k+\rlap/p) \gamma^\nu
\left( g_{\mu\nu} + \xi \frac{k_\mu k_\nu}{D_2} \right)\,.
\end{split}
\label{Electron:Sigma1}
\end{equation}
Using the ``multiplication table''~(\ref{Photon1:Mult})
we obtain
\begin{equation}
\begin{split}
N &{}= \frac{1}{4} \Tr \rlap/p \gamma_\mu (\rlap/k+\rlap/p) \gamma^\mu
+ \frac{\xi}{D_2}\,\frac{1}{4} \Tr \rlap/p \rlap/k (\rlap/k+\rlap/p) \rlap/k\\
&{}= - (d-2) (p^2 + p\cdot k)
+ \frac{\xi}{D_2} \left[ k^2 p\cdot k + 2 (p\cdot k)^2 - p^2 k^2 \right]\\
&{}\Rightarrow \frac{1}{2} \left[ d-2 + \xi \left( \frac{1}{D_2} - 1 \right) \right]\,,
\end{split}
\label{Electron:N}
\end{equation}
where terms with $D_1$ or $D_2$ in the numerator were omitted.
Restoring the power of $-p^2$, we have
\begin{equation}
\Sigma_V(p^2) = - \frac{e_0^2 (-p^2)^{-\varepsilon}}{(4\pi)^{d/2}}
\frac{1}{2} \left[ (d-2-\xi) G(1,1) + \xi G(1,2) \right]\,.
\label{Electron:Sigma2}
\end{equation}

It is easy to derive a nice property
\begin{equation}
\frac{G(n_1,n_2+1)}{G(n_1,n_2)} =
- \frac{(d-2n_1-2n_2)(d-n_1-n_2-1)}{n_2(d-2n_2-2)}
\label{Electron:Gratio}
\end{equation}
from the definition~(\ref{q1:G1}) of $G(n_1,n_2)$.
In particular,
\begin{equation}
\frac{G(1,2)}{G(1,1)} = - (d-3)\,,
\label{Electron:Gratio2}
\end{equation}
and we arrive at
\begin{equation}
\Sigma_V(p^2) = - \frac{e_0^2 (-p^2)^{-\varepsilon}}{(4\pi)^{d/2}}
\frac{d-2}{2} a_0 G_1\,.
\label{Electron:Res1}
\end{equation}
Note that the one-loop electron self-energy vanishes
in the Landau gauge $a_0=0$.
Recalling~(\ref{q1:g1}), we can rewrite the result as
\begin{equation}
\Sigma_V(p^2) = \frac{e_0^2 (-p^2)^{-\varepsilon}}{(4\pi)^{d/2}}
\frac{d-2}{(d-3)(d-4)} a_0 g_1\,.
\label{Electron:Res2}
\end{equation}

The electron propagator~(\ref{Electron:Dyson0}), with one-loop accuracy,
expressed via the renormalized quantities $\alpha(\mu)$~(\ref{QEDl:e2})
and $a(\mu)$~(\ref{QEDl:ren})
(we may take $a(\mu)=a_0$,
because it only appears in the $\alpha$ correction) is
\begin{equation}
\begin{split}
\rlap/p S(p) &{}= 1 + \frac{\alpha(\mu)}{4\pi} e^{-L\varepsilon} e^{\gamma\varepsilon}
g_1\,a(\mu) \frac{d-2}{(d-3)(d-4)}\\
&{}= 1 - \frac{\alpha(\mu)}{4\pi\varepsilon} a(\mu) e^{-L\varepsilon}
(1 + \varepsilon + \cdots)\,.
\end{split}
\label{Electron:Se}
\end{equation}
It should be equal to $Z_\psi(\alpha(\mu),a(\mu))\rlap/p S_r(p;\mu)$,
where $Z_\psi(\alpha)$ is a minimal~(\ref{QEDl:min}) renormalization constant,
and $S_r(p;\mu)$ is finite at $\varepsilon\to0$.
It is easy to see that
\begin{equation}
Z_\psi(\alpha,a) = 1 - a \frac{\alpha}{4\pi\varepsilon}
\label{Electron:Zpsi}
\end{equation}
and
\begin{equation}
\rlap/p S_r(p;\mu) = 1 + a(\mu) (L-1) \frac{\alpha(\mu)}{4\pi}\,.
\label{Electron:Sr}
\end{equation}
The bare propagator $S(p)=Z_\psi(\alpha(\mu))S_r(p;\mu)$
does not depend on $\mu$.
Differentiating it in $\log\mu$
we obtain the RG equation
\begin{equation}
\frac{\partial S_r(p;\mu)}{\partial\log\mu}
+ \gamma_\psi(\alpha(\mu),a(\mu)) S_r(p;\mu) = 0\,,
\label{Electron:RG}
\end{equation}
where the anomalous dimension
\begin{equation}
\gamma_\psi(\alpha(\mu),a(\mu))
= \frac{d\log Z_\psi(\alpha(\mu),a(\mu))}{d\log\mu}
\label{Electron:gamma}
\end{equation}
is
\begin{equation}
\gamma_\psi(\alpha,a) = 2 a \frac{\alpha}{4\pi} + \cdots
\label{Electron:gammapsi}
\end{equation}

\subsection{Vertex and charge renormalization}
\label{S:Vertex}

\begin{figure}[ht]
\begin{center}
\begin{picture}(50,24)
\put(15,10){\makebox(0,0){\includegraphics{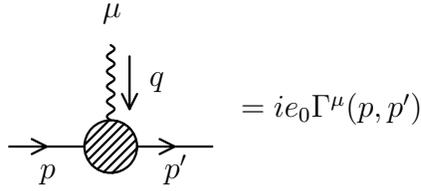}}}
\put(6.5,0){\makebox(0,0)[b]{$p$}}
\put(23.5,0){\makebox(0,0)[b]{$p'$}}
\put(20,13.5){\makebox(0,0)[l]{$q$}}
\put(15,24){\makebox(0,0)[t]{$\mu$}}
\put(31,10){\makebox(0,0)[l]{${}=i e_0 \Gamma^\mu(p,p')$}}
\end{picture}
\end{center}
\caption{Electron--photon vertex}
\label{F:Vertex}
\end{figure}

Let the sum of all one-particle-irreducible vertex diagrams,
i.e. diagrams which cannot be cut into disconnected pieces
by cutting a single electron or photon line,
not including the external propagators,
be the vertex $i e_0 \Gamma^\mu(p,p')$ (Fig.~\ref{F:Vertex}).
It can be written as
\begin{equation}
\Gamma^\mu(p,p') = \gamma^\mu + \Lambda^\mu(p,p')\,,
\label{Vertex:Gamma}
\end{equation}
where $\Lambda^\mu(p,p')$ starts from one loop.

When expressed via renormalized quantities,
the vertex should be equal to
\begin{equation}
\Gamma^\mu = Z_\Gamma \Gamma_r^\mu\,,
\label{Vertex:ZGamma}
\end{equation}
where $Z_\Gamma$ is a minimal~(\ref{QEDl:min})
renormalization constant,
and the renormalized vertex $\Gamma_r^\mu$
is finite at $\varepsilon\to0$.

\begin{figure}[ht]
\begin{center}
\begin{picture}(102,32)
\put(51,16){\makebox(0,0){\includegraphics{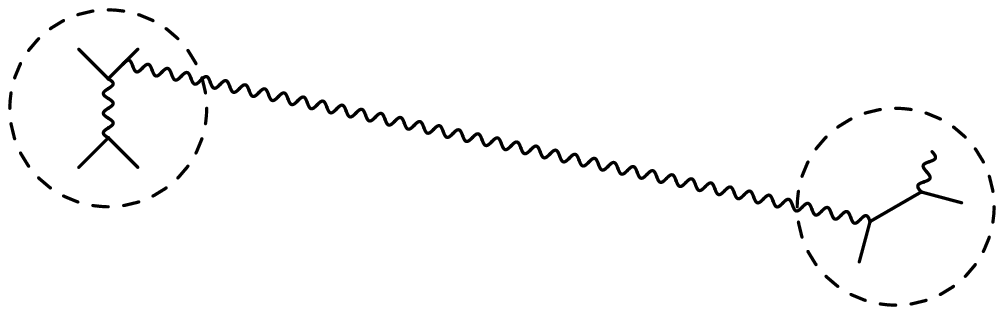}}}
\put(11,6){\makebox(0,0){\Large Far star}}
\put(91,26){\makebox(0,0){\Large Laboratory}}
\put(26,28){\makebox(0,0){\Large$Z_A^{1/2}$}}
\put(75,6){\makebox(0,0){\Large$Z_A^{1/2}$}}
\end{picture}
\end{center}
\caption{Scattering of a photon emitted in a far star}
\label{F:Star}
\end{figure}

In order to obtain a physical scattering amplitude
($S$-matrix element),
one should calculate the corresponding vertex
(one-particle-irreducible, external propagators are not included)
and multiply it by the spin wave functions of the external particles
and by the field renormalization constants $Z_i^{1/2}$
for each external particle $i$.
We can understand this rule in the following way.
In fact, there are no external legs, only propagators.
Suppose we are studying photon scattering in the laboratory.
This photon has been emitted somewhere.
Even if it was emitted in a far star (Fig.~\ref{F:Star}),
there is a photon propagator from the far star to the laboratory.
The (bare) propagator contains the factor $Z_A$.
We split it into $Z_A^{1/2}\cdot Z_A^{1/2}$,
and put one factor $Z_A^{1/2}$
into the emission process in the far star,
and the other factor $Z_A^{1/2}$
into the scattering process in the laboratory.

The physical matrix element $e_0 \Gamma Z_\psi Z_A^{1/2}$
of photon emission (or absorption) by an electron
must be finite at $\varepsilon\to0$.
Strictly speaking, it is not an $S$-matrix element,
because at least one particle must be off-shell,
but this does not matter.
This matrix element can be rewritten as
$e \Gamma_r Z_\alpha^{1/2} Z_\Gamma Z_\psi Z_A^{1/2}$.
The renormalized charge $e$
and the renormalized vertex $\Gamma_r$ are finite.
Therefore, the minimal~(\ref{QEDl:min}) renormalization constant
$Z_\alpha^{1/2} Z_\Gamma Z_\psi Z_A^{1/2}$
must be finite, too.
According to the definition,
the only minimal renormalization constant
which is finite at $\varepsilon\to0$ is 1.
Therefore,
\begin{equation}
Z_\alpha = (Z_\Gamma Z_\psi)^{-2} Z_A^{-1}\,.
\label{Vertex:Zalpha}
\end{equation}
In order to obtain the charge renormalization constant $Z_\alpha$,
one has to find the vertex renormalization constant $Z_\Gamma$
and the electron- and photon-field renormalization constants
$Z_\psi$ and $Z_A$.
In fact, the situation in QED is simpler,
because $Z_\Gamma Z_\psi=1$, due to the Ward identity
(Sect.~\ref{S:Ward}).

Starting from each diagram for $-i\Sigma(p)$,
we can construct a set of diagrams for $i e_0 \Lambda^\mu(p,p')$,
by attaching the external photon line
to each electron propagator in turn.
Let's calculate the contribution of this set
to $i e_0 \Lambda^\mu(p,p') q_\mu$
using the Ward identity of Fig.~\ref{F:Ward0}.
As an example, we consider all vertex diagrams generated
from a certain two-loop electron self-energy diagram:
\begin{equation}
\begin{split}
&\raisebox{-6.25mm}{\includegraphics{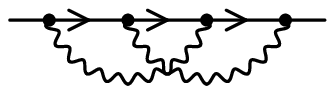}} \Rightarrow{}\\
&\raisebox{-6.25mm}{\includegraphics{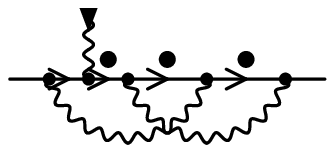}}
+ \raisebox{-6.25mm}{\includegraphics{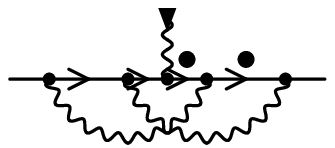}}
+ \raisebox{-6.25mm}{\includegraphics{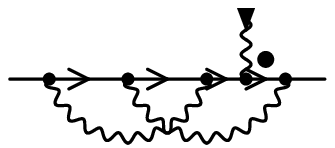}}\\
&{} = e_0 \left[ \raisebox{-6.25mm}{\includegraphics{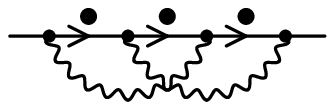}}
- \raisebox{-6.25mm}{\includegraphics{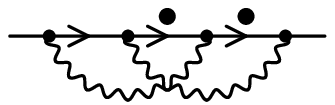}} \right.\\
&\hphantom{{}+e_0\Biggl[\Biggr.}
+ \raisebox{-6.25mm}{\includegraphics{w2.eps}}
- \raisebox{-6.25mm}{\includegraphics{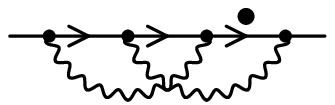}}\\
&\hphantom{{}+e_0\Biggl[\Biggr.}\left.{}
+ \raisebox{-6.25mm}{\includegraphics{w1.eps}}
- \raisebox{-6.25mm}{\includegraphics{w0.eps}} \right]\\
&{} = e_0 \left[ \raisebox{-6.25mm}{\includegraphics{w3.eps}}
- \raisebox{-6.25mm}{\includegraphics{w0.eps}} \right]\,.
\end{split}
\label{Vertex:Ward1}
\end{equation}
Of course, such cancellations happen for vertex diagrams
generated from any self-energy diagram
(and attaching the external photon line
to an electron loop gives 0).
Therefore, we arrive at the Ward--Takahashi identity
\begin{equation}
\Lambda^\mu(p,p') q_\mu = \Sigma(p) - \Sigma(p')\,,
\label{Vertex:Ward2}
\end{equation}
which can also be rewritten as
\begin{equation}
\Gamma^\mu(p,p') q_\mu = S^{-1}(p') - S^{-1}(p)\,.
\label{Vertex:Ward3}
\end{equation}
For $q\to0$, we obtain the Ward identity
\begin{equation}
\Lambda^\mu(p,p) = - \frac{\partial\Sigma(p)}{\partial p_\mu}
\quad\text{or}\quad
\Gamma^\mu(p,p) = \frac{\partial S^{-1}(p)}{\partial p_\mu}\,.
\label{Vertex:Ward4}
\end{equation}

Rewriting~(\ref{Vertex:Ward3}) via the renormalized quantities,
\begin{equation*}
Z_\Gamma \Gamma_r^\mu q_\mu
= Z_\psi^{-1} \left[ S_r^{-1}(p') - S_r^{-1}(p) \right]\,,
\end{equation*}
we see that $Z_\psi Z_\Gamma$ must be finite.
But the only minimal renormalization constant
finite at $\varepsilon\to0$ is 1:
\begin{equation}
Z_\psi Z_\Gamma = 1\,.
\label{Vertex:WardZ}
\end{equation}
Therefore, charge renormalization in QED
is determined by the photon field renormalization:
\begin{equation}
Z_\alpha = Z_A^{-1}\,.
\label{Vertex:ZalphaA}
\end{equation}

\begin{figure}[ht]
\begin{center}
\begin{picture}(26,18)
\put(13,9){\makebox(0,0){\includegraphics{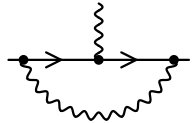}}}
\end{picture}
\end{center}
\caption{One-loop QED vertex}
\label{F:Vertex1}
\end{figure}

We know $Z_\Gamma$ at one loop from the Ward identity~(\ref{Vertex:WardZ})
and $Z_\psi$~(\ref{Electron:Zpsi}).
Nevertheless, let's also find it by a direct calculation.
This will be useful, because we'll have to do
several similar calculations in QCD.
We are only interested in the ultraviolet divergence
of the diagram in Fig.~\ref{F:Vertex1}.
This divergence is logarithmic.
We may nullify all external momenta,
because terms which depend on these momenta are convergent:
\begin{equation}
i e_0 \Lambda^\alpha = \int \frac{d^d k}{(2\pi)^d}
i e_0 \gamma^\mu i \frac{\rlap/k}{k^2} i e_0 \gamma^\alpha
i \frac{\rlap/k}{k^2} i e_0 \gamma^\nu
\frac{-i}{k^2} \left( g_{\mu\nu} - \xi \frac{k_\mu k_\nu}{k^2} \right)\,.
\label{Vertex:Lambda1}
\end{equation}
Of course, we should introduce some infrared regularization,
otherwise this diagram vanishes.
We have
\begin{equation}
\Lambda^\alpha = - i e_0^2 \int \frac{d^d k}{(2\pi)^d}
\frac{\gamma_\mu \rlap/k \gamma^\alpha \rlap/k \gamma^\mu - \xi k^2 \gamma^\alpha}%
{(k^2)^2}\,.
\label{Vertex:Lambda2}
\end{equation}
Averaging over $k$ directions:
\begin{equation*}
\rlap/k \gamma^\alpha \rlap/k
\to \frac{k^2}{d} \gamma_\nu \gamma^\alpha \gamma^\nu\,,
\end{equation*}
we obtain (4-dimensional $\gamma$-matrix algebra may be used)
\begin{equation}
\Lambda^\alpha = - i e_0^2 a_0 \gamma^\alpha \int \frac{d^d k}{(2\pi)^d}
\frac{1}{(-k^2)^2}\,.
\label{Vertex:Lambda3}
\end{equation}

Now let's find the ultraviolet divergence ($1/\varepsilon$)
of this integral,
introducing a sharp infrared cutoff into the Euclidean integral
(we may use $\Omega_4$~(\ref{V1:Omega1}) here)
\begin{equation}
\left. \int \frac{d^d k}{(2\pi)^d} \frac{1}{(-k^2)^2} \right|_{UV}
= \frac{i}{8\pi^2} \int_\lambda^\infty \ke^{-1-2\varepsilon} d\ke
= \frac{i\lambda^{-2\varepsilon}}{(4\pi)^2\varepsilon}
= \frac{i}{(4\pi)^2} \frac{1}{\varepsilon}\,.
\label{Vertex:UV}
\end{equation}
Any infrared regularization can be used;
instead of a cut-off, we could insert a non-zero mass, for example:
\begin{equation}
\left. \int \frac{d^d k}{(2\pi)^d} \frac{1}{(-k^2)^2} \right|_{UV}
= \int \frac{d^d k}{(2\pi)^d} \frac{1}{(m^2-k^2)^2}
= \frac{i m^{-2\varepsilon}}{(4\pi)^2} \Gamma(\varepsilon)
= \frac{i}{(4\pi)^2} \frac{1}{\varepsilon}
\label{Vertex:UVm}
\end{equation}
(see~(\ref{V1:def}), (\ref{V1:res})).

Therefore, the $1/\varepsilon$ part of $\Lambda^\alpha$ is
\begin{equation}
\Lambda^\alpha = a(\mu) \frac{\alpha(\mu)}{4\pi\varepsilon} \gamma^\alpha
\label{Vertex:LambdaRes}
\end{equation}
(it does not depend on momenta).
This calculation can be viewed as the simplest case
of the infrared rearrangement~\cite{V:80}.
We obtain the renormalization constant
\begin{equation}
Z_\Gamma = 1 + a \frac{\alpha}{4\pi\varepsilon}\,,
\label{Vertex:ZGammaRes}
\end{equation}
in agreement with~(\ref{Vertex:WardZ}), (\ref{Electron:Zpsi}).

The bare charge $e_0=Z_\alpha^{1/2}(\alpha(\mu))e(\mu)$ does not depend on $\mu$.
Differentiating~(\ref{QEDl:e2}) in $d\log\mu$,
we obtain the RG equation for $\alpha(\mu)$:
\begin{equation}
\frac{d\log\alpha(\mu)}{d\log\mu} = - 2 \varepsilon - 2 \beta(\alpha(\mu))\,,
\label{Vertex:RG}
\end{equation}
where $\beta$ function is defined by
\begin{equation}
\beta(\alpha_s(\mu))
= \frac{1}{2} \frac{d\log Z_\alpha(\alpha_s(\mu))}{d\log\mu}\,.
\label{Vertex:beta}
\end{equation}
With one-loop accuracy, differentiating
\begin{equation*}
Z_\alpha(\alpha) = 1 + z_1 \frac{\alpha}{4\pi\varepsilon} + \cdots
\end{equation*}
we can retain only the leading term $-2\varepsilon$ in~(\ref{Vertex:RG}):
\begin{equation*}
\beta(\alpha) = \beta_0 \frac{\alpha}{4\pi} + \cdots
= - z_1 \frac{\alpha}{4\pi} + \cdots
\end{equation*}
In other words, at one loop
\begin{equation}
Z_\alpha(\alpha) = 1 - \beta_0 \frac{\alpha}{4\pi\varepsilon} + \cdots
\label{Vertex:Z1}
\end{equation}
We have obtained (see~(\ref{Vertex:ZalphaA}), (\ref{PhotonZ:ZA}))
\begin{equation}
Z_\alpha = Z_A^{-1}
= 1 + \frac{4}{3} \frac{\alpha}{4\pi\varepsilon} + \cdots
\label{Vertex:ZalphaRes}
\end{equation}
Therefore, the QED $\beta$ function is
\begin{equation}
\beta(\alpha) = \beta_0 \frac{\alpha}{4\pi} + \cdots
= - \frac{4}{3} \frac{\alpha}{4\pi} + \cdots
\label{Vertex:betaRes}
\end{equation}

When we consider the renormalized $\alpha(\mu)$
after taking the physical limit $\varepsilon\to0$,
we may omit $-2\varepsilon$ in~(\ref{Vertex:RG}):
\begin{equation}
\frac{d\log\alpha(\mu)}{d\log\mu} = - 2 \beta(\alpha(\mu))\,.
\label{Vertex:RG4}
\end{equation}
In QED $\beta_0=-4/3$,
and the $\beta$-function~(\ref{Vertex:betaRes}) is negative,
at least at sufficiently small $\alpha$,
where perturbation theory is valid.
Therefore, the running $\alpha(\mu)$ grows with $\mu$.
This corresponds to the physical picture of charge screening:
at larger distances (smaller $\mu$)
the charge becomes smaller.

The RG equation~(\ref{Vertex:RG4}) in the one-loop approximation,
\begin{equation}
\frac{d}{d\log\mu}\;\frac{\alpha(\mu)}{4\pi}
= - 2 \beta_0 \left(\frac{\alpha(\mu)}{4\pi}\right)^2\,,
\label{Vertex:RG1}
\end{equation}
can be rewritten as
\begin{equation}
\frac{d}{d\log\mu}\;\frac{4\pi}{\alpha(\mu)}
= 2 \beta_0\,,
\label{Vertex:RG3}
\end{equation}
and its solution is
\begin{equation*}
\frac{4\pi}{\alpha(\mu')} - \frac{4\pi}{\alpha(\mu)}
= 2 \beta_0 \log \frac{\mu'}{\mu}\,.
\end{equation*}
Finally,
\begin{equation}
\alpha(\mu') = \frac{\alpha(\mu)}{\displaystyle
1 + 2 \beta_0 \frac{\alpha(\mu)}{4\pi} \log \frac{\mu'}{\mu}}\,.
\label{Vertex:RG2}
\end{equation}

\subsection{Electron mass}
\label{S:Mass}

Until now we treated electrons as massless.
This is a good approximation if characteristic energies are large
(in the $Z$ region at LEP, say).
Now let's recall that they have mass.
The QED Lagrangian~(\ref{QEDl:L}) now is
\begin{equation}
L = \bar{\psi}_0 \left(i\D - m_0\right) \psi_0 + \cdots
\label{Mass:L}
\end{equation}
and the free electron propagator~(\ref{QEDl:S0}) becomes
\begin{equation}
S_0(p) = \frac{1}{\rlap/p-m_0} = \frac{\rlap/p+m_0}{p^2-m_0^2}\,.
\label{Mass:S0}
\end{equation}
In addition to~(\ref{QEDl:ren}), we have also mass renormalization
\begin{equation}
m_0 = Z_m(\alpha(\mu))\,m(\mu)\,.
\label{Mass:ren}
\end{equation}
The electron self-energy has two $\gamma$-matrix structures,
because helicity is no longer conserved:
\begin{equation}
\Sigma(p) = \rlap/p \Sigma_V(p^2) + m_0 \Sigma_S(p^2)\,,
\label{Mass:Sigma}
\end{equation}
and the electron propagator is
\begin{equation}
S(p) = \frac{1}{\rlap/p - m_0 - \rlap/p \Sigma_V(p^2) - m_0 \Sigma_S(p^2)}
= \frac{1}{1-\Sigma_V(p^2)}\;
\frac{1}{\displaystyle
\rlap/p - \frac{1+\Sigma_S(p^2)}{1-\Sigma_V(p^2)} m_0}\,.
\label{Mass:S}
\end{equation}
It should be equal to $Z_\psi S_r(p;\mu)$.
The renormalization constants are determined by the conditions
\begin{equation}
(1-\Sigma_V) Z_\psi = \text{finite}\,,\quad
\frac{1+\Sigma_S}{1-\Sigma_V} Z_m =\text{finite}\,.
\label{Mass:Z}
\end{equation}
The first of them is the same as in the massless case;
the second one, defining $Z_m$, can be rewritten as
\begin{equation}
(1+\Sigma_S) Z_\psi Z_m = \text{finite}\,.
\label{Mass:Zm}
\end{equation}

The one-loop electron self-energy (Fig.~\ref{F:Electron1}) is
\begin{equation}
-i\Sigma(p) = \int \frac{d^d k}{(2\pi)^d} i e_0 \gamma^\mu
i \frac{\rlap/k+\rlap/p+m_0}{(k+p)^2-m_0^2} i e_0 \gamma^\nu
\frac{-i}{k^2} \left( g_{\mu\nu} - \xi \frac{k_\mu k_\nu}{k^2} \right)\,.
\label{Mass:Sigma1}
\end{equation}
In order to single out $\Sigma_S$, we should retain the $m_0$ term
in the numerator of the electron propagator
(it flips helicity).
At large $p^2$,
\begin{equation}
\begin{split}
\Sigma_S(p^2) &{}= - i e_0^2 \int \frac{d^d k}{(2\pi)^d}
\frac{\gamma^\mu \gamma^\nu}{(k+p)^2 k^2}
\left( g_{\mu\nu} - \xi \frac{k_\mu k_\nu}{k^2} \right)\\
&{}= - i e_0^2 (d-\xi) \int \frac{d^d k}{(2\pi)^d}
\frac{1}{(k+p)^2 k^2}\,.
\end{split}
\label{Mass:Sigma2}
\end{equation}
Retaining only the ultraviolet divergence~(\ref{Vertex:UV})
and re-expressing via renormalized quantities
(this is trivial at this order), we obtain
\begin{equation}
\Sigma_S = (3+a(\mu)) \frac{\alpha(\mu)}{4\pi\varepsilon}\,.
\label{Mass:SigmaS}
\end{equation}
From~(\ref{Mass:Zm}) we see that $a$ terms cancel:
\begin{equation*}
(1+\Sigma_S) Z_\psi Z_m = \left(1 + (3+a) \frac{\alpha}{4\pi\varepsilon}\right)
\left(1 - a \frac{\alpha}{4\pi\varepsilon}\right) Z_m = 1\,,
\end{equation*}
and $Z_m$ is gauge-independent:
\begin{equation}
Z_m = 1 - 3 \frac{\alpha}{4\pi\varepsilon} + \cdots
\label{Mass:ZmRes}
\end{equation}

The bare mass $m_0=Z_m(\alpha(\mu))m(\mu)$ does not depend on $\mu$:
\begin{equation}
\frac{d m(\mu)}{d\log\mu} + \gamma_m(\alpha(\mu)) m(\mu) = 0\,,
\label{Mass:RG}
\end{equation}
where the mass anomalous dimension is
\begin{equation}
\gamma_m(\alpha(\mu)) = \frac{d\log Z_m(\alpha(\mu))}{d\log\mu}\,.
\label{Mass:gamma}
\end{equation}
From~(\ref{Mass:ZmRes}) we obtain
\begin{equation}
\gamma_m(\alpha) = \gamma_{m0} \frac{\alpha}{4\pi} + \cdots
= 6 \frac{\alpha}{4\pi} + \cdots
\label{Mass:gammaRes}
\end{equation}
In order to solve the RG equation~(\ref{Mass:RG}) for $m(\mu)$,
let's write it down together with the RG equation~(\ref{Vertex:RG})
for $\alpha(\mu)$:
\begin{equation*}
\frac{d\log\alpha}{d\log\mu} = - 2 \beta(\alpha)\,,\quad
\frac{d\log m}{d\log\mu} = - \gamma_m(\alpha)\,,
\end{equation*}
and divide the second equation by the first one:
\begin{equation}
\frac{d\log m}{d\log\alpha} = \frac{\gamma_m(\alpha)}{2\beta(\alpha)}\,.
\label{Mass:RGalpha}
\end{equation}
The solution is
\begin{equation}
m(\mu') = m(\mu)\,
\exp \int_{\alpha(\mu)}^{\alpha(\mu')}
\frac{\gamma_m(\alpha)}{2\beta(\alpha)} \frac{d\alpha}{\alpha}\,.
\label{Mass:RGsol}
\end{equation}
At one loop, we obtain from~(\ref{Vertex:betaRes}) and~(\ref{Mass:gammaRes})
\begin{equation}
m(\mu') = m(\mu)
\left(\frac{\alpha(\mu')}{\alpha(\mu)}\right)%
^{\gamma_{m0}/(2\beta_0)}
= m(\mu)
\left(\frac{\alpha(\mu')}{\alpha(\mu)}\right)^{-9/4}\,.
\label{Mass:RGsol2}
\end{equation}
The running electron mass $m(\mu)$ decreases with $\mu$.

\section{QCD at one loop}
\label{S:QCD}

\subsection{Lagrangian and Feynman rules}
\label{S:QCDl}

The Lagrangian of QCD with $n_f$ massless flavours is
\begin{equation}
L = \sum_i \bar{q}_{0i} i\D q_{0i} - \frac{1}{4} G^a_{0\mu\nu} G_0^{a\mu\nu}\,,
\label{QCDl:L}
\end{equation}
where $q_{0i}$ are the quark fields,
\begin{equation}
D_\mu q_0 = \left(\partial_\mu - i g_0 A_{0\mu}\right) q_0\,,\quad
A_{0\mu} = A^a_{0\mu} t^a
\label{QCDl:Dq}
\end{equation}
are their covariant derivatives,
$A^a_{0\mu}$ is the gluon field,
$t^a$ are the generators of the colour group,
and the field strength tensor is defined by
\begin{equation}
[D_\mu,D_\nu] q_0 = - i g_0 G_{0\mu\nu} q_0\,,\quad
G_{0\mu\nu} = G^a_{0\mu\nu} t^a\,.
\label{QCDl:G}
\end{equation}
It is given by
\begin{equation}
G^a_{0\mu\nu} = \partial_\mu A^a_{0\nu} - \partial_\nu A^a_{0\mu}
+ g_0 f^{abc} A^b_{0\mu} A^c_{0\nu}\,,
\label{QCDl:Ga}
\end{equation}
where the structure constants $f^{abc}$ are defined by
\begin{equation}
[t^a,t^b] = i f^{abc} t^c\,.
\label{QCDl:f}
\end{equation}

Due to the gauge invariance,
it is not possible to obtain the gluon propagator from this Lagrangian.
We should fix the gauge.
The most popular way (called the covariant gauge)
requires to add the gauge-fixing term (with the parameter $a_0$)
and the ghost term to the Lagrangian~(\ref{QCDl:L}):
\begin{equation}
\Delta L = - \frac{1}{2a_0} \left(\partial_\mu A_0^{a\mu}\right)^2
+ (\partial^\mu\bar{c}_0^a)(D_\mu c_0^a)\,.
\label{QCDl:DL}
\end{equation}
Here $c_0^a$ is the ghost field --- a scalar field obeying Fermi statistics,
with the colour index $a$ (like the gluon).
Its covariant derivative is
\begin{equation}
D_\mu c_0^a = \left(\partial_\mu \delta^{ab} - i g_0 A_{0\mu}^{ab}\right) c_0^b\,,\quad
A_{0\mu}^{ab} = A_{0\mu}^c (t^c)^{ab}\,,
\label{QCDl:Dc}
\end{equation}
where
\begin{equation}
(t^c)^{ab} = i f^{acb}
\label{QCDl:fa}
\end{equation}
are the generators of the colour group in the adjoint representation.

\begin{figure}[ht]
\begin{center}
\begin{picture}(50,30)
\put(14,25){\makebox(0,0){\includegraphics{pf.eps}}}
\put(14,23){\makebox(0,0)[t]{$p$}}
\put(30,25){\makebox(0,0)[l]{${}=i S_0(p)$}}
\put(14,15){\makebox(0,0){\includegraphics{pp.eps}}}
\put(14,13){\makebox(0,0)[t]{$p$}}
\put(1.5,13){\makebox(0,0){$\mu$}}
\put(26,13){\makebox(0,0){$\nu$}}
\put(1.5,17){\makebox(0,0){$a$}}
\put(26,17){\makebox(0,0){$b$}}
\put(30,15){\makebox(0,0)[l]{${}=-i \delta^{ab} D^0_{\mu\nu}(p)$}}
\put(14,5){\makebox(0,0){\includegraphics{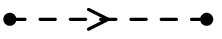}}}
\put(14,3){\makebox(0,0)[t]{$p$}}
\put(30,5){\makebox(0,0)[l]{${}=i \delta^{ab} G_0(p)$}}
\put(1.5,7){\makebox(0,0){$a$}}
\put(26,7){\makebox(0,0){$b$}}
\end{picture}
\end{center}
\caption{Propagators in QCD}
\label{F:QCDprop}
\end{figure}

The quadratic part of the QCD Lagrangian gives the propagators
shown in Fig.~\ref{F:QCDprop}.
The quark propagator is $S_0(p)$~(\ref{QEDl:S0});
the unit colour matrix is assumed.
The gluon propagator is $D^0_{\mu\nu}(p)$~(\ref{QEDl:D0});
the unit matrix in the adjoint representation $\delta^{ab}$
is written down explicitly.
The ghost propagator is
\begin{equation}
G_0(p) = \frac{1}{p^2}
\label{QCDl:G0}
\end{equation}
times $\delta^{ab}$.

\begin{figure}[ht]
\begin{center}
\begin{picture}(50,16)
\put(14,7){\makebox(0,0){\includegraphics{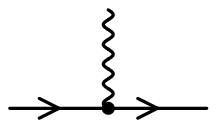}}}
\put(12,14){\makebox(0,0){$\mu$}}
\put(16,14){\makebox(0,0){$a$}}
\put(30,2){\makebox(0,0)[l]{${}=t^a \times i g_0 \gamma^\mu$}}
\end{picture}
\end{center}
\caption{Quark--gluon vertex}
\label{F:qg}
\end{figure}

The quark--gluon vertex is shown in Fig.~\ref{F:qg}.
The problem of calculation of a QCD Feynman diagram
reduces to two separate subproblems:
\begin{itemize}
\item calculation of the colour factor;
\item calculation of the ``colourless'' diagram.
\end{itemize}
The first subproblem can be formulated as
calculation of the \emph{colour diagram}.
It looks like the original QCD diagram,
where quark lines and gluon lines mean the unit matrices
in the fundamental representation and in the adjoint one;
ghost lines can be replaced by gluon lines in colour diagrams.
The quark--gluon vertex in a colour diagram
gives $t^a$, the first factor in Fig.~\ref{F:qg}.
When calculating the ``colourless'' diagram,
all factors except the colour structure are included;
the quark--gluon vertex gives $i g_0 \gamma^\mu$,
the second factor in Fig.~\ref{F:qg}.

\begin{figure}[ht]
\begin{center}
\begin{picture}(80,32)
\put(18,17){\makebox(0,0){\includegraphics{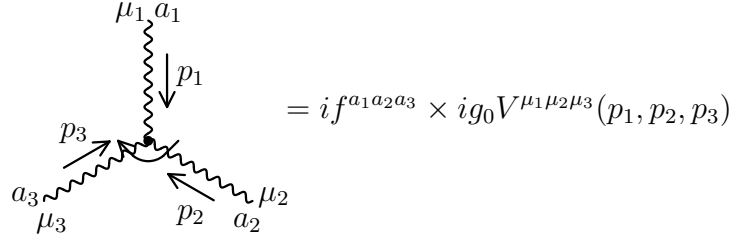}}}
\put(15.5,30){\makebox(0,0){$\mu_1$}}
\put(20.5,30){\makebox(0,0){$a_1$}}
\put(23.5,21.5){\makebox(0,0){$p_1$}}
\put(34.5,5.5){\makebox(0,0){$\mu_2$}}
\put(31,2){\makebox(0,0){$a_2$}}
\put(5,2){\makebox(0,0){$\mu_3$}}
\put(1.5,5.5){\makebox(0,0){$a_3$}}
\put(23.5,3){\makebox(0,0){$p_2$}}
\put(8,14){\makebox(0,0){$p_3$}}
\put(35,17){\makebox(0,0)[l]{$\displaystyle
{}= i f^{a_1 a_2 a_3} \times i g_0 V^{\mu_1\mu_2\mu_3}(p_1,p_2,p_3)$}}
\end{picture}
\end{center}
\caption{3--gluon vertex}
\label{F:g3}
\end{figure}

The 3--gluon vertex is shown in Fig.~\ref{F:g3}.
When separating it into the colour structure
(which appears in the colour diagram)
and the Lorentz structure
(which appears in the ``colourless'' diagram),
we have to choose some rotation direction.
These rotation directions must be the same
in the colour diagram and in the ``colourless'' one.
The colour structure is $i f^{a_1 a_2 a_3}$,
where the colour indices are written in the chosen order
(in Fig.~\ref{F:g3}, clockwise).
The Lorentz structure is
\begin{equation}
V^{\mu_1\mu_2\mu_3}(p_1,p_2,p_3)
= (p_3-p_2)^{\mu_1} g^{\mu_2\mu_3}
+ (p_1-p_3)^{\mu_2} g^{\mu_3\mu_1}
+ (p_2-p_1)^{\mu_3} g^{\mu_1\mu_2}\,,
\label{QCDl:V}
\end{equation}
where the polarization indices and the momenta
are numbered in the same way, here clockwise.
Of course, the full 3--gluon vertex does not know
about the rotation direction:
if we reverse it,
both the colour factor and the $V$ tensor change sign,
and their product does not change.

\begin{figure}[ht]
\begin{center}
\begin{picture}(92,22)
\put(8,11){\makebox(0,0){\includegraphics{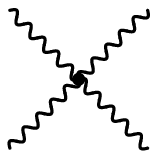}}}
\put(30,11){\makebox(0,0){\includegraphics{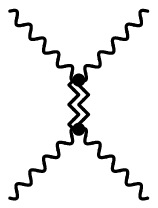}}}
\put(54.5,11){\makebox(0,0){\includegraphics{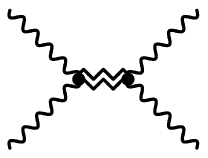}}}
\put(81.5,11){\makebox(0,0){\includegraphics{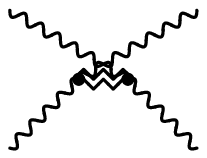}}}
\put(19,11){\makebox(0,0){$\Rightarrow$}}
\put(41,11){\makebox(0,0){$+$}}
\put(68,11){\makebox(0,0){$+$}}
\end{picture}
\end{center}
\caption{4--gluon vertex}
\label{F:g4}
\end{figure}

\begin{figure}[ht]
\begin{center}
\begin{picture}(50,10)
\put(9,5){\makebox(0,0){\includegraphics{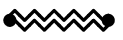}}}
\put(1,5){\makebox(0,0){$a$}}
\put(4,2.5){\makebox(0,0){$\mu$}}
\put(4,7.5){\makebox(0,0){$\nu$}}
\put(17,5){\makebox(0,0){$b$}}
\put(14,2.5){\makebox(0,0){$\alpha$}}
\put(14,7.5){\makebox(0,0){$\beta$}}
\put(19,5){\makebox(0,0)[l]{$\displaystyle{}= i \delta^{ab}
(g^{\mu\alpha}g^{\nu\beta}-g^{\mu\beta}g^{\nu\alpha})$}}
\end{picture}
\end{center}
\caption{Propagator of the auxiliary field}
\label{F:auxProp}
\end{figure}

\begin{figure}[ht]
\begin{center}
\begin{picture}(50,24)
\put(10.5,12){\makebox(0,0){\includegraphics{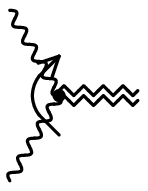}}}
\put(19.5,12){\makebox(0,0){$c$}}
\put(17,14.5){\makebox(0,0){$\beta$}}
\put(17,9.5){\makebox(0,0){$\alpha$}}
\put(5.5,1.5){\makebox(0,0){$\mu$}}
\put(5.5,22.5){\makebox(0,0){$\nu$}}
\put(2,4.5){\makebox(0,0){$a$}}
\put(2,19.5){\makebox(0,0){$b$}}
\put(22,12){\makebox(0,0)[l]{${}= i f^{abc} \times
g_0 g^{\mu\alpha} g^{\nu\beta}$}}
\end{picture}
\end{center}
\caption{Vertex of the auxiliary field interaction with gluons}
\label{F:auxVert}
\end{figure}

The 4-gluon vertex contains terms with 3 different colour structures.
Therefore, any diagram containing at least one 4-gluon vertex
cannot be factorized into the colour diagram and the ``colourless'' one.
This is very inconvenient for programs automating diagram calculations.
The authors of (at least) two such programs (CompHEP and GEFICOM)
invented the same trick to ensure the possibility of calculating
the colour factor as a separate subproblem.
Let's say that there is no 4--gluon vertex in QCD,
but there is a new field interacting with gluons
(this field is shown as a double zigzag line in Fig.~\ref{F:g4}).
This is an antisymmetric tensor field;
its propagator is shown in Fig.~\ref{F:auxProp},
and the vertex of its interaction with gluons --- in Fig.~\ref{F:auxVert}.
Its propagator in the momentum space does not depend on $p$.
Therefore, in the coordinate space, it contains $\delta(x)$ ---
this particle does not propagate,
and two vertices of its interaction with gluons (Fig.~\ref{F:g4})
are at the same point.
In the colour diagram, lines of this particle can be replaces by gluon lines,
and vertices --- by 3--gluon vertices.
In ``colourless'' diagrams, the second factor in Fig.~\ref{F:auxVert}
is used for the vertices.
The rotation directions must agree, as usual.

\begin{figure}[ht]
\begin{center}
\begin{picture}(50,18)
\put(14,9){\makebox(0,0){\includegraphics{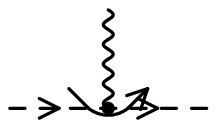}}}
\put(19,1){\makebox(0,0){$p$}}
\put(11.5,15.5){\makebox(0,0){$\mu$}}
\put(16.5,16){\makebox(0,0){$b$}}
\put(3,5.5){\makebox(0,0){$c$}}
\put(25,5.5){\makebox(0,0){$a$}}
\put(28,9){\makebox(0,0)[l]{${}= i f^{abc} \times i g_0 p^\mu$}}
\end{picture}
\end{center}
\caption{Ghost--gluon vertex}
\label{F:ghg}
\end{figure}

Finally, the ghost--gluon vertex is shown in Fig.~\ref{F:ghg}.
It contains only the outgoing ghost momentum,
but not the incoming ghost momentum.
This is because its Lagrangian~(\ref{QCDl:DL})
contains the covariant derivative of $c$
but the ordinary derivative of $\bar{c}$.
In colour diagrams, we replace ghost lines by gluon lines,
and vertices --- by 3--gluon vertices;
their rotation direction is fixed:
the incoming ghost $\to$ the outgoing ghost $\to$ the gluon
(Fig.~\ref{F:ghg}).

Calculation of colour diagrams is discussed in Appendix~\ref{S:Colour}.

The renormalized fields and parameters are related to the bare ones,
similarly to QED~(\ref{QEDl:ren}), by
\begin{equation}
q_{i0} = Z_q^{1/2} q_i\,,\quad
A_0 = Z_A^{1/2} A\,,\quad
c_0 = Z_c^{1/2} c\,,\quad
a_0 = Z_A a\,,\quad
g_0 = Z_\alpha^{1/2} g\,;
\label{QCDl:ren}
\end{equation}
the QCD running coupling $\alpha_s(\mu)$ is
\begin{equation}
\frac{\alpha_s(\mu)}{4\pi} = \mu^{-2\varepsilon}
\frac{g^2}{(4\pi)^{d/2}} e^{-\gamma\varepsilon}\,,
\quad
\frac{g_0^2}{(4\pi)^{d/2}} = \mu^{2\varepsilon}
\frac{\alpha_s(\mu)}{4\pi} Z_\alpha(\alpha(\mu)) e^{\gamma\varepsilon}\,.
\label{QCDl:alpha}
\end{equation}

\subsection{Quark propagator}
\label{S:Quark}

\begin{figure}[ht]
\begin{center}
\begin{picture}(32,10)
\put(16,5){\makebox(0,0){\includegraphics{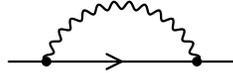}}}
\end{picture}
\end{center}
\caption{One-loop quark self-energy}
\label{F:Quark1}
\end{figure}

The quark propagator in QCD has the same structure~(\ref{Electron:Dyson1})
as the electron propagator in QED (Sect.~\ref{S:Electron}).
In massless QCD, the quark self-energy has the form
\begin{equation}
\Sigma(p) = \rlap/p \Sigma_V(p^2)\,.
\label{Quark:SigmaV}
\end{equation}
The one-loop diagram (Fig.~\ref{F:Quark1})
differs from the QED one~(\ref{Electron:Res1})
only by the replacement $e_0\to g_0$
and by the colour factor $C_F$~(\ref{Colour:CFdef}):
\begin{equation}
\Sigma_V(p^2) = - C_F \frac{g_0^2 (-p^2)^{-\varepsilon}}{(4\pi)^{d/2}}
\frac{d-2}{2} a_0 G_1\,.
\label{Quark:Res1}
\end{equation}
Therefore, the quark field renormalization constant with one-loop accuracy is
\begin{equation}
Z_q = 1 - C_F a \frac{\alpha_s}{4\pi\varepsilon} + \cdots
\label{Quark:Zq}
\end{equation}
and the quark-field anomalous dimension is
\begin{equation}
\gamma_q = 2 C_F a \frac{\alpha_s}{4\pi} + \cdots
\label{Quark:gammaq}
\end{equation}

If the quark flavour $q_i$ has a non-zero mass,
its self-energy has also the helicity-flipping structure $\Sigma_S(p^2)$.
The calculation of its one-loop ultraviolet-divergent part
(Sect.~\ref{S:Mass}) applies in QCD practically unchanged,
with the substitution $\alpha\to\alpha_s$ and the extra colour factor $C_F$:
\begin{equation}
\Sigma_S = C_F (3+a) \frac{\alpha_s}{4\pi\varepsilon} + \cdots
\label{Quark:SigmaS}
\end{equation}
Therefore, the quark-mass renormalization constant is
\begin{equation}
Z_m = 1 - 3 C_F \frac{\alpha_s}{4\pi\varepsilon} + \cdots
\label{Quark:Zm}
\end{equation}
and the mass anomalous dimension is
\begin{equation}
\gamma_m = 6 C_F \frac{\alpha_s}{4\pi} + \cdots
\label{Quark:gammam}
\end{equation}

\subsection{Gluon propagator}
\label{S:Gluon}

Ward identities in QCD are more complicated than in QED.
Nevertheless, the gluon self-energy $i \delta^{ab} \Pi_{\mu\nu}(p)$
is transverse:
\begin{equation}
\Pi_{\mu\nu}(p) p^\mu = 0\,,\quad
\Pi_{\mu\nu}(p) = (p^2 g_{\mu\nu} - p_\mu p_\nu) \Pi(p^2)\,,
\label{Gluon:Ward}
\end{equation}
as in QED.
Explanation of this is beyond the level of these lectures.

\begin{figure}[ht]
\begin{center}
\begin{picture}(102,18)
\put(51,9){\makebox(0,0){\includegraphics{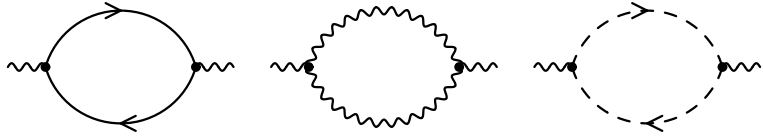}}}
\end{picture}
\end{center}
\caption{One-loop gluon self-energy}
\label{F:Gluon1}
\end{figure}

At one loop, the gluon self-energy is given by 3 diagrams
(Fig.~\ref{F:Gluon1}).
The quark-loop contribution differs
from the QED result~(\ref{Photon1:Res1})
only by the substitution $e_0\to g_0$,
and the colour factor $T_F n_f$:
\begin{equation}
\Pi_q(p^2) = - T_F n_f \frac{g_0^2 (-p^2)^{-\varepsilon}}{(4\pi)^{d/2}}
2 \frac{d-2}{d-1} G_1\,.
\label{Gluon:Piq}
\end{equation}

\begin{figure}[ht]
\begin{center}
\begin{picture}(64,44)
\put(32,22){\makebox(0,0){\includegraphics{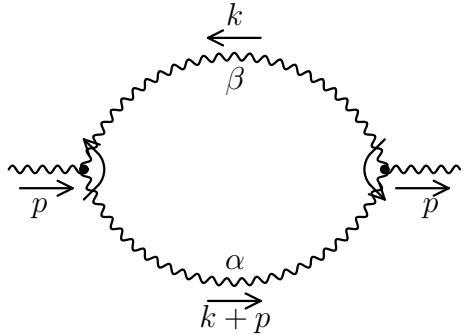}}}
\put(32,0){\makebox(0,0)[b]{$k+p$}}
\put(32,44){\makebox(0,0)[t]{$k$}}
\put(6,18){\makebox(0,0)[t]{$p$}}
\put(58,18){\makebox(0,0)[t]{$p$}}
\put(32,35.5){\makebox(0,0)[t]{$\beta$}}
\put(32,8.5){\makebox(0,0)[b]{$\alpha$}}
\end{picture}
\end{center}
\caption{Gluon-loop contribution}
\label{F:Gluong}
\end{figure}

The gluon-loop contribution (Fig.~\ref{F:Gluong})
has the symmetry factor $\frac{1}{2}$
and the colour factor $C_A$~(\ref{Colour:CA}).
In the Feynman gauge $a_0=1$,
\begin{equation}
\begin{split}
&\Pi_1{}_\mu^\mu = - i \frac{1}{2} C_A g_0^2 \int \frac{d^d k}{(2\pi)^d}
\frac{N}{k^2 (k+p)^2}\,,\\
&N = V_{\mu\alpha\beta}(p,-k-p,k) V^{\mu\beta\alpha}(-p,-k,k+p)\,.
\end{split}
\label{Gluon:Pig0}
\end{equation}
The Lorentz part of the 3-gluon vertex~(\ref{QCDl:V}) here is
\begin{equation*}
V_{\mu\alpha\beta}(p,-k-p,k) = (2k+p)_\mu g_{\alpha\beta}
- (k-p)_\alpha g_{\beta\mu} - (k+2p)_\beta g_{\mu\alpha}\,.
\end{equation*}
$V^{\mu\beta\alpha}(-p,-k,k+p)$ coincides with it.
Therefore, the numerator in~(\ref{Gluon:Pig0}) is
\begin{equation*}
\begin{split}
N ={}& d \left[ (2k+p)^2 + (k-p)^2 + (k+2p)^2 \right]\\
&{} - 2 (2k+p)\cdot(k-p) - 2 (2k+p)\cdot(k+2p)
+ 2 (k+2p)\cdot(k-p)\,.
\end{split}
\end{equation*}
Using the ``multiplication table''~(\ref{Photon1:Mult})
and omitting terms with $D_{1,2}$ in the numerator
(which give vanishing integrals), we have
\begin{equation*}
\begin{split}
&p^2 = -1\,,\quad
k^2 = -D_2 \Rightarrow 0\,,\quad
p\cdot k = \frac{1}{2} (1-D_1+D_2) \Rightarrow \frac{1}{2}\,,\\
&N \Rightarrow - 3 (d-1)\,.
\end{split}
\end{equation*}
Finally, we arrive at
\begin{equation}
\Pi_1{}_\mu^\mu = - \frac{3}{2} C_A
\frac{g_0^2 (-p^2)^{1-\varepsilon}}{(4\pi)^{d/2}}
G_1 (d-1)\,.
\label{Gluon:Pig}
\end{equation}

\begin{figure}[ht]
\begin{center}
\begin{picture}(64,38)
\put(32,19){\makebox(0,0){\includegraphics{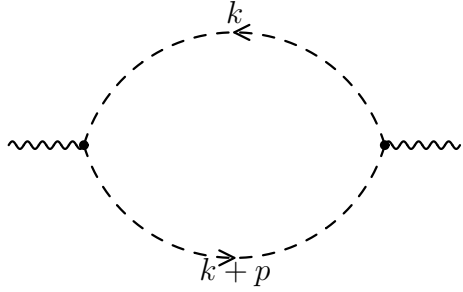}}}
\put(32,0){\makebox(0,0)[b]{$k+p$}}
\put(32,38){\makebox(0,0)[t]{$k$}}
\end{picture}
\end{center}
\caption{Ghost-loop contribution}
\label{F:Gluonc}
\end{figure}

The ghost-loop contribution (Fig.~\ref{F:Gluonc})
has the fermion-loop factor $-1$
and the colour factor $C_A$~(\ref{Colour:CA}):
\begin{equation}
\Pi_2{}_\mu^\mu = i C_A g_0^2 \int \frac{d^d k}{(2\pi)^d}
\frac{k\cdot(k+p)}{k^2 (k+p)^2}
= - \frac{1}{2} C_A \frac{g_0^2 (-p^2)^{1-\varepsilon}}{(4\pi)^{d/2}} G_1\,.
\label{Gluon:Pic}
\end{equation}

These two contributions, taken separately, are not transverse.
Their sum is transverse, and hence has the structure~(\ref{Gluon:Ward}):
\begin{equation}
\Pi_g(p^2) = - \frac{\Pi_1{}_\mu^\mu+\Pi_2{}_\mu^\mu}{(d-1)(-p^2)}
= C_A \frac{g_0^2 (-p^2)^{-\varepsilon}}{(4\pi)^{d/2}}
G_1 \frac{3d-2}{2(d-1)}\,.
\label{Gluon:Pig1}
\end{equation}
In an arbitrary covariant gauge,
\begin{equation}
\begin{split}
\Pi_g(p^2) ={}& C_A \frac{g_0^2 (-p^2)^{-\varepsilon}}{(4\pi)^{d/2}}
\frac{G_1}{2(d-1)}\\
&{}\times\left[3d-2 + (d-1)(2d-7)\xi - \frac{1}{4} (d-1)(d-4)\xi^2\right]
\end{split}
\label{Gluon:Piga}
\end{equation}
(here $\xi=1-a_0$~(\ref{Electron:xi});
we leave the derivation as an exercise for the reader).

The transverse part of the gluon propagator with one-loop accuracy,
expressed via renormalized quantities and expanded in $\varepsilon$, is
\begin{equation}
\begin{split}
p^2 D_\bot(p^2) ={}& 1 + \frac{\alpha_s(\mu)}{4\pi\varepsilon}
e^{-L\varepsilon}
\Biggl[ - \frac{1}{2} \left( a - \frac{13}{3} \right) C_A
- \frac{4}{3} T_F n_f\\
&{} + \left( \frac{9 a^2 + 18 a + 97}{36} C_A
- \frac{20}{9} T_F n_f \right) \varepsilon \Biggr]\,,
\end{split}
\label{Gluon:Da}
\end{equation}
where $L=\log (-p^2)/\mu^2$.
Therefore, the gluon field renormalization constant is
\begin{equation}
Z_A = 1 - \frac{\alpha_s}{4\pi\varepsilon}
\left[ \frac{1}{2} \left( a - \frac{13}{3} \right) C_A
+ \frac{4}{3} T_F n_f \right] + \cdots
\label{Gluon:ZA}
\end{equation}
We don't write down the expression for the renormalized gluon propagator;
it is easy to do this.
The gluon field anomalous dimension is
\begin{equation}
\gamma_A = \left[ \left(a - \frac{13}{3}\right) C_A + \frac{8}{3} T_F n_f \right]
\frac{\alpha_s}{4\pi} + \cdots
\label{Gluon:gammaA}
\end{equation}
Making the substitutions $\alpha_s\to\alpha$, $C_A\to0$, $T_F n_f\to1$,
we reproduce the QED results~(\ref{PhotonZ:ZA}), (\ref{PhotonZ:gammaA}).

\subsection{Ghost propagator}
\label{S:Ghost}

The ghost propagator has the structure
\begin{equation}
i G(p) = i G_0(p) + i G_0(p) (-i) \Sigma i G_0(p)
+ i G_0(p) (-i) \Sigma i G_0(p) (-i) \Sigma i G_0(p)
+ \cdots
\label{Ghost:Dyson1}
\end{equation}
where the ghost self-energy $-i\Sigma$ is a scalar function of $p^2$,
and the free ghost propagator $G_0(p)$ is given by~(\ref{QCDl:G0}).
Therefore,
\begin{equation}
G(p) = \frac{1}{p^2 - \Sigma(p^2)}\,.
\label{Ghost:Dyson2}
\end{equation}

\begin{figure}[ht]
\begin{center}
\begin{picture}(64,25.5)
\put(32,12.75){\makebox(0,0){\includegraphics{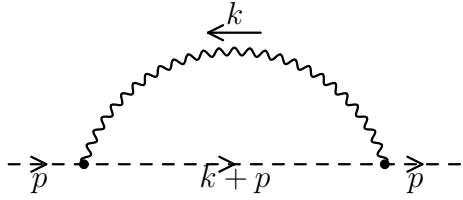}}}
\put(32,0){\makebox(0,0)[b]{$k+p$}}
\put(6,0){\makebox(0,0)[b]{$p$}}
\put(56,0){\makebox(0,0)[b]{$p$}}
\put(32,25.5){\makebox(0,0)[t]{$k$}}
\end{picture}
\end{center}
\caption{One-loop ghost self-energy}
\label{F:Ghost1}
\end{figure}

At one loop (Fig.~\ref{F:Ghost1}),
\begin{equation}
\begin{split}
\Sigma(p^2) &{}= - i C_A g_0^2 \int \frac{d^d k}{(2\pi)^d}
\frac{p^\mu (k+p)^\nu}{k^2 (k+p)^2}
\left( g_{\mu\nu} - \xi \frac{k_\mu k_\nu}{k^2} \right)\\
&{}= C_A \frac{g_0^2 (-p^2)^{1-\varepsilon}}{(4\pi)^{d/2}}
\left[ - \frac{1}{2} G(1,1) + \frac{\xi}{4} G(1,2) \right]\\
&{}= - \frac{1}{4} C_A \frac{g_0^2 (-p^2)^{1-\varepsilon}}{(4\pi)^{d/2}}
G_1 \left[ d-1 - (d-3) a_0 \right]\,.
\end{split}
\label{Ghost:Sigma}
\end{equation}
The propagator, expressed via renormalized quantities
and expanded in $\varepsilon$, is
\begin{equation}
G(p) = \frac{1}{p^2} \left[ 1 + C_A \frac{\alpha_s(\mu)}{4\pi\varepsilon}
e^{-L\varepsilon} \frac{3-a+4\varepsilon}{4} \right]\,.
\label{Ghost:G}
\end{equation}
Therefore, the ghost field renormalization constant is
\begin{equation}
Z_c = 1 + C_A \frac{3-a}{4} \frac{\alpha_s}{4\pi\varepsilon} + \cdots
\label{Ghost:Zc}
\end{equation}
and the anomalous dimension is
\begin{equation}
\gamma_c = - C_A \frac{3-a}{2} \frac{\alpha_s}{4\pi} + \cdots
\label{Ghost:gammac}
\end{equation}

\subsection{Quark--gluon vertex}
\label{S:QuarkGluon}

\begin{figure}[ht]
\begin{center}
\begin{picture}(62,18)
\put(13,9){\makebox(0,0){\includegraphics{v1.eps}}}
\put(49,9){\makebox(0,0){\includegraphics{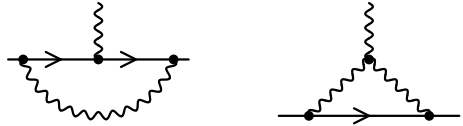}}}
\end{picture}
\end{center}
\caption{Quark--gluon vertex at one loop}
\label{F:qg1}
\end{figure}

Let's find the ultraviolet divergence of the quark--gluon vertex
at one loop (Fig.~\ref{F:qg1}).
The first diagram differs from the QED one
only by the colour factor $C_F-C_A/2$~(\ref{Colour:v1t}),
and
\begin{equation}
\Lambda_1^\alpha = \left( C_F - \frac{C_A}{2} \right)
\frac{\alpha_s}{4\pi\varepsilon} \gamma^\alpha\,.
\label{QuarkGluon:Lambda1}
\end{equation}

\begin{figure}[ht]
\begin{center}
\begin{picture}(50,38)
\put(25,19){\makebox(0,0){\includegraphics{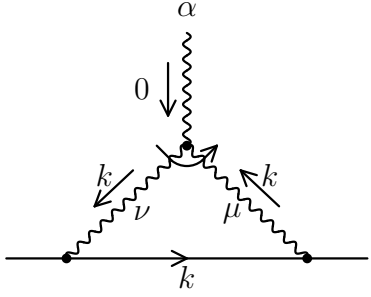}}}
\put(25,0){\makebox(0,0)[b]{$k$}}
\put(25,38){\makebox(0,0)[t]{$\alpha$}}
\put(20,26.5){\makebox(0,0)[r]{$0$}}
\put(36,15){\makebox(0,0){$k$}}
\put(14,15){\makebox(0,0){$k$}}
\put(31,10){\makebox(0,0){$\mu$}}
\put(19,10){\makebox(0,0){$\nu$}}
\end{picture}
\end{center}
\caption{The non-abelian diagram}
\label{F:qg2}
\end{figure}

The second diagram (Fig.~\ref{F:qg2})
has the colour factor $C_A/2$~(\ref{Colour:v2t}).
It has a logarithmic ultraviolet divergence.
Therefore, as in Sect.~\ref{S:Vertex},
in order to find the $1/\varepsilon$ term
we may nullify all the external momenta
and provide some infrared cutoff.
In the Feynman gauge ($a_0=1$),
\begin{equation}
\Lambda_2^\alpha = i \frac{C_A}{2} g_0^2 \int \frac{d^d k}{(2\pi)^d}
\frac{\gamma_\mu\rlap/k\gamma_\nu}{(k^2)^3} V^{\alpha\nu\mu}(0,-k,k)\,,
\label{QuarkGluon:Lambda2a}
\end{equation}
where the Lorentz part of the 3-gluon vertex~(\ref{QCDl:V}) is
\begin{equation*}
V^{\alpha\nu\mu}(0,-k,k) = 2 k^\alpha g^{\mu\nu}
- k^\mu g^{\nu\alpha} - k^\nu g^{\mu\alpha}\,.
\end{equation*}
Therefore,
\begin{equation*}
\Lambda_2^\alpha = i \frac{C_A}{2} g_0^2 \int \frac{d^d k}{(2\pi)^d}
\frac{2\gamma_\mu\rlap/k\gamma^\mu k^\alpha-2k^2\gamma^\alpha}{(k^2)^3}\,.
\end{equation*}
Averaging over $k$ directions:
\begin{equation*}
\rlap/k k^\alpha \to \frac{k^2}{d} \gamma^\alpha
\end{equation*}
and using the ultraviolet divergence~(\ref{Vertex:UV}),
we obtain
\begin{equation}
\Lambda_2^\alpha = \frac{3}{2} C_A \frac{\alpha_s}{4\pi\varepsilon} \gamma^\alpha\,.
\label{QuarkGluon:Lambda2b}
\end{equation}
In the arbitrary covariant gauge
\begin{equation}
\Lambda_2^\alpha = \frac{3}{4} (1+a) C_A \frac{\alpha_s}{4\pi\varepsilon} \gamma^\alpha
\label{QuarkGluon:Lambda2}
\end{equation}
(derive this result!).

\begin{sloppypar}
The $1/\varepsilon$ term (the ultraviolet divergence)
of the one-loop quark--gluon vertex is,
from~(\ref{QuarkGluon:Lambda1}) and~(\ref{QuarkGluon:Lambda2}),
\begin{equation}
\Lambda^\alpha = \left( C_F a + C_A \frac{a+3}{4} \right)
\frac{\alpha_s}{4\pi\varepsilon} \gamma^\alpha\,.
\label{QuarkGluon:Lambda}
\end{equation}
Therefore, the quark--gluon vertex renormalization constant is
\begin{equation}
Z_\Gamma = 1 + \left( C_F a + C_A \frac{a+3}{4} \right)
\frac{\alpha_s}{4\pi\varepsilon} + \cdots
\label{QuarkGluon:ZGamma}
\end{equation}
\end{sloppypar}

\subsection{Coupling constant renormalization}
\label{S:alphas}

The coupling renormalization constant is
\begin{equation}
Z_\alpha = (Z_\Gamma Z_q)^{-2} Z_A^{-1}\,,
\label{alphas:ZalphaDef}
\end{equation}
similarly to Sect.~\ref{S:Vertex}.
The product $Z_\Gamma Z_q$, with one-loop accuracy, is,
from~(\ref{QuarkGluon:ZGamma}) and~(\ref{Quark:Zq}),
\begin{equation}
Z_\Gamma Z_q = 1 + C_A \frac{a+3}{4} \frac{\alpha_s}{4\pi\varepsilon} + \cdots
\label{alphas:ZGammaq}
\end{equation}
In QED, it was equal to 1~(\ref{Vertex:WardZ}),
due to the Ward identity (Sect.~\ref{S:Vertex}).
Ward identities in QCD are more complicated.
Making the replacements $\alpha_s\to\alpha$, $C_F\to1$, $C_A\to0$,
we reproduce the simple QED result.

The coupling renormalization constant, with one-loop accuracy, is,
from~(\ref{alphas:ZGammaq}) and~(\ref{Gluon:ZA}),
\begin{equation}
Z_\alpha = 1 - \left( \frac{11}{3} C_A - \frac{4}{3} T_F n_f \right)
\frac{\alpha_s}{4\pi\varepsilon} + \cdots
\label{alphas:Zalpha}
\end{equation}
It is gauge-invariant.
This is a strong check of our calculations%
\footnote{We have actually presented detailed calculations
of $Z_A$ and $Z_\Gamma$ only in the Feynman gauge $a_0=1$;
this is enough for obtaining~(\ref{alphas:Zalpha}).}.
The $\beta$-function~(\ref{Vertex:beta}) in QCD is,
with one-loop accuracy,
\begin{equation}
\beta(\alpha_s) = \beta_0 \frac{\alpha_s}{4\pi} + \cdots\qquad
\beta_0 = \frac{11}{3} C_A - \frac{4}{3} T_F n_f\,.
\label{alphas:beta0}
\end{equation}
If there are not too many quark flavours
(namely, for $N_c=3$, if $n_f\le16$),
then $\beta_0>0$,
in contrast to the QED case $\beta_0=-4/3$.

The RG equation
\begin{equation}
\frac{d\log\alpha_s(\mu)}{d\log\mu} = - 2 \beta(\alpha_s(\mu))
\label{alphas:RG}
\end{equation}
shows that $\alpha_s(\mu)$ decreases when $\mu$ increases.
This behaviour (opposite to screening) is called
\emph{asymptotic freedom}.
Keeping only the leading (one-loop) term in the $\beta$-function,
we can rewrite the RG equation in the form~(\ref{Vertex:RG3}).
The running coupling at the renormalization scale $\mu'$
is related to that at $\mu$ by~(\ref{Vertex:RG2}),
but now with positive $\beta_0$.
The solution of the equation~(\ref{Vertex:RG3}) can also be written as
\begin{equation}
\alpha_s(\mu) = \frac{2\pi}{\displaystyle\beta_0\log\frac{\mu}{\Lambda_{\MS}}}\,,
\label{alphas:Lambda}
\end{equation}
where $\Lambda_{\MS}$ plays the role of the integration constant.
It sets the characteristic energy scale of QCD;
the coupling becomes small at $\mu\gg\Lambda_{\MS}$.

Coupling-constant renormalization in the non-abelian gauge theory
was first considered in~\cite{VT:65},
and the sign of the $\beta$-function corresponding to asymptotic freedom
was obtained.
The magnitude was not quite right
(it is right for a spontaneously-broken gauge theory with a higgs).
The first correct calculation of the one-loop $\beta$-function
in the non-abelian gauge theory has been published in~\cite{Kh:69}.
It was done in the Coulomb gauge, which is ghost-free.
The contribution of the transverse-gluon loop
has the same sign as that of the quark loop,
i.e., leads to screening.
However, there is another contribution,
that of the loop with an instantaneous Coulomb gluon.
It has the opposite sign, and outweights the first contribution.
This $\beta$-function has also been calculated by 't~Hooft~\cite{tH:71},
but not published (mentioned after Symanzik's talk at a meeting
in Marseilles in 1972).
Later it was calculated in the famous papers~\cite{GW:73,P:73}.
The authors of these papers applied asymptotic freedom
to explain the observed behaviour of deep inelastic
electron--proton scattering.
This was the real beginning of QCD as a theory of strong interactions.
The authors of~\cite{GW:73,P:73} received the Nobel prize in 2004.

\subsection{Ghost--gluon vertex}
\label{S:GhostGluon}

All QCD vertices contain just one coupling constant.
Its renormalization can be found from renormalization of any vertex:
quark--gluon (Sect.~\ref{S:QuarkGluon}, \ref{S:alphas}),
ghost--gluon, 3-gluon, or 4-gluon.
Here we shall derive it again by calculating the ghost--gluon vertex
at one loop.

\begin{figure}[ht]
\begin{center}
\begin{picture}(54,41)
\put(27,20.5){\makebox(0,0){\includegraphics{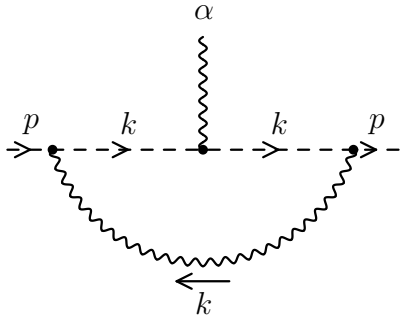}}}
\put(27,0){\makebox(0,0)[b]{$k$}}
\put(27,41){\makebox(0,0)[t]{$\alpha$}}
\put(4,24){\makebox(0,0)[b]{$p$}}
\put(50,24){\makebox(0,0)[b]{$p$}}
\put(17,24){\makebox(0,0)[b]{$k$}}
\put(37,24){\makebox(0,0)[b]{$k$}}
\end{picture}
\end{center}
\caption{One-loop ghost--gluon vertex (diagram 1)}
\label{F:cg1}
\end{figure}

The first diagram (Fig.~\ref{F:cg1})
has the colour factor $C_A/2$~(\ref{Colour:v3g}).
The ultraviolet $1/\varepsilon$ divergence of this diagram
is proportional to the bare vertex
(i.e., to the outgoing ghost momentum $p^\alpha$);
the coefficient diverges logarithmically.
We may nullify the external momenta except $p$, and use~(\ref{Vertex:UV}):
\begin{equation}
\begin{split}
\Lambda_1^\alpha &{}= - i \frac{C_A}{2} g_0^2
\int \frac{d^d k}{(2\pi)^d}
\frac{k^\alpha p^\mu k^\nu}{(k^2)^3}
\left(g_{\mu\nu} - \xi \frac{k_\mu k_\nu}{k^2}\right)\\
&{}= - i \frac{C_A}{2} g_0^2 a_0
\int \frac{d^d k}{(2\pi)^d}
\frac{k^\alpha p\cdot k}{(k^2)^3}\\
&{}= - i \frac{C_A}{2} g_0^2 a_0 \frac{1}{4} p^\alpha
\int \frac{d^d k}{(2\pi)^d} \frac{1}{(k^2)^2}
= \frac{1}{8} C_A a \frac{\alpha_s}{4\pi\varepsilon} p^\alpha\,.
\end{split}
\label{GhostGluon:Lambda1}
\end{equation}

\begin{figure}[ht]
\begin{center}
\begin{picture}(50,38)
\put(25,19){\makebox(0,0){\includegraphics{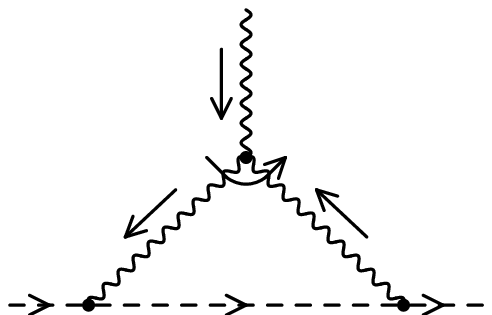}}}
\put(25,0){\makebox(0,0)[b]{$k$}}
\put(25,38){\makebox(0,0)[t]{$\alpha$}}
\put(20,26.5){\makebox(0,0)[r]{$0$}}
\put(5,0){\makebox(0,0)[b]{$p$}}
\put(45,0){\makebox(0,0)[b]{$p$}}
\put(37,16){\makebox(0,0){$k$}}
\put(13,16){\makebox(0,0){$k$}}
\put(31,10){\makebox(0,0){$\mu$}}
\put(19,10){\makebox(0,0){$\nu$}}
\end{picture}
\end{center}
\caption{One-loop ghost--gluon vertex (diagram 2)}
\label{F:cg2}
\end{figure}

The second diagram (Fig.~\ref{F:cg2}) has the same colour factor;
in the Feynman gauge ($a_0=1$),
\begin{equation}
\begin{split}
\Lambda_2^\alpha &{}= i \frac{C_A}{2} g_0^2
\int \frac{d^d k}{(2\pi)^d}
\frac{p_\mu k_\nu}{(k^2)^3} V^{\alpha\nu\mu}(0,-k,k)\\
&{}= i \frac{C_A}{2} g_0^2
\int \frac{d^d k}{(2\pi)^d}
\frac{p\cdot k\,k^\alpha - k^2 p^\alpha}{(k^2)^3}\\
&{}= - \frac{3}{4} i \frac{C_A}{2} g_0^2 p^\alpha
\int \frac{d^d k}{(2\pi)^d} \frac{1}{(k^2)^2}
= \frac{3}{8} C_A \frac{\alpha_s}{4\pi\varepsilon} p^\alpha\,.
\end{split}
\label{GhostGluon:Lambda2F}
\end{equation}
In the arbitrary covariant gauge,
\begin{equation}
\Lambda_2^\alpha = \frac{3}{8} C_A a \frac{\alpha_s}{4\pi\varepsilon} p^\alpha
\label{GhostGluon:Lambda2}
\end{equation}
(derive this result!).

The full result for the $1/\varepsilon$ term in the ghost--gluon vertex
at one loop is
\begin{equation}
\Lambda^\alpha = \frac{1}{2} C_A a \frac{\alpha_s}{4\pi\varepsilon} p^\alpha\,.
\label{GhostGluon:Lambda}
\end{equation}
Therefore, the vertex renormalization factor is
\begin{equation}
Z_{\Gamma c} = 1 + \frac{1}{2} C_A a \frac{\alpha_s}{4\pi\varepsilon}
+ \cdots
\label{GhostGluon:ZGamma}
\end{equation}
The coupling renormalization constant
\begin{equation}
Z_\alpha = (Z_{\Gamma c} Z_c)^{-2} Z_A^{-1}
\label{GhostGluon:Zalpha}
\end{equation}
is, from
\begin{equation*}
Z_{\Gamma c} Z_c = 1 + C_A \frac{3+a}{4} \frac{\alpha_s}{4\pi\varepsilon}
\end{equation*}
(which follows from~(\ref{Ghost:Zc}) and~(\ref{GhostGluon:ZGamma})),
equal to~(\ref{alphas:Zalpha}).
Thus we have re-derived $\beta_0$~(\ref{alphas:beta0});
this derivation is, probably, slightly easier than that
from the quark propagator and quark--gluon vertex.

\section{Two-loop corrections in QED and QCD}
\label{S:2L}

\subsection{Massless propagator diagram}
\label{S:q2}

\begin{figure}[ht]
\begin{center}
\begin{picture}(64,32)
\put(32,18){\makebox(0,0){\includegraphics{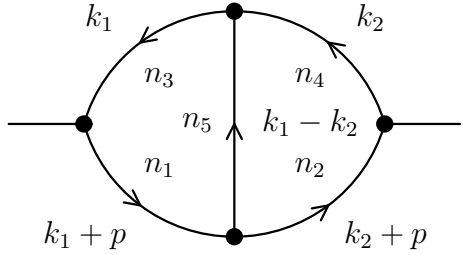}}}
\put(14,32){\makebox(0,0){$k_1$}}
\put(50,32){\makebox(0,0){$k_2$}}
\put(12,3){\makebox(0,0){$k_1+p$}}
\put(52,3){\makebox(0,0){$k_2+p$}}
\put(42,18){\makebox(0,0){$k_1-k_2$}}
\put(22,12){\makebox(0,0){$n_1$}}
\put(42,12){\makebox(0,0){$n_2$}}
\put(22,24){\makebox(0,0){$n_3$}}
\put(42,24){\makebox(0,0){$n_4$}}
\put(27,18){\makebox(0,0){$n_5$}}
\end{picture}
\end{center}
\caption{Two-loop massless propagator diagram}
\label{F:q2}
\end{figure}

Let's consider the two-loop massless propagator diagram (Fig.~\ref{F:q2}),
\begin{align}
&\int \frac{d^d k_1\,d^d k_2}{D_1^{n_1} D_2^{n_2} D_3^{n_3} D_4^{n_4} D_5^{n_5}} =
- \pi^d (-p^2)^{d-\sum n_i} G(n_1,n_2,n_3,n_4,n_5)\,,
\label{q2:G2def}\\
&D_1=-(k_1+p)^2\,,\quad
D_2=-(k_2+p)^2\,,\quad
D_3=-k_1^2\,,\quad
D_4=-k_2^2\,,\quad
D_5=-(k_1-k_2)^2\,.
\nonumber
\end{align}
The power of $-p^2$ is evident from dimensionality.
Our aim is to calculate the dimensionless function $G(n_1,n_2,n_3,n_4,n_5)$.
It is symmetric with respect to the interchanges
$(1\leftrightarrow2,3\leftrightarrow4)$ and $(1\leftrightarrow3,2\leftrightarrow4)$.
It vanishes when indices of two adjacent lines are non-positive integers,
because then it contains a no-scale subdiagram.

\begin{figure}[ht]
\begin{center}
\begin{picture}(52,26)
\put(26,13){\makebox(0,0){\includegraphics{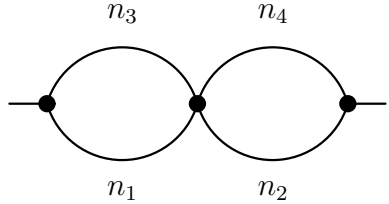}}}
\put(16,0){\makebox(0,0)[b]{$n_1$}}
\put(36,0){\makebox(0,0)[b]{$n_2$}}
\put(16,26){\makebox(0,0)[t]{$n_3$}}
\put(36,26){\makebox(0,0)[t]{$n_4$}}
\end{picture}
\end{center}
\caption{Trivial case $n_5=0$}
\label{F:n50}
\end{figure}

When one of the indices is zero, the problem becomes trivial.
If $n_5=0$, it is the product of two one-loop diagrams (Fig.~\ref{F:n50}):
\begin{equation}
G(n_1,n_2,n_3,n_4,0) = G(n_1,n_3) G(n_2,n_4)\,.
\label{q2:n50}
\end{equation}
For integer $n_i$, the result is proportional to $G_1^2=G(1,1,1,1,0)$,
see~(\ref{q1:g1}).

\begin{figure}[ht]
\begin{center}
\begin{picture}(116,32)
\put(12,17.5){\makebox(0,0){\includegraphics{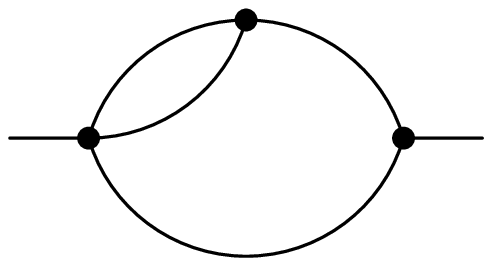}}}
\put(63,17.5){\makebox(0,0){\includegraphics{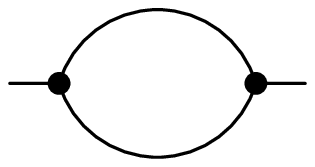}}}
\put(105,17.5){\makebox(0,0){\includegraphics{q1.eps}}}
\put(42,17.5){\makebox(0,0){${}={}$}}
\put(84,17.5){\makebox(0,0){${}\times{}$}}
\put(12,0){\makebox(0,0)[b]{$n_2$}}
\put(2,30){\makebox(0,0){$n_3$}}
\put(22,30){\makebox(0,0){$n_4$}}
\put(11,18.5){\makebox(0,0){$n_5$}}
\put(63,4.5){\makebox(0,0)[b]{$n_5$}}
\put(63,30.5){\makebox(0,0)[t]{$\vphantom{d}n_3$}}
\put(105,4.5){\makebox(0,0)[b]{$n_2$}}
\put(100,30.5){\makebox(0,0)[t]{$n_4+n_3+n_5-d/2$}}
\end{picture}
\end{center}
\caption{Trivial case $n_1=0$}
\label{F:n10}
\end{figure}

\begin{figure}[ht]
\begin{center}
\begin{picture}(32,18)
\put(16,9){\makebox(0,0){\includegraphics{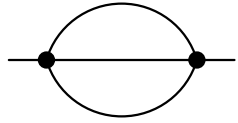}}}
\end{picture}
\end{center}
\caption{Basis integral}
\label{F:b2}
\end{figure}

If $n_1=0$ (Fig.~\ref{F:n10}),
then the inner loop gives
\begin{equation*}
\frac{G(n_3,n_5)}{(-k_2^2)^{n_3+n_5-d/2}}\,,
\end{equation*}
and hence
\begin{equation}
G(0,n_2,n_3,n_4,n_5) = G(n_3,n_5) G(n_2,n_4+n_3+n_5-d/2)\,.
\label{q2:n10}
\end{equation}
The cases $n_2=0$, $n_3=0$, $n_4=0$ are given by the symmetric formulae.
For integer $n_i$, the result is proportional to
\begin{equation}
\begin{split}
&G_2 = G(0,1,1,0,1) = G(1,1) G(1,2-d/2)
= \frac{4 g_2}{(d-3)(d-4)(3d-8)(3d-10)}\,,\\
&g_2 = \frac{\Gamma(1+2\varepsilon)\Gamma^3(1-\varepsilon)}{\Gamma(1-3\varepsilon)}
\end{split}
\label{q2:g2}
\end{equation}
(Fig.~\ref{F:b2}).

But what can we do if all 5 indices are positive?
We shall use integration by parts~\cite{CT:81} ---
a powerful method based on the simple observation
that integrals of full derivatives vanish.
When applied to the integrand of~\ref{q2:G2def},
the derivative $\partial/\partial k_2$ acts as
\begin{equation*}
\frac{\partial}{\partial k_2} \to
\frac{n_2}{D_2}2(k_2+p) + \frac{n_4}{D_4}2k_2 + \frac{n_5}{D_5}2(k_2-k_1)\,.
\end{equation*}
Applying $(\partial/\partial k_2)\cdot(k_2-k_1)$
to this integrand and using
\begin{equation*}
\begin{split}
&(k_2-k_1)^2 = -D_5\,,\quad
2 k_2 \cdot (k_2-k_1) = D_3 - D_4 - D_5\,,\\
&2 (k_2+p) \cdot (k_2-k_1) = D_1 - D_2 - D_5\,,
\end{split}
\end{equation*}
we see that this operation is equivalent to inserting
\begin{equation*}
d-n_2-n_4-2n_5 + \frac{n_2}{D_2}(D_1-D_5) + \frac{n_4}{D_4}(D_3-D_5)
\end{equation*}
under the integral sign in~(\ref{q2:G2def})
(the term $d$ comes from differentiating $k_2$).
The resulting integral vanishes.

On the other hand, we can express it via $G$ with shifted indices.
Let's introduce the notation
\begin{equation}
\1\pm G(n_1,n_2,n_3,n_4,n_5) = G(n_1\pm1,n_2,n_3,n_4,n_5)\,,
\label{q2:pmdef}
\end{equation}
and similarly for $\2\pm$, etc.
Then the relation we have derived
(it is called the triangle relation)
takes the form
\begin{equation}
\bigl[d-n_2-n_4-2n_5
+ n_2\2+(\1--\5-) + n_4\4+(\3--\5-)\bigr] G = 0\,.
\label{q2:tri}
\end{equation}

\begin{figure}[ht]
\begin{center}
\begin{picture}(68,68)
\put(34,34){\makebox(0,0){\includegraphics{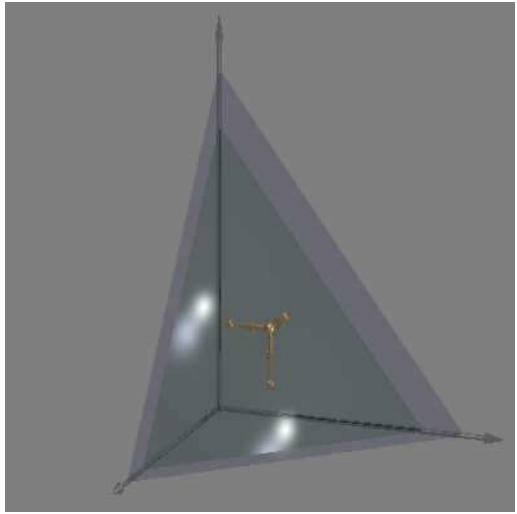}}}
\end{picture}
\end{center}
\caption{Integration by parts}
\label{F:IBP}
\end{figure}

From~(\ref{q2:tri}) we obtain
\begin{equation}
G = \frac{n_2\2+(\5--\1-) + n_4\4+(\5--\3-)}{d-n_2-n_4-2n_5} G\,.
\label{q2:sol}
\end{equation}
Each application of this relation reduces $n_1+n_3+n_5$ by 1
(Fig.~\ref{F:IBP}).
Therefore, sooner or later one of the indices $n_1$, $n_3$, $n_5$ will vanish,
and we'll get the trivial case~(\ref{q2:n50}), (\ref{q2:n10}),
or symmetric to it.

\begin{figure}[p]
\begin{center}
\begin{picture}(32,17)
\put(16,8.5){\makebox(0,0){\includegraphics{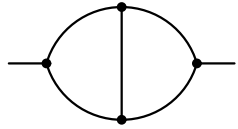}}}
\end{picture}
\end{center}
\caption{Two-loop propagator diagram}
\label{F:top}
\end{figure}

\begin{figure}[p]
\begin{center}
\begin{picture}(94,24)
\put(21,12){\makebox(0,0){\includegraphics{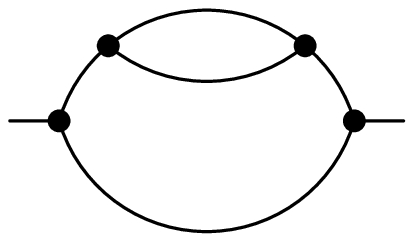}}}
\put(73,12){\makebox(0,0){\includegraphics{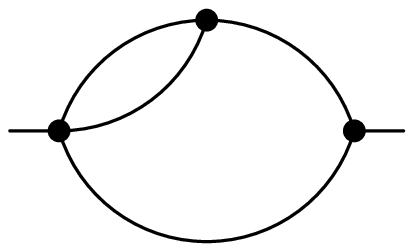}}}
\put(47,12){\makebox(0,0){{${}={}$}}}
\put(6,16){\makebox(0,0)[r]{{$n_1$}}}
\put(36,16){\makebox(0,0)[l]{{$n_2$}}}
\put(86,19){\makebox(0,0)[l]{{$n_1+n_2$}}}
\end{picture}
\end{center}
\caption{Insertion into a propagator}
\label{F:insert}
\end{figure}

\begin{figure}[p]
\begin{center}
\begin{picture}(104,17)
\put(21,8.5){\makebox(0,0){\includegraphics{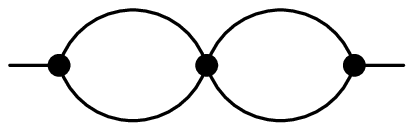}}}
\put(43,8.5){\makebox(0,0)[l]{${}=G_1^2$}}
\put(78,8.5){\makebox(0,0){\includegraphics{q2b2.eps}}}
\put(95,8.5){\makebox(0,0)[l]{${}=G_2$}}
\end{picture}
\end{center}
\caption{Basis integrals}
\label{F:basis}
\end{figure}

\begin{figure}[p]
\begin{center}
\begin{picture}(44,26)
\put(22,13){\makebox(0,0){\includegraphics{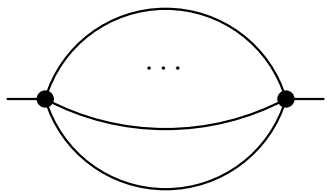}}}
\put(22,17){\makebox(0,0){$\cdots$}}
\end{picture}
\end{center}
\caption{Sunset diagrams}
\label{F:sunset}
\end{figure}

Let's summarize our achievements.
There is one generic topology of two-loop propagator diagrams (Fig.~\ref{F:top}).
This means that all other topologies are reduced cases of this one,
where some line (or lines) shrink(s) to a point,
i.e., the corresponding index (-ices) vanish%
\footnote{At first sight it seems that the diagram with a self-energy insertion
into a propagator (Fig.~\ref{F:insert}) is not a reduced case of Fig.~\ref{F:top}.
But this is not so: we can collect two identical denominators together,
and then it obviously becomes Fig.~\ref{F:top} with one line shrunk.}.
All Feynman integrals of this class,
with any integer indices $n_1$, $n_2$, $n_3$, $n_4$, $n_5$,
can be expressed, by a simple algorithm,
as linear combinations of two basis integrals (Fig.~\ref{F:basis}),
with coefficients being rational functions of $d$.
The basis integrals are $G_1^2$ and $G_2$, where
\begin{equation}
\begin{split}
&G_n = \frac{g_n}{\left(n+1-n\frac{d}{2}\right)_n
\left((n+1)\frac{d}{2}-2n-1\right)_n}\,,\\
&g_n = \frac{\Gamma(1+n\varepsilon)\Gamma^{n+1}(1-\varepsilon)}%
{\Gamma(1-(n+1)\varepsilon)}
\end{split}
\label{q2:sunset}
\end{equation}
is the $n$-loop sunset diagram (Fig.~\ref{F:sunset}).

Let's consider an example: Fig.~\ref{F:top} with all $n_i=1$,
i.e., $G(1,1,1,1,1)$.
Applying~(\ref{q2:sol}) to it, we obtain
\begin{equation}
\raisebox{-7.75mm}{\includegraphics{q2t.eps}}
= \frac{1}{\varepsilon} \left[ \raisebox{-7.75mm}{\includegraphics{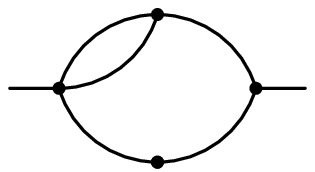}}
- \raisebox{-7.75mm}{\includegraphics{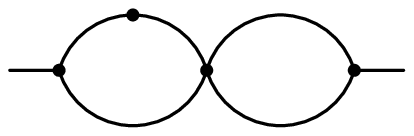}} \right]\,,
\label{q2:ex1}
\end{equation}
where the symmetry has been taken into account.
Here a dot on a line means that this denominator is squared.
We have
\begin{equation}
\begin{split}
G(1,1,1,1,1) &{}= \frac{2}{d-4} \left[ G(1,2,1,1,0) - G(0,2,1,1,1) \right]\\
&{}= \frac{2}{d-4} \left[ G(1,1) G(2,1) - G(1,1) G(2,3-d/2) \right]\\
&{}= \frac{2}{d-4}
\left[ \frac{G(2,1)}{G(1,1)} G_1^2 - \frac{G(2,-d/2+3)}{G(1,-d/2+2)} G_2 \right]\,,
\end{split}
\label{q2:ex2}
\end{equation}
because
\begin{equation*}
G_1 = G(1,1)\,,\quad
G_2 = G(0,1,1,0,1) = G(1,1) G(1,-d/2+2)\,.
\end{equation*}
Using~(\ref{Electron:Gratio}),
\begin{equation*}
\frac{G(2,3-d/2)}{G(1,2-d/2)} = - \frac{(3d-8)(3d-10)}{d-4}\,,
\end{equation*}
and we obtain
\begin{equation}
G(1,1,1,1,1) = \frac{2}{d-4}
\left[ - (d-3) G_1^2 + \frac{(3d-8)(3d-10)}{d-4} G_2 \right]
= \frac{8(g_2-g_1^2)}{(d-3)(d-4)^3}\,.
\label{q2:G11111}
\end{equation}
It is easy to expand this result in $\varepsilon$ using
\begin{equation}
\frac{g_2}{g_1^2}
= \frac{\Gamma(1+2\varepsilon)\Gamma^2(1-2\varepsilon)}%
{\Gamma^2(1+\varepsilon)\Gamma(1-\varepsilon)\Gamma(1-3\varepsilon)}
= 1 - 6 \zeta_3 \varepsilon^3 + \cdots
\label{q2:gratio}
\end{equation}
We arrive at
\begin{equation}
G(1,1,1,1,1) = 6 \zeta_3 + \cdots
\label{q2:z3}
\end{equation}

Three-loop massless propagator diagrams also can be calculated
using integration by parts~\cite{CT:81}.

\subsection{Photon self-energy}
\label{S:Photon2}

\begin{figure}[ht]
\begin{center}
\begin{picture}(106,17)
\put(16,8.5){\makebox(0,0){\includegraphics{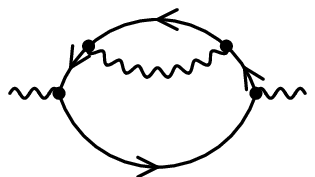}}}
\put(53,8.5){\makebox(0,0){\includegraphics{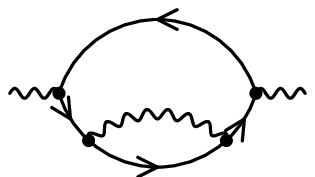}}}
\put(90,8.5){\makebox(0,0){\includegraphics{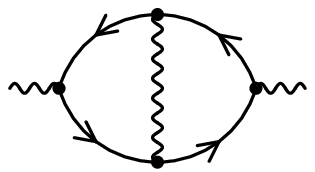}}}
\end{picture}
\end{center}
\caption{Two-loop photon self-energy}
\label{F:Photon2}
\end{figure}

The photon self-energy at two loops is given by 3 diagrams
(Fig.~\ref{F:Photon2}).
It is gauge-invariant, because there are no off-shell charged external particles.
Therefore, we may use any gauge;
the calculation in the Feynman gauge $a_0=1$ is easiest.

\begin{figure}[ht]
\begin{center}
\begin{picture}(70,40)
\put(35,20){\makebox(0,0){\includegraphics{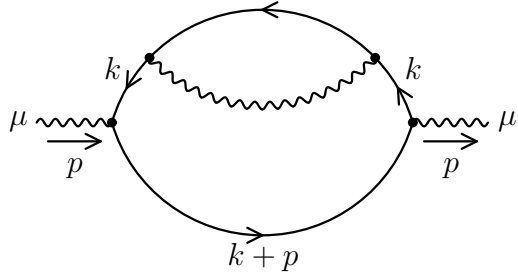}}}
\put(35,0){\makebox(0,0)[b]{$k+p$}}
\put(10,15){\makebox(0,0)[t]{$p$}}
\put(60,15){\makebox(0,0)[t]{$p$}}
\put(15,27){\makebox(0,0){$k$}}
\put(55,27){\makebox(0,0){$k$}}
\put(2.5,20){\makebox(0,0){$\mu$}}
\put(67.5,20){\makebox(0,0){$\mu$}}
\end{picture}
\end{center}
\caption{The first contribution}
\label{F:Photon2a}
\end{figure}

The first two diagrams contribute equally;
we only need to calculate one (Fig.~\ref{F:Photon2a})
and double the result.
This diagram contains one-loop electron self-energy subdiagram
$-i\rlap/p\Sigma_V$:
\begin{equation*}
i \Pi_1{}_\mu^\mu = - \int \frac{d^d k}{(2\pi)^d} \Tr
i e_0 \gamma_\mu i \frac{\rlap/k+\rlap/p}{(k+p)^2} i e_0 \gamma^\mu
i \frac{\rlap/k}{k^2} (-i) \rlap/k \Sigma_V(k^2) i \frac{\rlap/k}{k^2}\,.
\end{equation*}
Substituting the result~(\ref{Electron:Res1}) in the Feynman gauge $a_0=1$,
we obtain
\begin{equation*}
\begin{split}
\Pi_1{}_\mu^\mu &{}= - i \frac{e_0^4}{(4\pi)^{d/2}} G_1 \frac{d-2}{2}
\int \frac{d^d k}{(2\pi)^d}
\frac{\Tr\gamma_\mu(\rlap/k+\rlap/p)\gamma^\mu\rlap/k}%
{\left[-(k+p)^2\right] (-k^2)^{1+\varepsilon}}\\
&{}= i \frac{e_0^4}{(4\pi)^{d/2}} G_1 (d-2)^2
\int \frac{d^d k}{(2\pi)^d}
\frac{2 k\cdot(k+p)}{D_1 D_2^{1+\varepsilon}}\,.
\end{split}
\end{equation*}
Using the ``multiplication table''~(\ref{Photon1:Mult})
and omitting $D_1$ in the numerator, we have
\begin{equation*}
\Pi_1{}_\mu^\mu = - \frac{e_0^4 (-p^2)^{1-2\varepsilon}}{(4\pi)^d} (d-2)^2
G(1,1) \left[G(1,1+\varepsilon) - G(1,\varepsilon)\right]
\end{equation*}
Using the property~(\ref{Electron:Gratio}), we finally arrive at
\begin{equation}
\Pi_1{}_\mu^\mu = - \frac{e_0^4 (-p^2)^{1-2\varepsilon}}{(4\pi)^d}
G_2 \frac{2(d-2)^3}{d-4}\,.
\label{Photon2:Pi1}
\end{equation}

\begin{figure}[ht]
\begin{center}
\begin{picture}(70,40)
\put(35,20){\makebox(0,0){\includegraphics{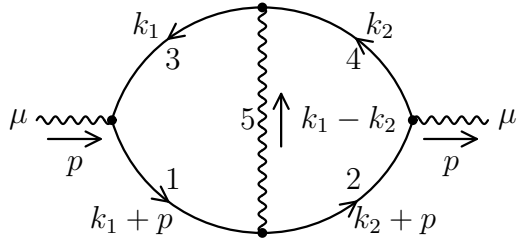}}}
\put(10,15){\makebox(0,0)[t]{$p$}}
\put(60,15){\makebox(0,0)[t]{$p$}}
\put(19.5,32.5){\makebox(0,0){$k_1$}}
\put(50.5,32.5){\makebox(0,0){$k_2$}}
\put(17.5,6.5){\makebox(0,0){$k_1+p$}}
\put(52.5,6.5){\makebox(0,0){$k_2+p$}}
\put(40,20){\makebox(0,0)[l]{$k_1-k_2$}}
\put(23,12){\makebox(0,0){1}}
\put(47,12){\makebox(0,0){2}}
\put(23,28){\makebox(0,0){3}}
\put(47,28){\makebox(0,0){4}}
\put(33,20){\makebox(0,0){5}}
\put(2.5,20){\makebox(0,0){$\mu$}}
\put(67.5,20){\makebox(0,0){$\mu$}}
\end{picture}
\end{center}
\caption{The second contribution}
\label{F:Photon2b}
\end{figure}

The second contribution (Fig.~\ref{F:Photon2b}) is more difficult.
It is a truly two-loop diagram (Fig.~\ref{F:q2}):
\begin{equation*}
i \Pi_2{}_\mu^\mu = - \int \frac{d^d k_1}{(2\pi)^d} \frac{d^d k_2}{(2\pi)^d}
\Tr i e_0 \gamma_\mu i \frac{\rlap/k_2+\rlap/p}{(k_2+p)^2} i e_0 \gamma_\nu
i \frac{\rlap/k_1+\rlap/p}{(k_1+p)^2}
i e_0 \gamma^\mu
i \frac{\rlap/k_1}{k_1^2} i e_0 \gamma^\nu i \frac{\rlap/k_2}{k_2^2}
\frac{-i}{(k_1-k_2)^2}\,,
\end{equation*}
or
\begin{equation*}
\begin{split}
&\Pi_2{}_\mu^\mu = - e_0^4 \int \frac{d^d k_1}{(2\pi)^d} \frac{d^d k_2}{(2\pi)^d}
\frac{N}{D_1 D_2 D_3 D_4 D_5}\,,\\
&N = \Tr \gamma_\mu (\rlap/k_2+\rlap/p)\gamma_\nu(\rlap/k_1+\rlap/p)
\gamma^\mu\rlap/k_1\gamma^\nu\rlap/k_2\,.
\end{split}
\end{equation*}
According to~(\ref{q2:G2def}), the ``multiplication table'' of the momenta is
\begin{equation}
\begin{split}
&p^2 = -1\,,\quad
k_1^2 = - D_3\,,\quad
k_2^2 = - D_4\,,\\
&p\cdot k_1 = \frac{1}{2} (1 + D_3 - D_1)\,,\quad
p\cdot k_2 = \frac{1}{2} (1 + D_4 - D_2)\,,\\
&k_1\cdot k_2 = \frac{1}{2} (D_5 - D_3 - D_4)\,.
\end{split}
\label{Photon2:mult}
\end{equation}
All products $D_i D_j$ in the numerator can be omitted
(produce vanishing integrals), except $D_1 D_4$ and $D_2 D_3$.
Calculating the trace with this simplification, we obtain
\begin{equation*}
\begin{split}
N \Rightarrow{}& 2 (d-2) \Bigl[ - (d-4) (D_1 D_4 + D_2 D_3)\\
&{} + 2 (D_1 + D_2 + D_3 + D_4)
- 2 D_5^2 + (d-8) D_5 - 2\Bigr]\,.
\end{split}
\end{equation*}
Therefore,
\begin{equation*}
\begin{split}
\Pi_2{}_\mu^\mu ={}& \frac{e_0^4 (-p^2)^{1-2\varepsilon}}{(4\pi)^d}
2 (d-2) \Bigl[ - 2 (d-4) G(0,1,1,0,1) + 8 G(0,1,1,1,1)\\
&{} - 2 G(1,1,1,1,-1) + (d-8) G(1,1,1,1,0) - 2 G(1,1,1,1,1) \Bigr]\,.
\end{split}
\end{equation*}

Some integrals here are trivial:
\begin{equation*}
G(1,1,1,1,0) = G_1^2\,,\quad
G(0,1,1,0,1) = G_2\,,
\end{equation*}
or simple:
\begin{equation*}
G(0,1,1,1,1) = G(1,1) G(1,3-d/2) = \frac{3d-8}{d-4} G_2
\end{equation*}
(here we used~(\ref{Electron:Gratio})).
The only truly two-loop integral, $G(1,1,1,1,1)$,
has been calculated in Sect.~\ref{S:q2},
and is given by~(\ref{q2:G11111}).
But what is $G(1,1,1,1,-1)$?
This integral would factor into two one-loop ones (Fig.~\ref{F:n50})
if it had no numerator.
In principle, it is possible to calculate such integrals
using integration-by-parts recurrence relations.
But we have not discussed the necessary methods.
Therefore, we shall calculate this integral in a straightforward manner:
\begin{equation*}
G(1,1,1,1,-1) = - \frac{1}{\pi^d}
\int \frac{d^d k_1}{D_1 D_3} \frac{d^d k_2}{D_2 D_4}
\left[-(k_1-k_2)^2\right]
= - \frac{2}{\pi^d}
\int \frac{k_1\,d^d k_1}{D_1 D_3} \cdot
\int \frac{k_2\,d^d k_2}{D_2 D_4}\,.
\end{equation*}
Both vector integrals are directed along $p$;
therefore, we may project them onto $p$:
\begin{equation}
G(1,1,1,1,-1) = - \frac{2}{\pi^d p^2}
\left( \int \frac{k_1\cdot p\,d^d k_1}{D_1 D_3}\right)^2
= - \frac{1}{2} G_1^2\,.
\label{Photon2:G1111m}
\end{equation}

Collecting all this together, we get
\begin{equation}
\Pi_2{}_\mu^\mu = \frac{e_0^4 (-p^2)^{1-2\varepsilon}}{(4\pi)^d}
2 \frac{d-2}{d-4} \left[ (d^2-7d+16) G_1^2
- 2 \frac{d^3-6d^2+20d-32}{d-4} G_2 \right]\,.
\label{Photon2:Pi2}
\end{equation}
Separate contributions to $\Pi_{\mu\nu}(p)$ are not transverse,
but their sum is, i.e., it has the structure~(\ref{Photon:Ward2}).
The full two-loop photon self-energy
\begin{equation*}
\Pi_2(p^2) = - \frac{2 \Pi_1{}_\mu^\mu + \Pi_2{}_\mu^\mu}{(d-1)(-p^2)}
\end{equation*}
is
\begin{equation}
\Pi_2(p^2) = \frac{e_0^4 (-p^2)^{-2\varepsilon}}{(4\pi)^d}
2 \frac{d-2}{(d-1)(d-4)}
\left[ - (d^2-7d+16) G_1^2
+ 4 \frac{(d-3)(d^2-4d+8)}{d-4} G_2 \right]\,.
\label{Photon2:Pi}
\end{equation}
The one-loop result (Sect.~\ref{S:Photon1}) is given by~(\ref{Photon1:Res1}).

\subsection{Photon field renormalization}
\label{S:PhotonZ2}

The transverse part of the photon propagator,
up to $e^4$, has the form
\begin{equation}
p^2 D_\bot(p^2) = \frac{1}{1-\Pi(p^2)}
= 1 + \frac{e_0^2 (-p^2)^{-\varepsilon}}{(4\pi)^{d/2}} f_1(\varepsilon)
+ \frac{e_0^4 (-p^2)^{-2\varepsilon}}{(4\pi)^d} f_2(\varepsilon)
+ \cdots
\label{PhotonZ2:form}
\end{equation}
From~(\ref{Photon1:Res1}) and~(\ref{Photon2:Pi}),
using~(\ref{q1:g1}) and~(\ref{q2:g2}), we have
\begin{equation*}
\begin{split}
\varepsilon f_1(\varepsilon) ={}&
- 4 \frac{1-\varepsilon}{(1-2\varepsilon)(3-2\varepsilon)} g_1\,,\\
\varepsilon^2 f_2 (\varepsilon) ={}&
4 \frac{1-\varepsilon}{\varepsilon(3-2\varepsilon)}
\left[ \frac{6-3\varepsilon+4\varepsilon^2-4\varepsilon^3}%
{(1-2\varepsilon)^2(3-2\varepsilon)} g_1^2
- 2 \frac{2-2\varepsilon+\varepsilon^2}{(1-3\varepsilon)(2-3\varepsilon)} g_2
\right]\,.
\end{split}
\end{equation*}
We can expand these functions in $\varepsilon$:
\begin{align*}
\varepsilon e^{\gamma\varepsilon} f_1(\varepsilon)
&{}= c_{10} + c_{11} \varepsilon + c_{12} \varepsilon^2 + \cdots\\
\varepsilon^2 e^{2\gamma\varepsilon} f_2(\varepsilon)
&{}= c_{20} + c_{21} \varepsilon + c_{22} \varepsilon^2 + \cdots
\end{align*}
At one loop, using
\begin{equation*}
g_1 = 1 - \frac{1}{2} \zeta_2 \varepsilon^2 + \cdots
\end{equation*}
we obtain
\begin{equation*}
\varepsilon e^{\gamma\varepsilon} f_1(\varepsilon)
= - \frac{4}{3} - \frac{20}{9} \varepsilon
+ \left( \frac{2}{3} \zeta_2 - \frac{112}{27} \right) \varepsilon^2
+ \cdots
\end{equation*}
At two loops, using~(\ref{q2:gratio}),
we see that $1/\varepsilon$ terms cancel, and
\begin{equation*}
\varepsilon^2 e^{2\gamma\varepsilon} f_2(\varepsilon)
= \frac{16}{9} + \frac{106}{27} \varepsilon
+ \left( 16 \zeta_3 - \frac{16}{9} \zeta_2 - \frac{7}{3} \right) \varepsilon^2
+ \cdots
\end{equation*}

In order to re-express it via $\alpha(\mu)$,
it is sufficient to use~(\ref{QEDl:e2})
with the one-loop renormalization constant~(\ref{Vertex:Z1}), (\ref{Vertex:ZalphaRes}):
\begin{equation}
\begin{split}
p^2 D_\bot(p^2) ={}& 1 + \frac{\alpha(\mu)}{4\pi\varepsilon}
e^{-L\varepsilon}
\left( 1 - \beta_0 \frac{\alpha}{4\pi\varepsilon} + \cdots \right)
\varepsilon e^{\gamma\varepsilon} f_1(\varepsilon)\\
&{} + \left(\frac{\alpha}{4\pi\varepsilon}\right)^2
e^{-2L\varepsilon} \varepsilon^2 e^{2\gamma\varepsilon} f_2(\varepsilon)
+ \cdots
\end{split}
\label{PhotonZ2:ren}
\end{equation}
where
\begin{equation*}
L = \log \frac{-p^2}{\mu^2}
\end{equation*}
(indicating the argument of $\alpha$ in the $\alpha^2$ term
would be beyond our accuracy).
Substituting the $\varepsilon$ expansions,
we obtain at $L=0$ (i.e.\ $\mu^2=-p^2$)
\begin{equation*}
\begin{split}
p^2 D_\bot(p^2) ={}& 1
+ \frac{\alpha(\mu)}{4\pi\varepsilon}
(c_{10} + c_{11} \varepsilon + c_{12} \varepsilon^2 + \cdots)\\
&{} + \left(\frac{\alpha}{4\pi\varepsilon}\right)^2
\left[c_{20} + c_{21} \varepsilon + c_{22} \varepsilon^2 + \cdots
- \beta_0 (c_{10} + c_{11} \varepsilon + c_{12} \varepsilon^2 + \cdots) \right]
+ \cdots
\end{split}
\end{equation*}
This should be equal to $Z_A(\alpha(\mu)) p^2 D_\bot^r(p^2;\mu)$,
where
\begin{equation}
Z_A(\alpha) = 1 + \frac{\alpha}{4\pi\varepsilon} z_1
+ \left(\frac{\alpha}{4\pi\varepsilon}\right)^2
(z_{20} + z_{21} \varepsilon)
+ \cdots
\label{PhotonZ2:ZAform}
\end{equation}
and
\begin{equation}
p^2 D_\bot^r(p^2;\mu) = 1
+ \frac{\alpha(\mu)}{4\pi} (r_1 + r_{11} \varepsilon + \cdots)
+ \left(\frac{\alpha}{4\pi}\right)^2
(r_2 + \cdots)
+ \cdots
\label{PhotonZ2:Drform}
\end{equation}
Equating $\alpha$ terms, we obtain
\begin{equation}
z_1 = c_{10}\,,\quad
r_1 = c_{11}\,,\quad
r_{11} = c_{12}\,.
\label{PhotonZ2:a1}
\end{equation}
Equating $\alpha^2$ terms, we obtain
\begin{equation}
z_{20} = c_{20} - \beta_0 c_{10}\,,\quad
z_{21} = c_{21} - (c_{10}+\beta_0) c_{11}\,,\quad
r_2 = c_{22} - (c_{10}+\beta_0) c_{12}\,.
\label{PhotonZ2:a2}
\end{equation}
Therefore, we arrive at the photon field renormalization constant
\begin{equation}
Z_A(\alpha) = 1 - \frac{4}{3} \frac{\alpha}{4\pi\varepsilon}
- 2 \varepsilon \left(\frac{\alpha}{4\pi\varepsilon}\right)^2
+ \cdots
\label{PhotonZ2:ZA}
\end{equation}
and the renormalized photon propagator at $\mu^2=-p^2$ (and $\varepsilon=0$)
\begin{equation}
p^2 D^r_\bot(p^2;\mu^2=-p^2) = 1 - \frac{20}{9} \frac{\alpha(\mu)}{4\pi}
+ \left( 16 \zeta_3 - \frac{55}{3} \right)
\left(\frac{\alpha}{4\pi}\right)^2
+ \cdots
\label{PhotonZ2:D0}
\end{equation}

The anomalous dimension~(\ref{PhotonZ:gamma}) of the photon field is
\begin{equation}
\begin{split}
\gamma_A &{}= \frac{d\log Z_A}{d\log\mu}
= \frac{d}{d\log\mu} \left[ z_1 \frac{\alpha}{4\pi\varepsilon}
+ \left( z_{20} - \frac{1}{2} z_1^2 + z_{21} \varepsilon \right)
\left(\frac{\alpha}{4\pi\varepsilon}\right)^2 \right]\\
&{}= z_1 \frac{\alpha}{4\pi\varepsilon}
\left( - 2 \varepsilon - 2 \beta_0 \frac{\alpha}{4\pi} \right)
+ 2 \left( z_{20} - \frac{1}{2} z_1^2 + z_{21} \varepsilon \right)
\left(\frac{\alpha}{4\pi\varepsilon}\right)^2 ( - 2 \varepsilon )\\
&{}= - 2 z_1 \frac{\alpha}{4\pi}
- 4 \left[ \left( z_{20} - \frac{1}{2} z_1^2 + \frac{1}{2} \beta_0 z_1 \right)
\frac{1}{\varepsilon} + z_{21} \right]
\left(\frac{\alpha}{4\pi}\right)^2\,.
\end{split}
\label{PhotonZ2:gamma}
\end{equation}
It must be finite at $\varepsilon\to0$.
Therefore, $z_{20}$,
the coefficient of $1/\varepsilon^2$ in the two-loop term in $Z_A$,
cannot be arbitrary.
It must satisfy
\begin{equation}
z_{20} = \frac{1}{2} z_1 (z_1 - \beta_0)\,.
\label{PhotonZ2:sc1}
\end{equation}
In other words, $c_{20}$,
the coefficient of $1/\varepsilon^2$ in the two-loop term
in $D_\bot(p^2)$~(\ref{PhotonZ2:form}), must satisfy
\begin{equation}
c_{20} = \frac{1}{2} c_{10} (c_{10} + \beta_0)\,.
\label{PhotonZ2:sc2}
\end{equation}
Then
\begin{equation*}
\gamma_{A0} = - 2 z_1\,,\quad
\gamma_{A1} = - 4 z_{21}\,,
\end{equation*}
i.e., the coefficients in the anomalous dimension
are determined by the coefficients of $1/\varepsilon$ in $Z_A$.
Therefore, $Z_A$ must have the form
\begin{equation}
Z_A = 1 - \frac{1}{2} \gamma_{A0} \frac{\alpha}{4\pi\varepsilon}
+ \frac{1}{8} \left[ \gamma_{A0} (\gamma_{A0}+2\beta_0) - 2 \gamma_{A1} \varepsilon\right]
\left(\frac{\alpha}{4\pi\varepsilon}\right)^2 + \cdots
\label{PhotonZ2:Z2}
\end{equation}
This is indeed so (see~(\ref{Vertex:betaRes}) for $\beta_0$),
and we obtain
\begin{equation}
\gamma_A(\alpha) = \frac{8}{3} \frac{\alpha}{4\pi}
+ 8 \left(\frac{\alpha}{4\pi}\right)^2 + \cdots
\label{PhotonZ2:gammaA}
\end{equation}

This can also be understood in a slightly different way.
The information contained in $Z_A$
is equivalent to that in $\gamma_A(\alpha)$.
This renormalization constant is gauge-invariant,
because $\Pi(p^2)$ is gauge-invariant in QED
(there are no off-shell external charged particles in it;
this is not so in QCD, where gluons are ``charged'').
Therefore, $Z_A$ depends on $\mu$ only via $\alpha(\mu)$
(there is no $a(\mu)$ in it).
Dividing~(\ref{PhotonZ:gamma}) by~(\ref{Vertex:RG}), we obtain
\begin{equation}
\frac{d\log Z_A}{d\log\alpha}
= - \frac{1}{2} \frac{\gamma_A(\alpha)}{\varepsilon+\beta(\alpha)}
= - \frac{\gamma_A(\alpha)}{2\varepsilon}
+ \frac{\beta(\alpha) \gamma_A(\alpha)}{2\varepsilon^2} + \cdots
\label{PhotonZ2:RG}
\end{equation}
Any minimal~(\ref{QEDl:min}) renormalization constant
can be represented as
\begin{equation}
Z_A = \exp \left( \frac{Z_1}{\varepsilon}
+ \frac{Z_2}{\varepsilon^2} + \cdots \right)\,,
\label{PhotonZ2:Zexp}
\end{equation}
where $Z_1$ starts from the order $\alpha$,
$Z_2$ --- from $\alpha^2$, and so on.
Then
\begin{equation*}
\frac{d Z_1}{d\log\alpha} = - \frac{1}{2} \gamma_A(\alpha)\,,\quad
\frac{d Z_2}{d\log\alpha} = \frac{1}{2} \beta(\alpha) \gamma_A(\alpha)\,,\ldots
\end{equation*}
and
\begin{equation}
\begin{split}
&Z_1 = - \frac{1}{2} \int_0^\alpha \gamma_A(\alpha) \frac{d\alpha}{\alpha}
= - \frac{1}{2} \gamma_{A0} \frac{\alpha}{4\pi}
- \frac{1}{4} \gamma_{A1} \left(\frac{\alpha}{4\pi}\right)^2 - \cdots\\
&Z_2 = \frac{1}{2} \int_0^\alpha \beta(\alpha) \gamma_A(\alpha) \frac{d\alpha}{\alpha}
= \frac{1}{4} \beta_0 \gamma_{A0} \left(\frac{\alpha}{4\pi}\right)^2 + \cdots\\
&\cdots
\end{split}
\label{PhotonZ2:Z12}
\end{equation}
One can obtain $\gamma_A(\alpha)$ from $Z_1$,
the coefficient of $1/\varepsilon$ in $Z_A$,
and vice versa.
Higher poles ($1/\varepsilon^2$, $1/\varepsilon^3$, \dots)
contain no new information:
at each order in $\alpha$,
their coefficients ($Z_2$, $Z_3$, \dots)
can be reconstructed from lower-loop results.
Up to two loops, this gives us~(\ref{PhotonZ2:Z2}).

Now let's return to~(\ref{PhotonZ2:ren}) with arbitrary $L$.
It should be equal to the product of~(\ref{PhotonZ2:ZAform})
and~(\ref{PhotonZ2:Drform}).
Equating $\alpha$ terms, we obtain
\begin{equation*}
z_1 = c_{10}\,,\quad
r_1(L) = c_{11} - c_{10} L\,,\quad
r_{11}(L) = c_{12} - c_{11} L + c_{10} \frac{L^2}{2}\,.
\end{equation*}
Equating $\alpha^2$ terms, we obtain
\begin{equation*}
\begin{split}
&z_{20} = c_{20} - \beta_0 c_{10}\,,\quad
z_{21} = c_{21} - 2 c_{20} L - (c_{10}+\beta_0) (c_{11}-c_{10}L)\,,\\
&r_2(L) = c_{22} - 2 c_{21} L + 2 c_{20} L^2
- (c_{10}+\beta_0) \left(c_{12} - c_{11} L + c_{10} \frac{L^2}{2}\right)\,.
\end{split}
\end{equation*}
But the renormalization constant $Z_A$ cannot depend
on kinematics of a specific process, i.e., on $L$.
Therefore, terms with $L$ in $z_{21}$ must cancel.
This is ensured by the consistency condition~(\ref{PhotonZ2:sc2}).
The renormalization constant is given by~(\ref{PhotonZ2:Z2}),
and the renormalized propagator at $\varepsilon=0$ --- by
\begin{equation}
\begin{split}
&p^2 D_\bot^r(p^2;\mu) = 1
+ \frac{\alpha(\mu)}{4\pi} \left( r_1(0) + \frac{1}{2} \gamma_{A0} L \right)\\
&{} + \left(\frac{\alpha}{4\pi}\right)^2
\left[ r_2(0)
+ \frac{1}{2} \left(\gamma_{A1}+r_1(0)\,(\gamma_{A0}-2\beta_0)\right) L
+ \frac{1}{8} \gamma_{A0} (\gamma_{A0}-2\beta_0) L^2 \right]
+ \cdots
\end{split}
\label{PhotonZ2:DL1}
\end{equation}

This result can be also obtained by solving
the RG equation~(\ref{PhotonZ:RGL}).
Taking into account
\begin{equation*}
\frac{d\alpha}{dL} = \beta(\alpha) \alpha\,,
\end{equation*}
we find that the coefficients obey the following equations
\begin{equation*}
\frac{d r_1(L)}{dL} = \frac{1}{2} \gamma_0\,,\quad
\frac{d r_2(L)}{dL} = \frac{1}{2} \gamma_1
+ \left( \frac{1}{2} \gamma_0 - \beta_0 \right) r_1(L)\,.
\end{equation*}
Solving them, we reproduce~(\ref{PhotonZ2:DL1}).

Substituting the coefficients, we obtain
\begin{equation}
p^2 D_\bot^r(p^2;\mu) = 1
+ \frac{\alpha(\mu)}{4\pi}
\left( \frac{4}{3} L - \frac{20}{9} \right)
+ \left(\frac{\alpha(\mu)}{4\pi}\right)^2
\left( \frac{16}{9} L^2 - \frac{52}{27} L + 10 \zeta_3 - \frac{55}{3} \right)
+ \cdots
\label{PhotonZ2:DL}
\end{equation}
This is a typical example of perturbative series for a quantity
with a single energy scale ($-p^2$).
In principle, we can choose the renormalization scale $\mu$ arbitrarily:
physical results don't depend on it.
However, if we choose it far away from the energy scale of the process,
the coefficients in the series contain powers
of the large logarithm $L$, and hence are large.
Therefore, truncating the series after some term
produces large errors.
It is better to choose the renormalization scale
of order of the characteristic energy scale,
then $|L|\lesssim1$.
The coefficients contain no large logarithm
and are just numbers (one hopes, of order 1).
The convergence is better.

For example, when describing QED processes at LEP,
at energies $\sim m_W$, it would be a very poor idea
to use the low-energy $\alpha$ with $\mu\sim m_e$:
coefficients of perturbative series would contain powers
of a huge logarithm $\log(m_W/m_e)$.
Using $\alpha(m_W)$, which is about $7\%$ larger,
makes the behaviour of perturbative series much better.

This renormalization-group improvement works for all processes
with a single characteristic energy scale.
If there are several widely separated scales,
no universal method exists.
For some classes of processes,
there are some specific methods,
such as, e.g., factorization.
But this depends on the process under consideration very much.

\subsection{Charge renormalization}
\label{S:Charge}

In QED, it is enough to know $Z_A$
to obtain charge renormalization~(\ref{Vertex:ZalphaA}).
With two-loop accuracy, we have from~(\ref{PhotonZ2:ZA}
\begin{equation}
Z_\alpha = Z_A^{-1} = 1 + \frac{4}{3} \frac{\alpha}{4\pi\varepsilon}
+ \left(\frac{16}{9} + 2 \varepsilon\right)
\left(\frac{\alpha}{4\pi\varepsilon}\right)^2\,.
\label{Charge:Zalpha}
\end{equation}
The $\beta$-function~(\ref{Vertex:beta}) is simply
\begin{equation}
\beta(\alpha) = - \frac{1}{2} \gamma_A(\alpha)\,,
\label{Charge:betaA}
\end{equation}
or (see~(\ref{PhotonZ2:gammaA}))
\begin{equation}
\beta(\alpha) = - \frac{4}{3} \frac{\alpha}{4\pi}
- 4 \left(\frac{\alpha}{4\pi}\right)^2 + \cdots
\label{Charge:beta}
\end{equation}

The information contained in $Z_\alpha$
is equivalent to that in $\beta(\alpha)$.
Dividing~(\ref{Vertex:beta}) by~(\ref{Vertex:RG}), we have
\begin{equation}
\frac{d\log Z_\alpha}{d\log\alpha}
= - \frac{\beta(\alpha)}{\varepsilon+\beta(\alpha)}
= - \frac{\beta(\alpha)}{\varepsilon}
+ \frac{\beta^2(\alpha)}{\varepsilon^2} + \cdots
\label{Charge:RG}
\end{equation}
Writing $Z_\alpha$ in the form~(\ref{PhotonZ2:Zexp}), we have
\begin{equation*}
\frac{d Z_1}{d\log\alpha} = - \beta(\alpha)\,,\quad
\frac{d Z_2}{d\log\alpha} = \beta^2(\alpha)\,,\ldots
\end{equation*}
and
\begin{equation}
\begin{split}
&Z_1 = - \int_0^\alpha \beta(\alpha) \frac{d\alpha}{\alpha}
= - \beta_0 \frac{\alpha}{4\pi}
- \beta_1 \left(\frac{\alpha}{4\pi}\right)^2 - \cdots\\
&Z_2 = \int_0^\alpha \beta^2(\alpha) \frac{d\alpha}{\alpha}
= \frac{1}{2} \beta_0^2 \left(\frac{\alpha}{4\pi}\right)^2 + \cdots\\
&\cdots
\end{split}
\label{Charge:Z12}
\end{equation}
One can obtain $\beta(\alpha)$ from $Z_1$,
the coefficient of $1/\varepsilon$ in $Z_\alpha$,
and vice versa.
Higher poles ($1/\varepsilon^2$, $1/\varepsilon^3$, \dots)
contain no new information:
at each order in $\alpha$,
their coefficients ($Z_2$, $Z_3$, \dots)
can be reconstructed from lower-loop results.
Up to two loops,
\begin{equation}
Z_\alpha = 1 - \beta_0 \frac{\alpha}{4\pi\varepsilon}
+ \left(\beta_0^2 - \frac{1}{2} \beta_1 \varepsilon\right)
\left(\frac{\alpha}{4\pi\varepsilon}\right)^2 + \cdots
\label{Charge:Z2}
\end{equation}
Comparing this with~(\ref{Charge:Zalpha}),
we again obtain~(\ref{Charge:beta}).

We shall later need $\beta(\alpha)$ for QED with $n_f$ massless lepton fields
(e.g., electrons and muons at energies $\gg m_\mu$).
The photon self-energy at one loop~(\ref{Photon1:Res1}) (Fig.~\ref{F:Photon1})
and at two loops~(\ref{Photon2:Pi}) (Fig.~\ref{F:Photon2})
gets the factor $n_f$ (at three loops, there are both $n_f$ and $n_f^2$ terms).
If $1-\Pi(p^2)$ is expressed via the renormalized $\alpha(\mu)$,
it should be equal to
\begin{equation*}
1 - \Pi(p^2) = Z_\alpha(\alpha(\mu))
\left[ p^2 D_\bot^r(p^2;\mu) \right]^{-1}\,,
\end{equation*}
where $Z_\alpha(\alpha)$ is minimal~(\ref{QEDl:min}),
and the second factor is finite at $\varepsilon\to0$.
Now we shall follow the derivation in Sect.~\ref{S:PhotonZ2}.
At one loop~(\ref{PhotonZ2:a1}), $z_1$ gets the factor $n_f$,
because $c_{10}$ gets it (from $\Pi_1(p^2)$).
Therefore, $\beta_0$ becomes $n_f$ times larger.
At two loops~(\ref{PhotonZ2:a2}),
$z_{21}$ also simply gets the factor $n_f$ from $c_{21}$,
because $c_{10}+\beta_0=z_1+\beta_0=0$ for $Z_\alpha$~(\ref{Charge:Z2}).
Therefore,
\begin{equation}
\beta(\alpha) = - \frac{4}{3} n_f \frac{\alpha}{4\pi}
- 4 n_f \left(\frac{\alpha}{4\pi}\right)^2 + \cdots
\label{Charge:betanf}
\end{equation}
At three loops, both $n_f$ and $n_f^2$ terms appear.
The photon field anomalous dimension is, from~(\ref{Charge:betaA}),
\begin{equation}
\gamma_A(\alpha) = \frac{8}{3} n_f \frac{\alpha}{4\pi}
+ 8 n_f \left(\frac{\alpha}{4\pi}\right)^2 + \cdots
\label{Charge:gammanf}
\end{equation}
It is easy to write down the renormalization constants
$Z_\alpha$~(\ref{Charge:Z2}) and $Z_A$~(\ref{PhotonZ2:Z2}).

\subsection{Electron self-energy}
\label{S:Electron2}

\begin{figure}[ht]
\begin{center}
\begin{picture}(106,13)
\put(16,6.5){\makebox(0,0){\includegraphics{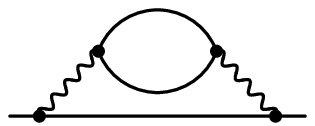}}}
\put(53,5){\makebox(0,0){\includegraphics{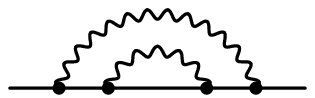}}}
\put(90,6){\makebox(0,0){\includegraphics{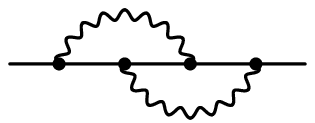}}}
\end{picture}
\end{center}
\caption{Two-loop electron self-energy}
\label{F:Electron2}
\end{figure}

The electron self-energy at two loops is given by 3 diagrams
(Fig.~\ref{F:Electron2}).

\begin{figure}[ht]
\begin{center}
\begin{picture}(52,24)
\put(26,14){\makebox(0,0){\includegraphics{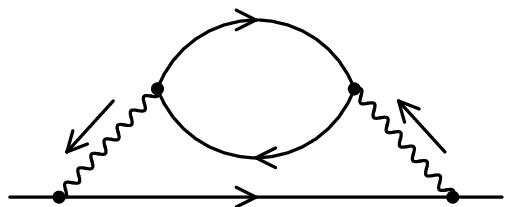}}}
\put(26,0){\makebox(0,0)[b]{$k+p$}}
\put(45,14){\makebox(0,0){$k$}}
\put(48,7){\makebox(0,0){$\mu$}}
\put(39,17){\makebox(0,0){$\mu$}}
\put(7,14){\makebox(0,0){$k$}}
\put(4,7){\makebox(0,0){$\nu$}}
\put(13,17){\makebox(0,0){$\nu$}}
\end{picture}
\end{center}
\caption{Photon self-energy insertion}
\label{F:Electron2a}
\end{figure}

\begin{sloppypar}
The first diagram (Fig.~\ref{F:Electron2a}) contains one-loop photon self-energy subdiagram
$i \left(k^2 g_{\mu\nu} - k_\mu k_\nu\right) \Pi(k^2)$.
This tensor is transverse, and hence longitudinal parts of the photon propagators
drop out.
Therefore, this diagram is gauge-invariant:
\begin{equation*}
- i \Sigma_{V1} \rlap/p
= \int \frac{d^d k}{(2\pi)^d} i e_0 \gamma^\mu i \frac{\rlap/k+\rlap/p}{(k+p)^2}
i e_0 \gamma^\nu \left(\frac{-i}{k^2}\right)^2
i \left(k^2 g_{\mu\nu} - k_\mu k_\nu\right) \Pi(k^2)\,.
\end{equation*}
We shall set $p^2=-1$;
the power of $-p^2$ will be restored in the result by dimensionality.
Taking $\frac{1}{4}\Tr\rlap/p$ of both sides
and substituting the result~(\ref{Photon1:Res1}), we obtain
\begin{align*}
&\Sigma_{V1} = i \frac{e_0^4}{(4\pi)^{d/2}} \frac{d-2}{d-1} G_1
\int \frac{d^d k}{(2\pi)^d} \frac{N}{D_1 D_2^{2+\varepsilon}}\,,\\
&N = \frac{1}{2} \Tr \rlap/p \gamma^\mu (\rlap/k+\rlap/p) \gamma^\nu
\cdot \left(k^2 g_{\mu\nu} - k_\mu k_\nu\right)\,.
\end{align*}
Using the ``multiplication table''~(\ref{Photon1:Mult})
and omitting $D_1$ in the numerator, we have
\begin{equation*}
N \Rightarrow (d-2) D_2^2 - (d-3) D_2 - 1\,,
\end{equation*}
and
\begin{equation*}
\Sigma_{V1} = \frac{e_0^4}{(4\pi)^d} \frac{d-2}{d-1} G_1
\left[ G(1,2+\varepsilon) + (d-3) G(1,1+\varepsilon) - (d-2) G(1,\varepsilon)
\right]\,.
\end{equation*}
Finally, using~(\ref{Electron:Gratio}) and restoring the power of $-p^2$,
we arrive at
\begin{equation}
\Sigma_{V1} = \frac{e_0^4 (-p^2)^{-2\varepsilon}}{(4\pi)^d}
G_2 \frac{2(d-2)^2}{d-6}\,.
\label{Electron2:Sigma1}
\end{equation}
\end{sloppypar}

\begin{figure}[ht]
\begin{center}
\begin{picture}(44,25)
\put(22,12.5){\makebox(0,0){\includegraphics{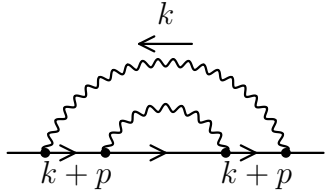}}}
\put(10,0){\makebox(0,0)[b]{$k+p$}}
\put(34,0){\makebox(0,0)[b]{$k+p$}}
\put(22,25){\makebox(0,0)[t]{$k$}}
\end{picture}
\end{center}
\caption{Electron self-energy insertion}
\label{F:Electron2b}
\end{figure}

\begin{sloppypar}
The second diagram (Fig.~\ref{F:Electron2b}) contains one-loop electron self-energy subdiagram
$-i\Sigma_V((k+p)^2)(\rlap/k+\rlap/p)$.
For simplicity, we shall calculate it in the Feynman gauge $a_0=1$:
\begin{equation*}
- i \Sigma_{V2} \rlap/p
= \int \frac{d^d k}{(2\pi)^d} i e_0 \gamma^\mu i \frac{\rlap/k+\rlap/p}{(k+p)^2}
(-i) \Sigma_V((k+p)^2) (\rlap/k+\rlap/p) i \frac{\rlap/k+\rlap/p}{(k+p)^2}
i e_0 \gamma_\mu \frac{-i}{k^2}\,.
\end{equation*}
Taking $\frac{1}{4}\Tr\rlap/p$ (with $p^2=-1$)
and substituting the result~(\ref{Electron:Res1}), we obtain
\begin{equation*}
\Sigma_{V2} = - i \frac{e_0^4}{(4\pi)^{d/2}} \frac{d-2}{4} G_1
\int \frac{d^d k}{(2\pi)^d} \frac{N}{D_1^{1+\varepsilon}D_2}\,,\quad
N = \frac{1}{2} \Tr \rlap/p \gamma^\mu (\rlap/k+\rlap/p) \gamma_\mu\,.
\end{equation*}
Using~(\ref{Photon1:Mult}) and omitting $D_2$ in the numerator, we have
$N \Rightarrow (d-2) (D_1+1)$, or
\begin{equation*}
\Sigma_{V2} = \frac{e_0^4}{(4\pi)^d} \frac{(d-2)^2}{4} G_1
\left[ G(1,1+\varepsilon) + G(1,\varepsilon) \right]\,.
\end{equation*}
Finally, we arrive at
\begin{equation}
\Sigma_{V2} = \frac{e_0^4 (-p^2)^{-2\varepsilon}}{(4\pi)^d} G_2
\frac{(d-2)^2 (d-3)}{d-4}\,.
\label{Electron2:Sigma2F}
\end{equation}
\end{sloppypar}

The result in the arbitrary covariant gauge is very simple:
\begin{equation}
\Sigma_{V2} = \frac{e_0^4 (-p^2)^{-2\varepsilon}}{(4\pi)^d} G_2
\frac{(d-2)^2 (d-3)}{d-4} a_0^2\,.
\label{Electron2:Sigma2}
\end{equation}
The reason is following.
It is not difficult to show (using~(\ref{Electron:Gratio}))
that the diagram of Fig.~\ref{F:Electron1} (see~(\ref{Electron:Sigma1}))
with the denominator $D_1$ raised to an arbitrary power $n$ instead of 1
is proportional to $a_0$ (prove this!).
This means that in order to calculate this diagram
with an arbitrary insertion(s) into the electron line,
we may take the upper photon propagator in the Feynman gauge,
and then multiply the result by $a_0$.

\begin{figure}[ht]
\begin{center}
\begin{picture}(52,36)
\put(26,18){\makebox(0,0){\includegraphics{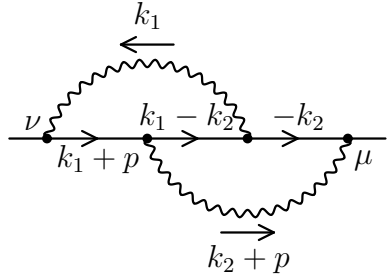}}}
\put(39.33333333,19){\makebox(0,0)[b]{$-k_2$}}
\put(12.66666667,17){\makebox(0,0)[t]{$k_1+p$}}
\put(24.5,19){\makebox(0,0)[b]{$k_1-k_2$}}
\put(19.33333333,36){\makebox(0,0)[t]{$k_1$}}
\put(32.66666667,0){\makebox(0,0)[b]{$k_2+p$}}
\put(48,15){\makebox(0,0){$\mu$}}
\put(4,20){\makebox(0,0){$\nu$}}
\end{picture}
\end{center}
\caption{Truly two-loop diagram}
\label{F:Electron2c}
\end{figure}

The third diagram (Fig.~\ref{F:Electron2c}) is truly two-loop;
it has the topology of Fig.~\ref{F:q2}.
In the Feynman gauge,
\begin{equation*}
- i \Sigma_{V3} \rlap/p = \int \frac{d^d k_1}{(2\pi)^d} \frac{d^d k_2}{(2\pi)^d}
i e_0 \gamma^\mu i \frac{-\rlap/k_2}{k_2^2} i e_0 \gamma^\nu
i \frac{\rlap/k_1-\rlap/k_2}{(k_1-k_2)^2} i e_0 \gamma_\mu
i \frac{\rlap/k_1+\rlap/p}{(k_1+p)^2} i e_0 \gamma_\nu
\frac{-i}{k_1^2} \frac{-i}{(k_2+p)^2}\,,
\end{equation*}
or
\begin{align*}
&\Sigma_{V3} = e_0^4 \int \frac{d^d k_1}{(2\pi)^d} \frac{d^d k_2}{(2\pi)^d}
\frac{N}{D_1 D_2 D_3 D_4 D_5}\,,\\
&N = \frac{1}{4} \Tr \rlap/p \gamma^\mu \rlap/k_2 \gamma^\nu (\rlap/k_1-\rlap/k_2)
\gamma_\mu (\rlap/k_1+\rlap/p) \gamma_\nu\,.
\end{align*}
Using the ``multiplication table''~(\ref{Photon2:mult})
and omitting all products $D_i D_j$ except $D_1 D_4$ and $D_2 D_3$, we have
\begin{equation*}
\begin{split}
N \Rightarrow{}& \frac{d-2}{2} \Bigl[ (d-4) D_1 D_4 - (d-8) D_2 D_3\\
&{} + 2 (D_1 - D_2 - D_3 + D_4) + 2 (D_2^2 + D_3^2) + (d-4) D_5 \Bigr]\,.
\end{split}
\end{equation*}
Therefore,
\begin{equation*}
\Sigma_{V3} = - \frac{e_0^4 (-p^2)^{-2\varepsilon}}{(4\pi)^d} \frac{d-2}{2}
\Bigl[ 4 G(0,1,1,0,1) + 4 G(1,-1,1,1,1) + (d-4) G(1,1,1,1,0) \Bigr]\,.
\end{equation*}

What is $G(1,-1,1,1,1)$?
It would be trivial to calculate first the inner loop integral ($d^d k_2$)
and then the outer one ($d^d k_1$), if there were no $D_2$ in the numerator.
In principle, it is possible to calculate such integrals using
integration-by-parts recurrence relations.
But we have not discussed the necessary methods.
Therefore, we shall calculate this integral in a straightforward manner:
\begin{equation*}
\begin{split}
G(1,-1,1,1,1) &{}= - \frac{1}{\pi^d}
\int \frac{d^d k_1}{D_1 D_3} \int \frac{d^d k_2}{D_4 D_5}
\left[-(k_2+p)^2\right]\\
&{}= - \frac{1}{\pi^d}
\int \frac{d^d k_1}{D_1 D_3} \int \frac{d^d k_2}{D_4 D_5}
\left(1 - 2 p \cdot k_2\right)\,.
\end{split}
\end{equation*}
In the second term, the inner vector integral is directed along $k_1$,
and we may substitute
\begin{equation*}
k_2 \to \frac{k_2\cdot k_1}{k_1^2} k_1 \to \frac{1}{2} k_1
\end{equation*}
(the terms with $D_4$ or $D_5$ in the numerator give vanishing integrals).
Therefore,
\begin{equation*}
\begin{split}
G(1,-1,1,1,1) &{}= - \frac{1}{\pi^d}
\int \frac{d^d k_1}{D_1 D_3} \left(1 - p \cdot k_1\right)
\int \frac{d^d k_2}{D_4 D_5}\\
&{}= - \frac{1}{\pi^d}
\int \frac{d^d k_1}{D_1 D_3} \frac{1}{2} (1 - D_3)
\int \frac{d^d k_2}{D_4 D_5}\\
&{}= \frac{1}{2} \left[ G(1,0,1,1,1) - G(1,0,0,1,1) \right]\,.
\end{split}
\end{equation*}
Using the trivial integrals
\begin{equation*}
G(1,0,0,1,1) = G_2\,,\quad
G(1,0,1,1,1) = \frac{3d-8}{d-4} G_2\,,
\end{equation*}
we obtain
\begin{equation}
G(1,-1,1,1,1) = \frac{d-2}{d-4} G_2\,.
\label{Electron2:G1m111}
\end{equation}

Substituting the integrals, we obtain the Feynman-gauge result
\begin{equation}
\Sigma_{V3} = - \frac{e_0^4 (-p^2)^{-2\varepsilon}}{(4\pi)^d} \frac{d-2}{2}
\left[ (d-4) G_1^2 + 8 \frac{d-3}{d-4} G_2 \right]\,.
\label{Electron2:Sigma3F}
\end{equation}
A similar, but more lengthy calculation in the arbitrary covariant gauge yields
\begin{equation}
\Sigma_{V3} = - \frac{e_0^4 (-p^2)^{-2\varepsilon}}{(4\pi)^d} \frac{d-2}{4}
\left[ \bigl((d-2) a_0^2 + d - 6\bigr) G_1^2
- 2 \frac{d-3}{d-4} \bigl((d-4) a_0^2 - d - 4\bigr) \right]\,.
\label{Electron2:Sigma3}
\end{equation}

We shall consider a more general case of QED with $n_f$ massless lepton fields;
the results for the usual QED can be easily obtained by setting $n_f=1$.
Then the diagram of Fig.~\ref{F:Electron2a} gives~(\ref{Electron2:Sigma1})
with the extra factor $n_f$;
the diagram of Fig.~\ref{F:Electron2b} gives~(\ref{Electron2:Sigma2}),
and that of Fig.~\ref{F:Electron2c} gives~(\ref{Electron2:Sigma3}).
The complete result for the two-loop term in $\Sigma_V(p^2)$ is
\begin{equation}
\begin{split}
\Sigma_{2V}(p^2) = \frac{e_0^4 (-p^2)^{-2\varepsilon}}{(4\pi)^d} (d-2)
\biggl[& 2 \frac{d-2}{d-6} G_2 n_f
- \frac{1}{4} \bigl((d-2) a_0^2 + d - 6\bigr) G_1^2\\
&{}+ \frac{1}{2} \frac{d-3}{d-4} \bigl((3d-8) a_0^2 - d - 4\bigr) G_2 \biggr]\,.
\end{split}
\label{Electron2:Sigma}
\end{equation}

\subsection{Electron field renormalization}
\label{S:ElectronZ2}

Now we proceed as in Sect.~\ref{S:PhotonZ2}.
The electron propagator $\rlap/p S(p)$ has the form similar to~(\ref{PhotonZ2:form}),
but now the coefficients are gauge-dependent:
\begin{equation*}
f_1(\varepsilon) = f_1'(\varepsilon) + f_1''(\varepsilon) a_0\,.
\end{equation*}
From~(\ref{Electron:Res1}) and~(\ref{Electron2:Sigma}),
using~(\ref{q1:g1}) and~(\ref{q2:g2}), we have
\begin{equation*}
\begin{split}
\varepsilon f_1(\varepsilon) ={}&
- \frac{1-\varepsilon}{1-2\varepsilon} g_1 a_0\,,\\
\varepsilon^2 f_2 (\varepsilon) ={}&
(1-\varepsilon) \biggl[
\frac{2\varepsilon(1-\varepsilon)}%
{(1+\varepsilon)(1-2\varepsilon)(1-3\varepsilon)(2-3\varepsilon)}
g_2 n_f\\
&\hphantom{(1-\varepsilon)\biggl[\biggr.}
+ \frac{1+\varepsilon}{(1-2\varepsilon)^2} g_1^2
- \frac{4-\varepsilon-(2-3\varepsilon)a_0^2}{2(1-3\varepsilon)(2-3\varepsilon)} g_2
\biggr]\,.
\end{split}
\end{equation*}
We can expand these functions in $\varepsilon$:
\begin{align*}
\varepsilon e^{\gamma\varepsilon} f_1'(\varepsilon)
&{}= c_{10}' + c_{11}' \varepsilon + c_{12}' \varepsilon^2 + \cdots\,,\quad
\varepsilon e^{\gamma\varepsilon} f_1''(\varepsilon)
= c_{10}'' + c_{11}'' \varepsilon + c_{12}'' \varepsilon^2 + \cdots\\
\varepsilon^2 e^{2\gamma\varepsilon} f_2(\varepsilon)
&{}= c_{20} + c_{21} \varepsilon + c_{22} \varepsilon^2 + \cdots
\end{align*}
In our particular case, $f_1'(\varepsilon)=0$,
\begin{equation*}
\begin{split}
\varepsilon e^{\gamma\varepsilon} f_1''(\varepsilon) &{}=
- 1 - \varepsilon - \left(2 - \frac{1}{2} \zeta_2\right) \varepsilon^2 + \cdots\\
\varepsilon^2 e^{2\gamma\varepsilon} f_2(\varepsilon) &{}=
\frac{a_0^2}{2} + \left( n_f + a^2 + \frac{3}{4} \right) \varepsilon
+ \left( \frac{7}{2} n_f + 3 a_0^2 - \frac{1}{2} \zeta_2 a_0^2 + \frac{5}{8} \right)
\varepsilon^2 + \cdots
\end{split}
\end{equation*}
The situation here is simpler than in Sect.~\ref{S:PhotonZ2}:
the terms with $g_1^2$ and with $g_2$ in $\varepsilon^2 f_2(\varepsilon)$
are separately finite at $\varepsilon\to0$,
so that we can put $g_2=g_1^2$ with the $\varepsilon^2$ accuracy,
and $\zeta_3$ does not appear.

The propagator expressed via the renormalized quantities $\alpha(\mu)$, $a(\mu)$
is, with the $\alpha^2$ accuracy,
\begin{equation*}
\rlap/p S(p) = 1
+ \frac{\alpha(\mu)}{4\pi\varepsilon} e^{-L\varepsilon} Z_\alpha
\varepsilon e^{\gamma\varepsilon}
\left[f_1'(\varepsilon) + f_1''(\varepsilon) Z_A a(\mu)\right]
+ \left(\frac{\alpha}{4\pi\varepsilon}\right)^2 e^{-2L\varepsilon}
\varepsilon^2 e^{2\gamma\varepsilon} f_2(\varepsilon)\,,
\end{equation*}
see~(\ref{PhotonZ2:ren}).
At $L=0$ we have
\begin{equation*}
\begin{split}
&\rlap/p S(p) = 1 + \frac{\alpha(\mu)}{4\pi\varepsilon}
\left[ c_{10}' + c_{10}'' a(\mu)
+ \left(c_{11}'+c_{11}''a(\mu)\right) \varepsilon
+ \left(c_{12}'+c_{12}''a(\mu)\right) \varepsilon^2
+ \cdots \right]\\
&{} + \left(\frac{\alpha}{4\pi\varepsilon}\right)^2
\biggl[ c_{20} + c_{21} \varepsilon + c_{22} \varepsilon^2 + \cdots
- \beta_0 \left(c_{10} + c_{11} \varepsilon + c_{12} \varepsilon^2 + \cdots\right)\\
&\hphantom{{}+\left(\frac{\alpha}{4\pi\varepsilon}\right)^2\biggl[\biggr.}
+ \frac{1}{2} \gamma_{A0} a
\left(c_{10}'' + c_{11}'' \varepsilon + c_{12}'' \varepsilon^2 + \cdots\right)
\biggr]\,.
\end{split}
\end{equation*}
Here $c_{1n}=c_{1n}'+c_{1n}''a(\mu)$;
in $c_{2n}$, we may substitute $a(\mu)$ instead of $a_0$.
This should be equal to $Z_\psi(\alpha(\mu),a(\mu)) \rlap/p S_r(p;\mu)$.
Equating $\alpha$ terms, we obtain~(\ref{PhotonZ2:a1});
equating $\alpha^2$ terms, we now have
\begin{equation*}
\begin{split}
z_{20} &{}= c_{20} - \beta_0 c_{10} - \frac{1}{2} \gamma_{A0} c_{10}'' a\,,\\
z_{21} &{}= c_{21} - (c_{10}+\beta_0) c_{11} - \frac{1}{2} \gamma_{A0} c_{11}'' a\,,\\
r_2 &{}= c_{22} - (c_{10}+\beta_0) c_{12} - \frac{1}{2} \gamma_{A0} c_{12}'' a\,.
\end{split}
\end{equation*}
There are extra terms as compared to the gauge-invariant case~(\ref{PhotonZ2:a2}).
In QED $\gamma_{A0}=-2\beta_0$.

We obtain
\begin{equation}
Z_\psi(\alpha,a) = 1 - a \frac{\alpha}{4\pi\varepsilon}
+ \left[ \frac{a^2}{2} + \left( n_f + \frac{3}{4} \right) \varepsilon \right]
\left(\frac{\alpha}{4\pi\varepsilon}\right)^2 + \cdots
\label{ElectronZ2:Z}
\end{equation}
It is also easy to write down $\rlap/p S_r(p;\mu)$ at $L=0$ (i.e., $\mu^2=-p^2$),
or even at arbitrary $\mu$, but we shall not do this here.

The anomalous dimension of the electron field is
\begin{equation*}
\begin{split}
\gamma_\psi &{}= \frac{d\log Z_\psi}{d\log\mu}
= \left[ - 2 (\varepsilon+\beta(\alpha)) \frac{\partial}{\partial\log\alpha}
- \gamma_A(\alpha) \frac{\partial}{\partial\log a} \right]\\
&\hphantom{{}=\frac{d\log Z_\psi}{d\log\mu}=}
\left[ (z_1'+z_1''a) \frac{\alpha}{4\pi\varepsilon}
+ \left( z_{20} - \frac{1}{2} z_1^2 + z_{21} \varepsilon \right)
\left(\frac{\alpha}{4\pi\varepsilon}\right)^2
\right]\\
&{} = - 2 z_1 \frac{\alpha}{4\pi}
- 4 \left[ \left( z_{20} - \frac{1}{2} z_1^2 + \frac{1}{2} \beta_0 z_1
+ \frac{1}{4} \gamma_{A0} z_1'' a \right) \frac{1}{\varepsilon}
+ z_{21} \right]
\left(\frac{\alpha}{4\pi}\right)^2\,,
\end{split}
\end{equation*}
where $z_1=z_1'+z_1''a$.
It must be finite at $\varepsilon\to0$;
therefore, the two-loop term must satisfy the self-consistency relation
\begin{equation}
z_{20} = \frac{1}{2} z_1 (z_1-\beta_0) - \frac{1}{4} \gamma_{A0} z_1'' a\,,
\label{ElectronZ2:sc1}
\end{equation}
or
\begin{equation}
c_{20} = \frac{1}{2} c_{10} (c_{10}+\beta_0) + \frac{1}{4} \gamma_{A0} c_{10}'' a\,.
\label{ElectronZ2:sc2}
\end{equation}
The renormalization constant of a non-gauge-invariant quantity
(such as the electron field) must have the form,
up to two loops,
\begin{equation}
Z_\psi = 1 - \frac{1}{2} \gamma_{\psi0} \frac{\alpha}{4\pi\varepsilon}
+ \frac{1}{8} \left[ \gamma_{\psi0} (\gamma_{\psi0}+2\beta_0)
+ \gamma_{A0} \gamma_{\psi0}'' a
- 2 \gamma_{\psi1} \varepsilon\right]
\left(\frac{\alpha}{4\pi\varepsilon}\right)^2 + \cdots
\label{ElectronZ2:Z2}
\end{equation}
where $\gamma_{\psi0}=\gamma_{\psi0}'+\gamma_{\psi0}''a$
(this formula generalizes the gauge-invariant case~(\ref{PhotonZ2:Z2})).

We see that the self-consistency condition is indeed satisfied, and
\begin{equation}
\gamma_\psi(\alpha,a) = 2 a \frac{\alpha}{4\pi}
- (4 n_f + 3) \left(\frac{\alpha}{4\pi}\right)^2 + \cdots
\label{ElectronZ2:gamma}
\end{equation}
In the normal QED, with just one charged lepton field,
\begin{equation}
\gamma_\psi(\alpha,a) = 2 a \frac{\alpha}{4\pi}
- 7 \left(\frac{\alpha}{4\pi}\right)^2 + \cdots
\label{ElectronZ2:gamma2}
\end{equation}

The electron mass anomalous dimension can be found in a similar way.
We can calculate $\Sigma_S(p^2)$ at two loops (neglecting $m^2$)
by retaining $m_0$ in the numerator of a single electron propagator
in Fig.~\ref{F:Electron2} and setting $m_0\to0$ in all the other places.
This single $m_0$ has to be somewhere along the electron line
which goes through all the diagrams,
not in the electron loop in the first diagram:
we need one helicity flip of the external electron,
and one helicity flip in a loop yields zero contribution.
Then we extract $Z_m$ from~(\ref{Mass:Z}).
It must be gauge-invariant, because $m(\mu)$ is gauge-invariant.
The result is
\begin{equation}
\gamma_m(\alpha) = 6 \frac{\alpha}{4\pi}
- \left( \frac{20}{3} n_f - 3 \right)
\left(\frac{\alpha}{4\pi}\right)^2 + \cdots
\label{ElectronZ2:gammam}
\end{equation}
In the normal QED, with just one charged lepton field,
\begin{equation}
\gamma_m(\alpha) = 6 \frac{\alpha}{4\pi}
- \frac{11}{3} \left(\frac{\alpha}{4\pi}\right)^2 + \cdots
\label{ElectronZ2:gammam2}
\end{equation}

\subsection{Two-loop corrections in QCD}
\label{S:QCD2}

The one-loop $\beta$-function in QCD is~(\ref{alphas:beta0});
the two-loop one is
\begin{equation}
\beta_1 = \frac{34}{3} C_A^2 - 4 C_F T_F n_f - \frac{20}{3} C_A T_F n_f\,.
\label{QCD2:beta1}
\end{equation}
The second term follows from the QED result~(\ref{Charge:beta});
non-abelian terms (containing $C_A$) are more difficult to derive.

How to express $\alpha_s(\mu')$ via $\alpha_s(\mu)$,
if the scales $\mu$ and $\mu'$ are not too widely separated?
We need to solve the RG equation~(\ref{alphas:RG})
with the initial condition at $\mu$.
Let's introduce short notation:
\begin{equation*}
a_s = \frac{\alpha_s(\mu)}{4\pi}\,,\quad
a_s' = \frac{\alpha_s(\mu')}{4\pi}\,,\quad
b_1 = \frac{\beta_1}{\beta_0}\,.
\end{equation*}
Then the integral of our equation is
\begin{equation}
- \int_{a_s}^{a_s'} \frac{1}{1 + b_1 a_s + \cdots} \frac{d a_s}{a_s^2}
= 2 \beta_0 \log \frac{\mu'}{\mu}\,.
\label{QCD2:int}
\end{equation}
It is natural to introduce notation
\begin{equation*}
l = 2 \beta_0 \log \frac{\mu'}{\mu}\,.
\end{equation*}
Expanding the left-hand side, we obtain
\begin{equation}
- \int_{a_s}^{a_s'} \left(1 - b_1 a_s + \cdots\right) \frac{d a_s}{a_s^2}
= \frac{1}{a_s'} - \frac{1}{a_s} + b_1 \log \frac{a_s'}{a_s} + \cdots = l\,.
\label{QCD2:intexp}
\end{equation}
The solution is a series:
\begin{equation*}
a_s' = a_s \left(1 + c_1 a_s + c_2 a_s^2 + \cdots\right)\,.
\end{equation*}
Substituting it into~(\ref{QCD2:intexp}), we obtain
\begin{equation*}
\frac{1}{a_s} \left[1 - c_1 a_s + (c_1^2-c_2) a_s^2 + \cdots\right]
- \frac{1}{a_s}
+ b_1 \left[c_1 a_s + \cdots\right] = l
\end{equation*}
Therefore, $c_1 = -l$ and $c_2 = l^2 - b_1 l$.
The final result is
\begin{equation}
a_s' = a_s \left[1 - l a_s + (l^2 - b_1 l) a_s^2 + \cdots\right]\,.
\label{QCD2:mumu}
\end{equation}
It is easy to find $a_s^3$, \dots{} corrections;
they will contain $b_2$, \dots{}

It is not possible to find the exact solution of the RG equation
for $a_s(\mu)$ in elementary functions,
if two (or more) terms in $\beta(a_s)$ are kept%
\footnote{With two terms, the solution can be written via the Lambert $W$-function.}.
However, an implicit solution can be obtained.
Separating variables in the RG equation~(\ref{alphas:RG}),
we can write it as
\begin{equation}
\frac{1}{2\beta(a_s)} \frac{d a_s}{a_s} = - d \log\mu\,.
\label{QCD2:sepvar}
\end{equation}
Let's subtract and add two first terms in the expansion of the integrand in $a_s$:
\begin{equation}
\int_0^{a_s(\mu)} \left(\frac{1}{2\beta(a_s)} - \frac{1}{2\beta_0 a_s}
+ \frac{\beta_1}{2\beta_0^2}\right) \frac{d a_s}{a_s}
- \frac{1}{2\beta_0 a_s(\mu)}
- \frac{\beta_1}{2\beta_0^2} \log a_s(\mu)
= - \log \frac{\mu}{\Lambda_{\MS}}\,.
\end{equation}
The added terms are explicitly integrated;
the difference behaves well at $a_a\to0$,
and can be integrated from $0$.
The integration constant $\Lambda_{\MS}$ has appeared here,
as in the one-loop case~(\ref{alphas:Lambda}).
We can solve for this constant:
\begin{equation}
\Lambda = \mu \exp \left( - \frac{1}{2 \beta_0 a_s(\mu)} \right)
a_s(\mu)^{-\beta_1/(2\beta_0^2)} K(a_s(\mu))\,,
\label{QCD2:Lambda}
\end{equation}
where
\begin{equation}
K(a_s) = \exp \int_0^{a_s}
\left(\frac{1}{2\beta(a_s)} - \frac{1}{2\beta_0 a_s}
+ \frac{\beta_1}{2\beta_0^2} \right) \frac{d a_s}{a_s}
= 1 + \cdots
\label{QCD2:K}
\end{equation}
This is a regular series in $a_s$;
its $a_s$ term contains $\beta_2$, and so on.

The quark mass anomalous dimension is, up to two loops,
\begin{equation}
\gamma_m = 6 C_F \frac{\alpha_s}{4\pi}
+ C_F \left( 3 C_F + \frac{97}{3} C_A - \frac{20}{3} T_F n_f \right)
\left(\frac{\alpha_s}{4\pi}\right)^2 + \cdots
\label{QCD2:gammam}
\end{equation}
Everything here, except the $C_A$ term in $\gamma_{m1}$,
can be obtained from QED result~(\ref{ElectronZ2:gammam})
by inserting obvious colour factors.
The solution~(\ref{Mass:RGsol}) of the RG equation
can be written in the following convenient form.
Let's subtract and add the first term of the expansion
of the integrand in $\alpha_s$;
the difference can be integrated from 0:
\begin{equation}
m(\mu') = m(\mu)
\left(\frac{\alpha(\mu')}{\alpha(\mu)}\right)^{\gamma_{m0}/(2\beta_0)}
K_{\gamma_m}(\alpha_s(\mu')) K_{\gamma_m}^{-1}(\alpha_s(\mu))\,,
\label{QCD2:RGsol}
\end{equation}
where for any anomalous dimension
\begin{equation*}
\gamma(\alpha_s) = \gamma_0 \frac{\alpha_s}{4\pi}
+ \gamma_1 \left(\frac{\alpha_s}{4\pi}\right)^2 + \cdots
\end{equation*}
we define
\begin{equation}
\begin{split}
K_\gamma(\alpha_s) &{}= \exp \int_0^{\alpha_s}
\left( \frac{\gamma(\alpha_s)}{2\beta(\alpha_s)} - \frac{\gamma_0}{2\beta_0} \right)
\frac{d\alpha_s}{\alpha_s}\\
&{}= 1 + \frac{\gamma_0}{2\beta_0}
\left( \frac{\gamma_1}{\gamma_0} - \frac{\beta_1}{\beta_0} \right)
\frac{\alpha_s}{4\pi} + \cdots
\end{split}
\label{QCD2:Kg}
\end{equation}
This function has the obvious properties
\begin{equation*}
K_0(\alpha_s) = 1\,,\quad
K_{-\gamma}(\alpha_s) = K_\gamma^{-1}(\alpha_s)\,,\quad
K_{\gamma_1+\gamma_2}(\alpha_s) = K_{\gamma_1}(\alpha_s) K_{\gamma_2}(\alpha_s)\,.
\end{equation*}
The solution~(\ref{QCD2:RGsol}) can also be rewritten as
\begin{equation}
m(\mu) = \hat{m}
\left(\frac{\alpha_s(\mu)}{4\pi}\right)^{\gamma_{m0}/(2\beta_0)}
K_{\gamma_m}(\alpha_s(\mu))\,,
\label{QCD2:RGsol2}
\end{equation}
where $\hat{m}$ is a renormalization group invariant
which characterizes the given quark flavour.
The running mass $m(\mu)$ decreases when $\mu$ increases,
because $\gamma_{m0}=6C_F>0$, $\beta_0>0$,
so that the exponent in~(\ref{QCD2:RGsol2}) is positive,
and $\alpha_s(\mu)$ decreases with $\mu$.

The QCD $\beta$- and $\gamma$-functions are also known
at three~\cite{TVZ:80,T:82,LV:93}
and four~\cite{RVL:97,C:97,VLR:97,CR:00,Ch:05,Cz:05} loops.

\section{On-shell renormalization scheme}
\label{S:OS}

\subsection{On-shell renormalization of photon field}
\label{S:OSPhoton}

We shall consider QED with non-zero electron mass~(\ref{Mass:L}).
Until now, we used \MS{} renormalization scheme.
In this scheme, the electron mass $m(\mu)$ and the coupling $\alpha(\mu)$
depend on the renormalization scale $\mu$.
It would be more exact to call \MS{} a one-parameter family
of renormalization schemes.
But what are the experimentally measured
electron mass $m\approx0.511\,\text{MeV}$
and coupling $\alpha\approx1/137$?
They are defined in another renormalization scheme ---
on-shell scheme.
It contains no parameters.
\MS{} scheme is most useful at high energies,
when the electron mass can be neglected
(or considered as a small correction);
$\mu$ should be of order of characteristic energies.
When energies are of order $m$,
the on-shell scheme is often more convenient.
We shall discuss it now.

The photon field renormalized in the on-shell scheme $A_{\text{os}}$
is related to the bare one $A_0$ by
\begin{equation}
A_0 = \left(Z_A^{\text{os}}\right)^{1/2} A_{\text{os}}\,,
\label{OSPhoton:A}
\end{equation}
where $Z_A^{\text{os}}$ is the renormalization constant
(it is not minimal in the sense~(\ref{QEDl:min})).
The renormalized propagator is related to the bare one by
\begin{equation}
D_\bot(p^2) = Z_A^{\text{os}} D_\bot^{\text{os}}(p^2)\,.
\label{OSPhoton:Dos}
\end{equation}
The bare photon propagator near the mass shell is
\begin{equation}
D_\bot(p^2) = \frac{1}{1 - \Pi(p^2)} \frac{1}{p^2}
= \frac{1}{1 - \Pi(0)} \frac{1}{p^2} + \cdots\,.
\label{OSPhoton:D}
\end{equation}
We require, by definition, that it behaves
as the free propagator $1/p^2$ near the mass shell.
Therefore, the photon-field renormalization constant
in the on-shell scheme is
\begin{equation}
Z_A^{\text{os}} = \frac{1}{1 - \Pi(0)}\,.
\label{OSPhoton:Z}
\end{equation}

How can we calculate $\Pi(0)$?
From~(\ref{Photon:Ward2}) we have
\begin{equation*}
\Pi_\mu^\mu(p) = (d-1) p^2 \Pi(p^2)\,,
\end{equation*}
and hence
\begin{equation}
\left. \frac{\partial}{\partial p_\nu} \frac{\partial}{\partial p^\nu}
\Pi_\mu^\mu(p) \right|_{p=0} = 2 d (d-1) \Pi(0)\,.
\label{OSPhoton:Pi0}
\end{equation}
The left-hand side has zero external momenta ---
it is a vacuum diagram,
with the only energy scale $m$.
It is not difficult to calculate a few more terms
of expansion of $\Pi(p^2)$ in $p^2$ in the same manner.

\begin{figure}[ht]
\begin{center}
\begin{picture}(32,25)
\put(16,12.5){\makebox(0,0){\includegraphics{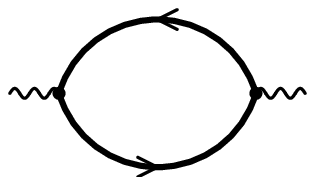}}}
\put(16,0){\makebox(0,0)[b]{$k+p$}}
\put(16,25){\makebox(0,0)[t]{$k$}}
\end{picture}
\end{center}
\caption{One-loop photon self-energy}
\label{F:Photon1m}
\end{figure}

At one loop (Fig.~\ref{F:Photon1m}),
setting $m_0=1$ (it will be restored by dimensionality),
we have
\begin{equation}
\Pi(0) = \frac{i e_0^2}{2d(d-1)}
\int \frac{d^d k}{(2\pi)^d} \Tr \gamma_\mu S_0(k) \gamma^\mu
\left[ \frac{\partial}{\partial p_\nu} \frac{\partial}{\partial p^\nu}
S_0(k+p) \right]_{p=0}\,.
\label{OSPhoton:Pi1}
\end{equation}
Expanding the free electron propagator $S_0(k+p)$~(\ref{Mass:S0}) in $p$,
\begin{equation*}
\begin{split}
S_0(k+p) ={}& \frac{\rlap/k+1}{k^2-1}
+ \frac{\rlap/p}{k^2-1}
- \frac{2p\cdot k\,(\rlap/k+1)}{(k^2-1)^2}\\
&{} - \frac{2p\cdot k\,\rlap/p}{(k^2-1)^2}
- \frac{p^2(\rlap/k+1)}{(k^2-1)^2}
+ \frac{(2p\cdot k)^2(\rlap/k+1)}{(k^2-1)^3}
+ \mathcal{O}(p^3)\,,
\end{split}
\end{equation*}
we obtain
\begin{equation*}
\left[ \frac{\partial}{\partial p_\nu} \frac{\partial}{\partial p^\nu}
S_0(k+p) \right]_{p=0}
= - \frac{2d(\rlap/k+1)+4\rlap/k}{(k^2-1)^2}
+ \frac{8k^2(\rlap/k+1)}{(k^2-1)^3}\,.
\end{equation*}
Using also~(\ref{Photon1:gamma1}), (\ref{Photon1:gamma0})
to simplify $\gamma_\mu S_0(k) \gamma^\mu$,
we can easily calculate the trace:
\begin{equation*}
\begin{split}
\Pi(0) &{}= \frac{4 i e_0^2}{d(d-1)}
\int \frac{d^d k}{(2\pi)^d}
\left[ \frac{8}{D^4} + \frac{4(d-3)}{D^3} + \frac{d^2-4d+4}{D^2} \right]\\
&{}= - 4 \frac{e_0^2}{(4\pi)^{d/2}}
\left[ 8 V(4) + 4 (d-3) V(3) + (d^2-4d+4) V(2) \right]\,,
\end{split}
\end{equation*}
where $D=1-k^2$, and the definition~(\ref{V1:def})
of the vacuum integrals was used.
Using~(\ref{V1:res}), we can express all $V(n)$
via $V(2)=\Gamma(\varepsilon)$;
also restoring the power of $m_0$ by dimensionality,
we finally arrive at
\begin{equation}
\Pi(0) = - \frac{4}{3} \frac{e_0^2 m_0^{-2\varepsilon}}{(4\pi)^{d/2}}
\Gamma(\varepsilon)\,.
\label{OSPhoton:Pi}
\end{equation}
The ultraviolet divergence of the diagram of Fig.~\ref{F:Photon1}
does not depend on masses and external momenta,
and is the same in~(\ref{OSPhoton:Pi}) as in~(\ref{Photon1:Res2}).

Therefore, the photon-field renormalization constant
in the on-shell scheme is, with one-loop accuracy,
\begin{equation}
Z_A^{\text{os}} = 1
- \frac{4}{3} \frac{e_0^2 m_0^{-2\varepsilon}}{(4\pi)^{d/2}}
\Gamma(\varepsilon)\,.
\label{OSPhoton:ZA}
\end{equation}
The renormalized photon propagator in the \MS{} scheme
and in the on-shell scheme are both ultraviolet-finite, therefore,
the ratio $Z_A^{\text{os}}/Z_A(\alpha(\mu))$ must be finite
at $\varepsilon\to0$ if both of them are expressed via renormalized quantities
(the on-shell photon propagator has no IR divergences
because the photon is neutral).
Therefore, we can reproduce
the \MS{} renormalization constant~(\ref{PhotonZ:ZA})
from~(\ref{OSPhoton:ZA}).

\subsection{One-loop massive on-shell propagator diagram}
\label{S:M1}

Before discussing the electron mass and field renormalization in the on-shell scheme,
we have to learn how to calculate the relevant diagrams.

\begin{figure}[ht]
\begin{center}
\begin{picture}(64,22)
\put(32,11){\makebox(0,0){\includegraphics{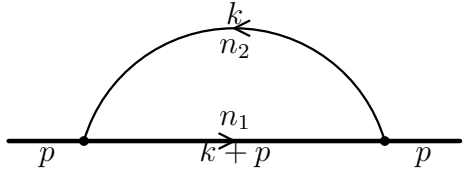}}}
\put(32,0){\makebox(0,0)[b]{$k+p$}}
\put(7,0){\makebox(0,0)[b]{$p$}}
\put(57,0){\makebox(0,0)[b]{$p$}}
\put(32,22){\makebox(0,0)[t]{$k$}}
\put(32,5){\makebox(0,0)[b]{$n_1$}}
\put(32,17){\makebox(0,0)[t]{$n_2$}}
\end{picture}
\end{center}
\caption{One-loop on-shell propagator diagram}
\label{F:M1}
\end{figure}

Let's consider the on-shell propagator integral (Fig.~\ref{F:M1}, $p^2=m^2$)
\begin{equation}
\begin{split}
&\int \frac{d^d k}{D_1^{n_1}D_2^{n_2}} = i \pi^{d/2} m^{d-2(n_1+n_2)} M(n_1,n_2)\,,\\
&D_1 = m^2 - (k+p)^2 = - k^2 - 2p\cdot k\,,\quad
D_2 = -k^2\,.
\end{split}
\label{M1:def}
\end{equation}
The power of $m$ is evident from the dimensional counting,
and our aim is to find the dimensionless function $M(n_1,n_2)$;
we can put $m=1$ to simplify the calculation.
It vanishes if $n_1$ is a non-positive integer.

Using Wick rotation and $\alpha$ parametrization~(\ref{V1:alpha}),
we rewrite the definition~(\ref{M1:def}) as
\begin{equation}
M(n_1,n_2) = \frac{\pi^{-d/2}}{\Gamma(n_1)\Gamma(n_2)}
\int e^{-\alpha_1(\ke^2+2\pe\cdot\ke)-\alpha_2\ke^2}
\alpha_1^{n_1-1} \alpha_2^{n_2-1} d\alpha_1\,d\alpha_2\,d^d\ke\,.
\label{M1:alpha}
\end{equation}
We want to separate a full square in the exponent;
to this end, we shift the integration momentum:
\begin{equation*}
\ke'=\ke+\frac{\alpha_1}{\alpha_1+\alpha_2}\pe
\end{equation*}
and obtain
\begin{equation*}
\begin{split}
M(n_1,n_2) &{}= \frac{\pi^{-d/2}}{\Gamma(n_1)\Gamma(n_2)}
\int \exp\left[-\frac{\alpha_1^2}{\alpha_1+\alpha_2}\right]
\alpha_1^{n_1-1} \alpha_2^{n_2-1} d\alpha_1\,d\alpha_2
\int e^{-(\alpha_1+\alpha_2)\ke^{\prime2}} d^d\ke'\\
&{} = \frac{1}{\Gamma(n_1)\Gamma(n_2)}
\int \exp\left[-\frac{\alpha_1^2}{\alpha_1+\alpha_2}\right]
(\alpha_1+\alpha_2)^{-d/2} \alpha_1^{n_1-1} \alpha_2^{n_2-1} d\alpha_1\,d\alpha_2\,.
\end{split}
\end{equation*}
Substituting $\eta=\alpha_1+\alpha_2$, $\alpha_1=\eta x$, $\alpha_2=\eta(1-x)$,
we get
\begin{equation*}
\begin{split}
M(n_1,n_2) &{}= \frac{1}{\Gamma(n_1)\Gamma(n_2)}
\int_0^1 x^{n_1-1} (1-x)^{n_2-1} dx
\int_0^\infty e^{-\eta x^2} \eta^{-d/2+n_1+n_2-1} d\eta\\
&{} = \frac{\Gamma(-d/2+n_1+n_2)}{\Gamma(n_1)\Gamma(n_2)}
\int_0^1 x^{d-n_1-2n_2-1} (1-x)^{n_2-1} dx\,.
\end{split}
\end{equation*}
Finally, we arrive at
\begin{equation}
M(n_1,n_2) = \frac{\Gamma(-d/2+n_1+n_2)\Gamma(d-n_1-2n_2)}%
{\Gamma(n_1)\Gamma(d-n_1-n_2)}\,.
\label{M1:res}
\end{equation}

The denominator in~(\ref{M1:def}) behaves as $(k^2)^{n_1+n_2}$
at $k\to\infty$.
Therefore, the integral diverges if $d\ge2(n_1+n_2)$.
At $d\to4$ this means $n_1+n_2\le2$.
This ultraviolet divergence shows itself as a $1/\varepsilon$ pole
of the first $\Gamma$ function in the numerator of~(\ref{M1:res})
(this $\Gamma$ function depends on $n_1+n_2$,
i.e., on the behaviour of the integrand at $k\to\infty$).
The integral~(\ref{M1:def}) can also have infrared divergences.
Its denominator behaves as $k^{n_1+2n_2}$ at $k\to0$,
and the integral diverges in this region if $d\le n_1+2n_2$.
At $d\to4$ this means $n_1+2n_2\ge4$.
This infrared divergence shows itself as a $1/\varepsilon$ pole
of the second $\Gamma$ function in the numerator of~(\ref{M1:res})
(this $\Gamma$ function depends on $n_1+2n_2$,
i.e., on the behaviour of the integrand at $k\to0$).

\begin{figure}[ht]
\begin{center}
\begin{picture}(32,11)
\put(16,5.5){\makebox(0,0){\includegraphics{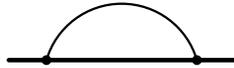}}}
\end{picture}
\end{center}
\caption{One-loop on-shell propagator diagram}
\label{F:topos}
\end{figure}

\begin{figure}[ht]
\begin{center}
\begin{picture}(14,12)
\put(7,6){\makebox(0,0){\includegraphics{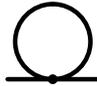}}}
\end{picture}
\end{center}
\caption{The basis integral}
\label{F:bos}
\end{figure}

Let's summarize.
There is one generic topology of one-loop massive on-shell propagator diagrams
in QED and QCD (Fig.~\ref{F:topos}).
All Feynman integrals of this class,
with any integer indices $n_1$, $n_2$, are proportional to $V_1$~(\ref{V1:V1})
(Fig.~\ref{F:bos}), with coefficients being rational functions of $d$.
For example, for $M(1,1)$,
\begin{equation}
\raisebox{-4.5mm}{\includegraphics{m1.eps}}
= - \frac{1}{2} \frac{d-2}{d-3}
\raisebox{-5mm}{\includegraphics{m1b.eps}}\,.
\label{M1:M11}
\end{equation}

Two-loop~\cite{GBGS:90,BGS:91,B:92} and three-loop~\cite{LR:96,MR:00}
massive on-shell diagrams can be calculated using integration by parts.

\subsection{On-shell renormalization of electron mass and field}
\label{S:OSm}

The electron mass $m$ in the on-shell renormalization scheme
is defined as the position of the pole of the electron propagator.
On-shell external electron lines have $p^2=m^2$;
it is convenient to use the free propagator containing the same quantity $m$
rather than $m_0$.
Therefore, we rewrite the Lagrangian~(\ref{Mass:L}) as
\begin{equation}
L = \bar{\psi}_0 (i\D - m) \psi_0 + \delta m \bar{\psi}_0 \psi_0\,,\quad
\delta m = m - m_0\,,
\label{OSm:L}
\end{equation}
and consider the mass counterterm not as a part of the unperturbed Lagrangian,
but as a perturbation.
Then the free electron propagator
\begin{equation}
S_0(p) = \frac{1}{\rlap/p-m} = \frac{\rlap/p+m}{p^2-m^2}
\label{OSm:S0}
\end{equation}
contains $m$, and the mass counterterm produces the vertex
\begin{equation}
\raisebox{-1.25mm}{\includegraphics{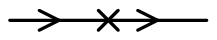}} = i\,\delta m\,.
\label{OSm:ctvert}
\end{equation}
The mass renormalization constant in the on-shell scheme is defined as usual:
\begin{equation}
m_0 = Z_m^{\text{os}} m\,.
\label{OSm:Zmdef}
\end{equation}

It is more convenient to write the electron self-energy~(\ref{Mass:Sigma})
in the form
\begin{equation}
\Sigma(p) = m \Sigma_1(p^2) + (\rlap/p-m) \Sigma_2(p^2)
\label{OSm:Sigma}
\end{equation}
now.
The electron propagator
\begin{equation}
S(p) = \frac{1}{[1-\Sigma_2(p^2)](\rlap/p-m)
+ \delta m - m \Sigma_1(p^2)}
\label{OSm:S}
\end{equation}
has a pole at $p^2=m^2$ if
\begin{equation}
\delta m = m \Sigma_1(m^2)\,,
\label{OSm:massren}
\end{equation}
or
\begin{equation}
Z_m^{\text{os}} = 1 - \frac{\delta m}{m} = 1 - \Sigma_1(m^2)\,.
\label{OSm:massren2}
\end{equation}
The equation~(\ref{OSm:massren}) can be solved for $\delta m$ by iterations
(at higher orders, its right-hand side contains $\delta m$
because of the vertex~(\ref{OSm:ctvert}).

Near the mass shell, we can expand $\Sigma_1(p^2)$ as
\begin{equation}
\Sigma_1(p^2) - \frac{\delta m}{m}
= \Sigma_1'(m^2) (p^2 - m^2) + \cdots
\label{OSm:Sigma1exp}
\end{equation}
so that
\begin{equation}
S(p) = \frac{1}{1 - \Sigma_2(m^2) - 2 m^2 \Sigma_1'(m^2)}\;
\frac{\rlap/p + m}{p^2 - m^2} + \cdots
\label{OSm:Smshell}
\end{equation}
The electron field renormalized in the on-shell scheme,
\begin{equation}
\psi_0 = \left(Z_\psi^{\text{os}}\right)^{1/2} \psi_{\text{os}}\,,\quad
S(p) = Z_\psi^{\text{os}} S_{\text{os}}(p)\,,
\label{OSm:psios}
\end{equation}
is defined in such a way that its propagator $S_{\text{os}}(p)$
behaves as $S_0(p)$~(\ref{OSm:S0}) near the mass shell:
\begin{equation}
Z_\psi^{\text{os}}
= \left[1 - \Sigma_2(m^2) - 2 m^2 \Sigma_1'(m^2)\right]^{-1}\,.
\label{OS:Zpsidef}
\end{equation}

In order to calculate $Z_m^{\text{os}}$ and $Z_\psi^{\text{os}}$,
it is convenient to introduce the function
\begin{equation}
T(t) = \frac{1}{4m} \Tr (\rlap/v+1) \Sigma(mv(1+t))
= \Sigma_1(m^2) + \left[\Sigma_2(m^2) + 2 m^2 \Sigma_1'(m^2)\right] t + \cdots
\label{OSm:T}
\end{equation}
so that
\begin{equation}
Z_m^{\text{os}} = 1 - T(0)\,,\quad
Z_\psi^{\text{os}} = \left[1 - T'(0)\right]^{-1}\,.
\label{OSm:T0}
\end{equation}

\begin{figure}[ht]
\begin{center}
\begin{picture}(64,25)
\put(32,12){\makebox(0,0){\includegraphics{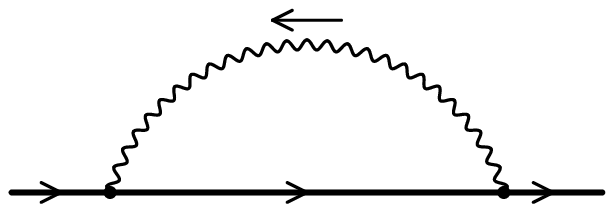}}}
\put(32,0){\makebox(0,0)[b]{$k+p$}}
\put(7,0){\makebox(0,0)[b]{$p$}}
\put(57,0){\makebox(0,0)[b]{$p$}}
\put(32,25){\makebox(0,0)[t]{$k$}}
\end{picture}
\end{center}
\caption{One-loop electron self-energy}
\label{F:Electron1m}
\end{figure}

Let's calculate it at one loop (Fig.~\ref{F:Electron1m}).
We put $m=1$; the power of $m$ will be restored by dimensionality:
\begin{equation}
T(t) = - i e_0^2 \int \frac{d^d k}{(2\pi)^d}\;\frac{1}{D_1(t)\,D_2}\;
\frac{1}{4} \Tr (\rlap/v+1) \gamma^\mu (\rlap/k+\rlap/p+1) \gamma^\nu
\left(g_{\mu\nu} + \xi \frac{k_\mu k_\nu}{D_2}\right)\,,
\label{OSm:T1}
\end{equation}
where
\begin{equation*}
p = v(1+t)\,,\quad
D_1(t) = 1 - (k+p)^2\,,\quad
D_2 = - k^2\,.
\end{equation*}
In calculating the numerator, we can express
\begin{equation*}
p\cdot k = \frac{1}{2} \left[ D_2 - D_1(t) + 1 - (1+t)^2 \right]
\end{equation*}
and omit terms with $D_1(t)$,
because the resulting integrals contain no scale.
We obtain
\begin{equation*}
T(t) = - i e_0^2 \int \frac{d^d k}{(2\pi)^d} \frac{1}{D_1(t)}
\left[\frac{2}{D_2} - \frac{d-2}{2} (1-t) + \mathcal{O}(t^2) \right]\,.
\end{equation*}
This result is gauge-independent.

Expanding
\begin{equation*}
D_1(t) = D_1 + (D_1-D_2-2) t + \mathcal{O}(t^2)\,,\quad
D_1 = 1 - (k+v)^2\,,
\end{equation*}
we obtain
\begin{equation*}
\begin{split}
T(t) = - i e_0^2 \int \frac{d^d k}{(2\pi)^d}
\biggl[& \frac{2(1-t)}{D_1 D_2} - \frac{(d-2)(1-2t)}{2 D_1} + \frac{4t}{D_1^2 D_2}
- \frac{(d-4)t}{D_1^2} - \frac{(d-2)D_2 t}{2 D_1^2}\\
&{} + \mathcal{O}(t^2) \biggr]\,.
\end{split}
\end{equation*}
Using~(\ref{M1:res}) and restoring the power of $m$,
we finally arrive at
\begin{equation}
T(t) = \frac{e_0^2 m^{-2\varepsilon}}{(4\pi)^{d/2}}
\Gamma(\varepsilon) \frac{d-1}{d-3} (1-t) + \mathcal{O}(t^2)\,.
\label{OSm:Tt}
\end{equation}
Therefore,
\begin{equation}
Z_m^{\text{os}} = Z_\psi^{\text{os}}
= 1 - \frac{e_0^2 m^{-2\varepsilon}}{(4\pi)^{d/2}}
\Gamma(\varepsilon) \frac{d-1}{d-3}\,.
\label{OSm:Z1}
\end{equation}
The fact that $Z_m^{\text{os}} = Z_\psi^{\text{os}}$ at one loop is a pure accident;
at two loops it is no longer so~\cite{GBGS:90,BGS:91}.
The on-shell mass $m$ is a measurable quantity,
and hence gauge-invariant;
therefore, $Z_m^{\text{os}}$ is gauge-invariant to all orders.
It has been proved that $Z_\psi^{\text{os}}$ is also gauge-invariant
in QED (but not in QCD).

What's the relation between the on-shell mass $m$
and the \MS{} mass $m(\mu)$?
\begin{equation*}
m_0 = Z_m(\alpha(\mu)) m(\mu) = Z_m^{\text{os}} m\,,
\end{equation*}
and therefore
\begin{equation}
m(\mu) = \frac{Z_m^{\text{os}}}{Z_m(\alpha(\mu))} m\,.
\label{OSm:mm}
\end{equation}
We have to re-express $Z_m^{\text{os}}$ via $\alpha(\mu)$;
then the ratio of renormalization constants is finite at $\varepsilon\to0$,
because both renormalized masses are finite.
At one loop
\begin{equation}
\frac{m(\mu)}{m} = 1 - 6 \frac{\alpha}{4\pi}
\left( \log \frac{\mu}{m} + \frac{2}{3} \right)\,.
\label{OSm:mm1}
\end{equation}
This formula is OK when $\mu\sim m$;
otherwise, it is much better to relate $m(m)$ to $m$,
and then to solve the RG equation with the initial condition $m(m)$.

\subsection{On-shell charge}
\label{S:OSe}

On the mass shell ($p^2=m^2$), $\Gamma^\mu(p,p)$ has only one
$\gamma$-matrix structure,
\begin{equation}
\Gamma^\mu(p,p) = Z_\Gamma^{\text{os}} \gamma^\mu\,,
\label{OSe:Gamma}
\end{equation}
if sandwiched between $\bar{u}_2\ldots u_1$
which satisfy the Dirac equation
(see Sect.~\ref{S:mu} for more details).
The physical matrix element of scattering of an electron
by a photon is
\begin{equation*}
e_0 \Gamma^\mu Z_\psi^{\text{os}} \left(Z_A^{\text{os}}\right)^{1/2}\,.
\end{equation*}
The only case when all 3 particles are on-shell is $p^2=m^2$,
and the photon momentum $q\to0$.
By definition, this matrix element is
\begin{equation}
e \gamma^\mu\,,
\label{OSe:e}
\end{equation}
where $e$ is the renormalized charge in the on-shell scheme.
It is related to the bare charge by
\begin{equation}
e_0 = \left(Z_\alpha^{\text{os}}\right)^{1/2} e\,.
\label{OSe:Z}
\end{equation}
Therefore,
\begin{equation}
Z_\alpha^{\text{os}} = \left(Z_\Gamma^{\text{os}} Z_\psi^{\text{os}}\right)^{-2}
\left(Z_A^{\text{os}}\right)^{-1}\,.
\label{OSe:Zalpha}
\end{equation}
On-shell electron charge is measured in macroscopic experiments
with smooth electromagnetic fields (having $q\to0$); it is
\begin{equation}
\alpha = \frac{e^2}{4\pi} \approx \frac{1}{137}\,.
\label{OSe:137}
\end{equation}

In QED, the situation is simplified by the Ward identity~(\ref{Vertex:Ward4}).
Near the mass shell
\begin{equation*}
S(p) = \frac{Z_\psi^{\text{os}}}{\rlap/p-m}\,,
\end{equation*}
and hence
\begin{equation*}
\Gamma^\mu(p,p) = \left(Z_\psi^{\text{os}}\right)^{-1} \gamma^\mu\,.
\end{equation*}
Therefore,
\begin {equation}
Z_\psi^{\text{os}} Z_\Gamma^{\text{os}} = 1\,,
\label{OSe:Ward}
\end{equation}
and
\begin{equation}
Z_\alpha^{\text{os}} = \left(Z_A^{\text{os}}\right)^{-1}\,.
\label{OSe:Ward2}
\end{equation}

At one loop, from~(\ref{OSPhoton:ZA}),
\begin{equation}
Z_\alpha^{\text{os}} = 1
+ \frac{4}{3} \frac{e_0^2 m_0^{-2\varepsilon}}{(4\pi)^{d/2}} \Gamma(\varepsilon)\,.
\label{OSe:Z1}
\end{equation}
The \MS{} (running) coupling $\alpha(\mu)$
is related to the on-shell coupling $\alpha$ by
\begin{equation*}
\alpha(\mu) = \alpha \frac{Z_\alpha^{\text{os}}}{Z_\alpha(\mu)}
= \alpha \left[1 + \frac{4}{3} \frac{\alpha(\mu)}{4\pi}
\left(\left(\frac{\mu}{m}\right)^{2\varepsilon}
e^{\gamma\varepsilon} \Gamma(\varepsilon) - \frac{1}{\varepsilon}\right)
\right]\,.
\end{equation*}
Therefore
\begin{equation}
\alpha(\mu) = \alpha \left[1 + \frac{8}{3} \frac{\alpha}{4\pi} \log \frac{\mu}{m}
\right]\,.
\label{OSe:ee1}
\end{equation}
We can always find the $\mu$-dependence from the RG equation,
the initial condition is $\alpha(m)=\alpha$ at one loop.

\subsection{Magnetic moment}
\label{S:mu}

Let's consider scattering of an on-shell electron in electromagnetic field.
The physical scattering amplitude is
\begin{equation}
e_0 Z_\psi^{\text{os}} \left(Z_A^{\text{os}}\right)^{1/2}\,
\bar{u}' \Gamma^\mu(p,p') u\,,
\label{mu:amp}
\end{equation}
where $p^2=m^2$, $p^{\prime2}=m^2$,
and the initial and final electron wave functions satisfy the Dirac equation
\begin{equation}
\rlap/p u = m u\,,\quad
\bar{u}' \rlap/p' = \bar{u}' m\,.
\label{mu:Dirac}
\end{equation}
We can substitute the on-shell charge
\begin{equation*}
e = e_0 \left(Z_A^{\text{os}}\right)^{1/2}\,.
\end{equation*}
Using the Ward identity~(\ref{Vertex:Ward3})
and the electron self-energy~(\ref{OSm:Sigma}),
we obtain
\begin{equation*}
\Gamma^\mu(p,p') q_\mu = (\rlap/p' - \rlap/p) (1 - \Sigma_2(m^2))\,,
\end{equation*}
where $q=p'-p$,
and hence for on-shell electron wave functions~(\ref{mu:Dirac})
\begin{equation}
\bar{u}' \Gamma^\mu q_\mu u = 0\,.
\label{mu:Ward}
\end{equation}

What $\gamma$-matrix structures can $\bar{u'} \Gamma^\mu u$ have?
Using the Dirac equations~(\ref{mu:Dirac}),
we can always eliminate $\rlap/p$ by anticommuting it to $u$,
and $\rlap/p'$ by anticommuting it to $\bar{u}'$.
For example,
\begin{equation}
\bar{u}' \sigma^{\mu\nu} q_\nu u =
i\,\bar{u}' \left( (p+p')^\mu - 2 m \gamma^\mu \right) u\,.
\label{mu:sigma}
\end{equation}
We are left with 3 structures: $\gamma^\mu$, $(p+p')^\mu$,
and $q^\mu=(p'-p)^\mu$.
The last one is excluded by~(\ref{mu:Ward}).
Therefore,
\begin{equation}
Z_\psi^{\text{os}}\,\bar{u}' \Gamma^\mu u =
\bar{u}' \left[ (F_1(q^2) + F_2(q^2)) \gamma^\mu
- F_2(q^2) \frac{(p+p')^\mu}{2m} \right] u\,.
\label{mu:FF1}
\end{equation}
Using~(\ref{mu:sigma}), we can also rewrite this as
\begin{equation}
\begin{split}
Z_\psi^{\text{os}}\,\bar{u}' \Gamma^\mu u &{}=
\bar{u}' \left[ F_1(q^2) \gamma^\mu
+ F_2(q^2) \frac{i \sigma^{\mu\nu} q_\nu}{2m} \right] u\\
&{}= \bar{u}' \left[ F_1(q^2) \frac{(p+p')^\mu}{2m}
+ (F_1(q^2) + F_2(q^2)) \frac{i \sigma^{\mu\nu} q_\nu}{2m} \right] u\,.
\end{split}
\label{mu:FF2}
\end{equation}

Let's rewrite~(\ref{mu:FF1}) as
\begin{equation*}
\bar{u}' \Gamma^\mu u = \bar{u}' \left( \sum f_i T_i^\mu \right) u\,,\quad
T_1^\mu = \frac{(p+p')^\mu}{2m}\,,\quad
T_2^\mu = \gamma^\mu\,,
\end{equation*}
then
\begin{equation}
F_1(q^2) = Z_\psi^{\text{os}} (f_1+f_2)\,,\quad
F_2(q^2) = - Z_\psi^{\text{os}} f_1\,.
\label{mu:Ff}
\end{equation}
If we introduce the traces
\begin{equation*}
y_i = \frac{1}{4 m^2} \Tr T'_{i\mu} (\rlap/p'+m) \Gamma^\mu (\rlap/p+m)\,,\quad
T'_{1\mu} = \frac{p_\mu}{m}\,,\quad
T'_{2\mu} = \gamma_\mu\,,
\end{equation*}
then
\begin{equation*}
y_i = M_{ij} f_j\,,\quad
M_{ij} = \frac{1}{4 m^2} \Tr T'_{i\mu} (\rlap/p'+m) T_j^\mu (\rlap/p+m)\,,
\end{equation*}
and we can find $f_i$ by solving the linear system:
\begin{equation*}
f_i = \left(M^{-1}\right)_{ij} y_j\,.
\end{equation*}
Let's introduce the notation
\begin{equation*}
t = - \frac{q^2}{4 m^2}\,.
\end{equation*}
Calculating the traces, we find
\begin{equation*}
M = 2 \left(
\begin{array}{cc}
(1+t)^2 & 1+t \\
1+t     & 1-(d-2)t
\end{array}
\right)\,,
\end{equation*}
and hence
\begin{equation*}
M^{-1} = - \frac{1}{2(d-2)t(1+t)^2} \left(
\begin{array}{cc}
1-(d-2)t & -(1+t)\\
-(1+t)   & (1+t)^2
\end{array}
\right)\,.
\end{equation*}
Finally, we obtain from~(\ref{mu:Ff})
\begin{align}
F_1(q^2) &{}= \frac{Z_\psi^{\text{os}}}{2(d-2)(1+t)^2}
\frac{1}{4} \Tr \bigl[ (d-1) v_\mu - (1+t) \gamma_\mu \bigr]
(\rlap/v'+1) \Gamma^\mu (\rlap/v+1)\,,
\label{mu:F1}\\
F_2(q^2) &{}= \frac{Z_\psi^{\text{os}}}{2(d-2)t(1+t)^2}
\frac{1}{4} \Tr \bigl[ \bigl(1-(d-2)t\bigr) v_\mu - (1+t) \gamma_\mu \bigr]
(\rlap/v'+1) \Gamma^\mu (\rlap/v+1)\,,
\label{mu:F2}
\end{align}
where $v=p/m$, $v'=p'/m$.
At the tree level, $\Gamma^\mu=\gamma^\mu$,
and, naturally, we obtain $F_1(q^2)=1$, $F_2(q^2)=0$.
When calculating loop corrections,
we can apply these projectors to integrands of vertex diagrams,
and express $F_{1,2}(q^2)$ via scalar integrals.

The Dirac form factor~(\ref{mu:F1}) at $q^2=0$ is
\begin{equation}
F_1(0) = Z_\psi^{\text{os}}\,\frac{1}{4} \Tr v_\mu \Gamma^\mu(mv,mv) (\rlap/v+1)\,.
\label{mu:F10}
\end{equation}
Due to the Ward identity~(\ref{Vertex:Ward4}),
\begin{equation}
F_1(0) = Z_\psi^{\text{os}}
\left[ 1 - v^\mu \left. \frac{\partial}{\partial p^\mu}
\frac{1}{4} \Tr \Sigma(p) (\rlap/v+1) \right|_{p=mv} \right]
= Z_\psi^{\text{os}} \left[1 - T'(0)\right] = 1\,,
\end{equation}
see~(\ref{OSm:T}), (\ref{OSm:T0}).
The total charge of electron is not changed by radiative corrections.

It may seem that the Pauli form factor~(\ref{mu:F2})
is singular at $q^2\to0$.
Of course, it is not.
Let's substitute the expansion
\begin{equation*}
\Gamma^\mu(mv,mv+q) = \Gamma_0^\mu + \Gamma_1^{\mu\nu} \frac{q_\nu}{m}
+ \cdots
\end{equation*}
into~(\ref{mu:F2}),
split $q=(q\cdot v)v+q_\bot$
(where $q\cdot v/m=2t$, $q_\bot^2/m^2=-4t(1+t)$),
and average over the directions of $q_\bot$
in the $(d-1)$-dimensional subspace orthogonal to $v$:
\begin{equation*}
\frac{\overline{q^\alpha}}{m} = 2 t v^\alpha\,,\quad
\frac{\overline{q^\alpha q^\beta}}{m^2} = - \frac{4t}{d-1}
\left[ (1+t) g^{\alpha\beta} - (1+dt) v^\alpha v^\beta \right]\,.
\end{equation*}
We obtain
\begin{equation}
\begin{split}
F_2(0) = \frac{Z_\psi^{\text{os}}}{d-2}
\biggl[& \frac{1}{4} \Tr (\gamma_\mu-dv_\mu) \Gamma_0^\mu (\rlap/v+1)\\
&{} + \frac{2}{d-1} \frac{1}{4} \Tr
\left( \gamma_\mu \gamma_\nu + \gamma_\mu v_\nu - \gamma_\nu v_\mu - v_\mu v_\nu \right)
\Gamma_1^{\mu\nu} (\rlap/v+1) \biggr]\,.
\end{split}
\label{mu:F20}
\end{equation}
This means that in order to calculate the anomalous magnetic moment
we need the vertex and its first derivative in $q$ at $q=0$.

\begin{figure}[ht]
\begin{center}
\begin{picture}(54,41)
\put(27,20.5){\makebox(0,0){\includegraphics{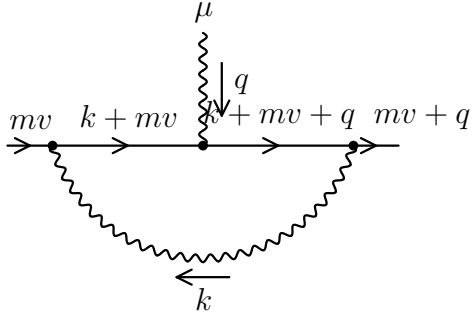}}}
\put(27,0){\makebox(0,0)[b]{$k$}}
\put(27,41){\makebox(0,0)[t]{$\mu$}}
\put(4,24){\makebox(0,0)[b]{$mv$}}
\put(56,24){\makebox(0,0)[b]{$mv+q$}}
\put(17,24){\makebox(0,0)[b]{$k+mv$}}
\put(37,24){\makebox(0,0)[b]{$k+mv+q$}}
\put(31,30){\makebox(0,0)[l]{$q$}}
\end{picture}
\end{center}
\caption{One-loop anomalous magnetic moment}
\label{F:mu}
\end{figure}

As already mentioned,
the tree diagram ($\Gamma_0^\mu=\gamma^\mu$, $\Gamma_1^{\mu\nu}=0$)
does not contribute to $F_2(0)$.
The first contributing diagram is one-loop (Fig.~\ref{F:mu}).
All charged external lines are on-shell,
therefore, this vertex diagram is gauge-invariant.
We shall use Feynman gauge to simplify the calculation,
and put $m=1$ (the power of $m$ will be restored by dimensionality).
Then the one-loop vertex is
\begin{equation*}
- i e_0^2 \int \frac{d^d k}{(2\pi)^d} \frac{1}{k^2}
\gamma_\alpha S(k+v+q) \gamma^\mu S(k+v) \gamma^\alpha\,.
\end{equation*}

We need to expand it in $q$ up to the linear term.
In this diagram, only one propagator depends on $q$.
Using
\begin{equation}
S(p+q) = S(p) + \frac{\rlap/q}{p^2-m^2} - \frac{\rlap/p+m}{(p^2-m^2)^2}\,2 p\cdot q
+ \cdots
\label{mu:S}
\end{equation}
for $p=k+mv$, we obtain ($m=1$)
\begin{align*}
&\Gamma_0^\mu = i e_0^2 \int \frac{d^d k}{(2\pi)^d}
\frac{\gamma_\alpha (\rlap/k+\rlap/v+1) \gamma^\mu (\rlap/k+\rlap/v+1) \gamma^\alpha}%
{D_1^2 D_2}\,,\\
&\Gamma_1^{\mu\nu} = i e_0^2 \int \frac{d^d k}{(2\pi)^d}
\frac{\gamma_\alpha \left[ D_1 \gamma^\nu + 2 (k+v)^\nu (\rlap/k+\rlap/v+1) \right]
\gamma^\mu (\rlap/k+\rlap/v+1) \gamma^\alpha}%
{D_1^3 D_2}\,,
\end{align*}
where $D_1=1-(k+v)^2$, $D_2=-k^2$ (see~(\ref{M1:def})).
We may replace $Z_\psi^{\text{os}}\to1$ in~(\ref{mu:F20}),
because corrections are beyond our accuracy;
the two contributions to $F_2(0)$ are
\begin{align*}
&\frac{i e_0^2}{d-2} \int \frac{d^d k}{(2\pi)^d} \frac{N_0}{D_1^2 D_2}\,,\\
&N_0 = \frac{1}{4} \Tr (\gamma_\mu-dv_\mu) \gamma_\alpha
(\rlap/k+\rlap/v+1) \gamma^\mu (\rlap/k+\rlap/v+1) \gamma^\alpha (\rlap/v+1)\,,
\end{align*}
and
\begin{align*}
&\frac{2 i e_0^2}{(d-1)(d-2)} \int \frac{d^d k}{(2\pi)^d} \frac{N_1}{D_1^3 D_2}\,,\\
&N_1 = \frac{1}{4} \Tr
(\gamma_\mu \gamma_\nu + \gamma_\mu v_\nu - \gamma_\nu v_\mu - v_\mu v_\nu)
\gamma_\alpha\\
&\hphantom{N_1= \frac{1}{4}\Tr}\times
\left[ D_1 \gamma^\nu + 2 (k+v)^\nu (\rlap/k+\rlap/v+1) \right]
\gamma^\mu (\rlap/k+\rlap/v+1) \gamma^\alpha (\rlap/v+1)\,.
\end{align*}
When calculating $N_0$, we may omit terms with $D_1^2$:
\begin{equation*}
N_0 \Rightarrow - d (d-2) D_1 D_2 - (d-1) (d-4) D_1 + \frac{1}{2} d (d-2) D_2^2
+ d (d-3) D_2 - 4 (d-1)\,.
\end{equation*}
When calculating $N_1$, we may omit terms with $D_1^3$:
\begin{equation*}
\begin{split}
N_1 \Rightarrow \frac{1}{2} \Bigl[& 3 (d-2) D_1^2 D_2 + (d^3-8d^2+17d-14) D_1^2
- 3 (d-2) D_1 D_2^2\\
&{} - (d^3-8d^2+21d-26) D_1 D_2
+ 4 (d-1)^2 D_1 + (d-2) D_2^3 + 4 (d-3) D_2^2 - 16 D_2 \Bigr]\,.
\end{split}
\end{equation*}
Collecting all this together, we obtain
\begin{equation*}
\begin{split}
F_2(0) = - \frac{e_0^2}{(4\pi)^{d/2}} \frac{1}{d-2}
\Bigl\{& - d (d-1) M(1,0) - (d-1) (d-4) M(1,1)\\
&{} + \frac{1}{2} d (d-2) M(2,-1) + d (d-3) M(2,0) - 4 (d-1) M(2,1)\\
&{} + \frac{1}{d-1} \Bigl[ 3 (d-2) M(1,0) + (d^3-8d^2+17d-14) M(1,1)\\
&\hphantom{{}+\frac{1}{d-1}\Bigl[\Bigr.}
- 3 (d-2) M(2,-1) - (d^3-8d^2+21d-26) M(2,0)\\
&\hphantom{{}+\frac{1}{d-1}\Bigl[\Bigr.}
+ 4 (d-1)^2 M(2,1) + (d-2) M(3,-2)\\
&\hphantom{{}+\frac{1}{d-1}\Bigl[\Bigr.}
+ 4 (d-3) M(3,-1) - 16 M(3,0)
\Bigr] \Bigr\}\,.
\end{split}
\end{equation*}

Using~(\ref{M1:res}) and restoring the power of $m$,
we finally arrive at the anomalous magnetic moment
\begin{equation}
F_2(0) = \frac{e_0^2 m^{-2\varepsilon}}{(4\pi)^{d/2}} \Gamma(\varepsilon)
\frac{(d-4)(d-5)}{d-3} + \cdots
\label{mu:mud}
\end{equation}
As expected, it is finite at $\varepsilon\to0$:
\begin{equation}
F_2(0) = \frac{\alpha}{2\pi} + \cdots
\end{equation}
As we can see from the second line in~(\ref{mu:FF2}),
the total magnetic moment of electron is
\begin{equation*}
F_1(0) + F_2(0) = 1 + \frac{\alpha}{2\pi} + \cdots
\end{equation*}
(in units of Bohr magneton).
It is not difficult to calculate the two-loop correction to the magnetic moment,
if one has a program implementing the integration-by-parts evaluation
of two-loop on-shell propagator integrals.
The three-loop calculation~\cite{LR:96} was a major breakthrough.

\subsection{Two-loop massive vacuum diagram}
\label{S:V2}

\begin{figure}[ht]
\begin{center}
\begin{picture}(22,28)
\put(11,14){\makebox(0,0){\includegraphics{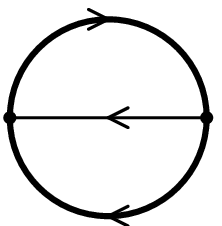}}}
\put(11,28){\makebox(0,0)[t]{$k_1$}}
\put(11,0){\makebox(0,0)[b]{$k_2$}}
\put(11,15){\makebox(0,0)[b]{$k_1-k_2$}}
\put(11,23){\makebox(0,0)[t]{$n_1$}}
\put(11,5){\makebox(0,0)[b]{$n_2$}}
\put(11,13){\makebox(0,0)[t]{$n_3$}}
\end{picture}
\end{center}
\caption{Two-loop massive vacuum diagram}
\label{F:V2}
\end{figure}

Let's consider the two-loop massive vacuum diagram (Fig.~\ref{F:V2}):
\begin{equation}
\begin{split}
&\int \frac{d^d k_1\,d^d k_2}{D_1^{n_1} D_2^{n_2} D_3^{n_3}}
= - \pi^d m^{2(d-n_1-n_2-n_3)} V(n_1,n_2,n_3)\,,\\
&D_1 = m^2 - k_1^2\,,\quad
D_2 = m^2 - k_2^2\,,\quad
D_3 = - (k_1-k_2)^2\,.
\end{split}
\label{V2:def}
\end{equation}
The power of $m$ is evident from the dimensional counting,
and our aim is to find the dimensionless function $V(n_1,n_2,n_3)$;
we can put $m=1$ to simplify the calculation.
It is symmetric with respect to $n_1\leftrightarrow n_2$;
it vanishes if $n_1$ or $n_2$ is a non-positive integer.
Using Wick rotation and $\alpha$ parametrization~(\ref{V1:alpha}),
we rewrite the definition~(\ref{V2:def}) as
\begin{equation}
\begin{split}
V(n_1,n_2,n_3) ={}& \frac{\pi^{-d}}{\Gamma(n_1)\Gamma(n_2)\Gamma(n_3)}
\int e^{-\alpha_1(\ke_1^2+1)-\alpha_2(\ke_2^2+1)-\alpha_3(\ke_1-\ke_2)^2}\\
&{}\times\alpha_1^{n_1-1} \alpha_2^{n_2-1} \alpha_3^{n_3-1}
d\alpha_1\,d\alpha_2\,d\alpha_3\,d^d\ke_1\,d^d\ke_2\,.
\end{split}
\label{V2:alpha}
\end{equation}

The $d$-dimensional integral of the exponent of the quadratic form
is the product of $d$ one-dimensional integrals:
\begin{equation}
\int e^{-A_{ij} \ke_i\cdot\ke_j} d^d\ke_1\,d^d\ke_2
= \left[ \int e^{-A_{ij} \ke_{ix} \ke_{jx}} d\ke_{1x}\,d\ke_{2x} \right]^d
= \frac{\pi^d}{\left[\det A\right]^{d/2}}\,.
\label{V2:gauss}
\end{equation}
This is the definition of such $d$-dimensional integration;
note that the result contains $d$ as a symbol.
In our case,
\begin{equation*}
A = \left(
\begin{array}{cc}
\alpha_1+\alpha_3 & -\alpha_3\\
-\alpha_3 & \alpha_2+\alpha_3
\end{array}
\right)\,,
\end{equation*}
and we obtain
\begin{equation*}
\begin{split}
V(n_1,n_2,n_3) ={}& \frac{1}{\Gamma(n_1)\Gamma(n_2)\Gamma(n_3)}
\int e^{-\alpha_1-\alpha_2}\\
&{} \times (\alpha_1\alpha_2+\alpha_1\alpha_3+\alpha_2\alpha_3)^{-d/2}
\alpha_1^{n_1-1} \alpha_2^{n_2-1} \alpha_3^{n_3-1}
d\alpha_1\,d\alpha_2\,d\alpha_3\,.
\end{split}
\end{equation*}

To calculate this integral, it is most convenient to choose
the ``radial'' variable $\eta=\alpha_1+\alpha_2$ (Sect.~\ref{S:V1}),
and substitute $\alpha_i=\eta x_i$.
The integral in $\eta$ is trivial:
\begin{equation*}
\begin{split}
&V(n_1,n_2,n_3) = \frac{1}{\Gamma(n_1)\Gamma(n_2)\Gamma(n_3)}
\int_0^\infty e^{-\eta} \eta^{-d+n_1+n_2+n_3-1} d\eta\\
&\qquad{} \times \int (x_1 x_2+x_1 x_3+x_2 x_3)^{-d/2}
x_1^{n_1-1} x_2^{n_2-1} x_3^{n_3-1}
\delta(x_1+x_2-1) d x_1\,d x_2\,d x_3\\
&{} = \frac{\Gamma(-d+n_1+n_2+n_3)}{\Gamma(n_1)\Gamma(n_2)\Gamma(n_3)}
\int \left[x_1(1-x_1) + x_3\right]^{-d/2}
x_1^{n_1-1} (1-x_1)^{n_2-1} x_3^{n_3-1} d x_1\,d x_3\,.
\end{split}
\end{equation*}
Substituting $x_3=x_1(1-x_1)y$, we get
\begin{equation*}
\begin{split}
V(n_1,n_2,n_3) ={}& \frac{\Gamma(-d+n_1+n_2+n_3)}{\Gamma(n_1)\Gamma(n_2)\Gamma(n_3)}
\int_0^\infty \frac{y^{n_3-1} d y}{(y+1)^{d/2}}\\
&{} \times \int_0^1 x^{-d/2+n_1+n_3-1} (1-x)^{-d/2+n_2+n_3-1} d x
\end{split}
\end{equation*}
The integral
\begin{equation*}
\int_0^\infty \frac{y^{n_3-1} d y}{(y+1)^{d/2}}
= \frac{\Gamma(n_3)\Gamma(d/2-n_3)}{\Gamma(d/2)}
\end{equation*}
is easily calculated using the substitution $z=1/(y+1)$;
the second integral is the Euler $B$-function.
We arrive at the result~\cite{V:80}
\begin{equation}
V(n_1,n_2,n_3)
= \frac{\Gamma\left(\frac{d}{2}-n_3\right)
\Gamma\left(n_1+n_3-\frac{d}{2}\right)\Gamma\left(n_2+n_3-\frac{d}{2}\right)
\Gamma(n_1+n_2+n_3-d)}%
{\Gamma\left(\frac{d}{2}\right)\Gamma(n_1)\Gamma(n_2)\Gamma(n_1+n_2+2n_3-d)}\,.
\label{V2:res}
\end{equation}
This is the only class of two-loop diagrams for which a general formula
for arbitrary $n_i$ (not necessarily integer) is known.

\begin{figure}[ht]
\begin{center}
\begin{picture}(22,22)
\put(11,11){\makebox(0,0){\includegraphics{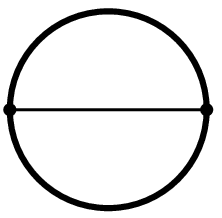}}}
\end{picture}
\end{center}
\caption{Two-loop vacuum diagram}
\label{F:topm}
\end{figure}

\begin{figure}[ht]
\begin{center}
\begin{picture}(22,22)
\put(11,11){\makebox(0,0){\includegraphics{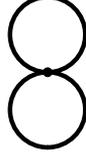}}}
\end{picture}
\end{center}
\caption{The basis integral}
\label{F:bm}
\end{figure}

Let's summarize.
There is one generic topology of two-loop vacuum diagrams
in QED and QCD (Fig.~\ref{F:topm}).
All Feynman integrals of this class,
with any integer indices $n_i$, are proportional to $V_1^2$~(\ref{V1:V1})
(Fig.~\ref{F:bm}), with coefficients being rational functions of $d$.

Three-loop massive vacuum diagrams can be calculated
using integration by parts~\cite{B:92}.

\subsection{On-shell renormalization of photon field and charge at two loops}
\label{S:OSPhoton2}

Using the results of the previous Section,
it is not (very) difficult to calculate
the photon self-energy $\Pi(0)$ at two loops.
We apply~(\ref{OSPhoton:Pi0}) to the diagrams of Fig.~\ref{F:Photon2},
calculate the derivatives in $p$,
and reduce the problem to the vacuum integrals~(\ref{V2:def}).
The result is
\begin{equation}
\Pi_2(0) = - \frac{2}{3} \frac{e_0^4 m_0^{-4\varepsilon}}{(4\pi)^d}
\Gamma^2(\varepsilon) \frac{(d-4)(5d^2-33d+34)}{d(d-5)}\,.
\label{OSPhoton2:Pi}
\end{equation}
Therefore, for the on-shell charge $e$ we obtain
\begin{equation}
\begin{split}
\frac{e_0^2}{e^2} &{}= Z_\alpha^{\text{os}}
= \left(Z_A^{\text{os}}\right)^{-1} = 1 - \Pi(0)\\
&{}= 1 + \frac{4}{3} \frac{e_0^2 m_0^{-2\varepsilon}}{(4\pi)^{d/2}}
\Gamma(\varepsilon)
- \frac{4}{3} \varepsilon
\frac{9+7\varepsilon-10\varepsilon^2}{(2-\varepsilon)(1+2\varepsilon)}
\left(\frac{e_0^2 m_0^{-2\varepsilon}}{(4\pi)^{d/2}}\Gamma(\varepsilon)\right)^2
+ \cdots
\end{split}
\label{OSPhoton2:Zalpha}
\end{equation}
where $\Pi(0)$ at one~(\ref{OSPhoton:Pi})) and two~(\ref{OSPhoton2:Pi}) was used.
In the one-loop term, we have to substitute
$e_0^2 = Z_\alpha^{\text{os}} e^2$ and $m_0 = Z_m^{\text{os}} m$,
with one-loop $Z_\alpha^{\text{os}}$~(\ref{OSe:Z1})
and $Z_m^{\text{os}}$~(\ref{OSm:Z1}).
This results in
\begin{equation}
\begin{split}
\frac{e^2}{e_0^2} &{}= \left(Z_\alpha^{\text{os}}\right)^{-1}\\
&{}= 1 - \frac{4}{3} \frac{e^2 m^{-2\varepsilon}}{(4\pi)^{d/2}} \Gamma(\varepsilon)
- 4 \varepsilon
\frac{1+7\varepsilon-4\varepsilon^3}{(2-\varepsilon)(1-2\varepsilon)(1+2\varepsilon)}
\left(\frac{e^2 m^{-2\varepsilon}}{(4\pi)^{d/2}} \Gamma(\varepsilon)\right)^2
+ \cdots\\
&{}= 1 - \frac{4}{3} \frac{e^2 m^{-2\varepsilon}}{(4\pi)^{d/2}} \Gamma(\varepsilon)
- \varepsilon (2+15\varepsilon+\cdots)
\left(\frac{e^2 m^{-2\varepsilon}}{(4\pi)^{d/2}} \Gamma(\varepsilon)\right)^2
+ \cdots
\end{split}
\label{OSPhoton2:Z2}
\end{equation}

On the other hand, for the \MS{} charge we have
\begin{equation}
\frac{e^2(\mu)}{e_0^2} = Z_\alpha^{-1}
= 1 + z_1 \frac{\alpha(\mu)}{4\pi\varepsilon}
+ (z_{20}+z_{21}\varepsilon)
\left(\frac{\alpha(\mu)}{4\pi\varepsilon}\right)^2
+ \cdots
\label{OSPhoton2:ZMS}
\end{equation}
(let's pretend for a moment that we don't know $Z_\alpha$ yet).
The ratio $e^2(\mu)/e^2$ must be finite at $\varepsilon\to0$.
At one loop, this requirement gives $z_1=-4/3$.
Setting $\mu=m$, we can substitute
\begin{equation*}
\frac{\alpha(m)}{4\pi\varepsilon}
= \frac{e^2 m^{-2\varepsilon}}{(4\pi)^{d/2}} \Gamma(\varepsilon)
= \frac{\alpha}{4\pi\varepsilon}
\end{equation*}
in the two-loop term, because the differences are of a higher order
(here $\alpha=e^2/(4\pi)$).
Thus we obtain $z_{20}=0$, $z_{21}=-2$.
We have reproduced $Z_\alpha^{-1}$ in \MS~(\ref{Charge:Zalpha}),
(\ref{PhotonZ2:ZA}) from our on-shell calculation.
Finally, we arrive at the relation
between the \MS{} $\alpha(m)$ and the on-shell $\alpha$:
\begin{equation}
\alpha(m) = \alpha \left[1
+ 15 \left(\frac{\alpha}{4\pi}\right)^2 + \cdots\right]\,.
\label{OSe:ee2}
\end{equation}
The three-loop correction has been calculated in~\cite{B:92}.
If we need $\alpha(\mu)$ for $\mu\neq m$,
we can solve the RG equation~(\ref{Vertex:RG4})
with this initial condition.

\subsection{On-shell renormalization in QCD}
\label{S:OSQCD}

QCD perturbation theory is only applicable at large momenta
(or small distances).
Therefore, it makes no sense to renormalize the gluon field,
the coupling,
as well as light-quark masses and fields, in the on-shell scheme.
However, it is possible (and often convenient)
to use this scheme for renormalizing heavy-quark masses and fields,
at the same time using \MS{} for $\alpha_s(\mu)$,
the gluon field and light-quark masses and fields.
One-loop renormalization constants can be trivially obtained
from the QED results~(\ref{OSm:Z1}):
\begin{equation}
Z_m^{\text{os}} = Z_Q^{\text{os}}
= 1 - C_F \frac{g_0^2 m^{-2\varepsilon}}{(4\pi)^{d/2}}
\Gamma(\varepsilon) \frac{d-1}{d-3}\,.
\label{OSQCD:Z1}
\end{equation}
Similarly, the relation between the \MS{} mass $m(m)$
and the on-shell mass $m$ is, from~(\ref{OSm:mm1}),
\begin{equation}
\frac{m(m)}{m} = 1 - 4 C_F \frac{\alpha_s(m)}{4\pi} + \cdots
\label{OSQCD:mm1}
\end{equation}
The two-loop correction has been found in~\cite{GBGS:90}
and the three-loop one --- in~\cite{MR:00}%
\footnote{It had been found numerically~\cite{CS:99}
before the analytical result~\cite{MR:00} was obtained.}.
The quark-field renormalization constant $Z_Q^{\text{os}}$
at two loops~\cite{BGS:91} is gauge-invariant,
but at three loops~\cite{MR:00} not (unlike in QED).


\section{Decoupling of heavy-particle loops}
\label{S:Dec}

\subsection{Photon field}
\label{S:DPhoton}

Let's consider QED with massless electrons
and muons having mass $M$.
When we consider processes with characteristic energies $E\ll M$,
the existence of muons is not important.
Everything can be described by an effective low-energy theory,
in which there are no muons.
In other words, muons only exist in loops of size $\sim1/M$;
if we are interested in processes having characteristic distances
much larger than $1/M$, such loops can be replaced by
local interactions of electrons and photons.

The effective low-energy theory contains the light fields ---
electrons and photons.
The Lagrangian of this theory, describing interactions of these fields
at low energies, contains all operators constructed from these fields
which are allowed by the symmetries.
Operators with dimensionalities $d_i>4$ are multiplied by coefficients
having negative dimensionalities;
these coefficients contain $1/M^{d_i-4}$.
Therefore, this Lagrangian can be viewed as an expansion in $1/M$:
\begin{equation}
L' = \bar{\psi}'_0 i\D' \psi'_0
- \frac{1}{4} F'_{0\mu\nu} F_0^{\prime\mu\nu}
- \frac{1}{2a'_0} \left(\partial_\mu A_0^{\prime\mu}\right)^2
+ \frac{c_0}{M} \bar{\psi}'_0 F'_{0\mu\nu} \sigma^{\mu\nu} \psi'_0
+ \mathcal{O}\left(\frac{1}{M^2}\right)
\label{DPhoton:L}
\end{equation}
(primed quantities are those in the effective theory).
The coefficients in this Lagrangian ($c$, \dots) are fixed by matching ---
equating $S$-matrix elements up to some power of $p_i/M$.

\begin{figure}[ht]
\begin{center}
\begin{picture}(52,28)
\put(26,14){\makebox(0,0){\includegraphics{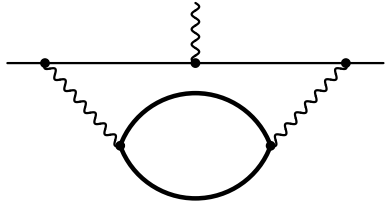}}}
\end{picture}
\end{center}
\caption{The muon contribution to the electron magnetic moment}
\label{F:muon}
\end{figure}

Let's imagine for a moment that we only know electrons and photons.
There are two ways to search for ``new physics'':
\begin{itemize}
\item
To raise energies of our accelerators
in the hope to produce real new particles (e.g., muons);
\item To measure low-energy quantities
(such as the electron magnetic moment) with a high precision
in the hope to find effects of higher terms in the effective Lagrangian
caused by loops of virtual new particles
(e.g., the last term in~(\ref{DPhoton:L})).
\end{itemize}
If we find such a deviation from predictions of pure QED
(with only electrons and photons),
we can estimate the scale of ``new physics'', i.e.\ the muon mass $M$.
If there is an unknown heavy charged particle,
then it interacts with photon and gives a contribution $c\sim e^5$
(Fig.~\ref{F:muon})%
\footnote{We suppose that it does not interact with electrons directly.}.

Operators of the full theory are also expansions in $1/M$,
in terms of all operators of the effective theory
with appropriate quantum numbers.
In particular, the bare electron and the photon fields of the full theory are,
up to $1/M^2$ corrections,
\begin{equation}
\psi_0 = \left(\zeta_\psi^0\right)^{1/2} \psi'_0\,,\quad
A_0 = \left(\zeta_A^0\right)^{1/2} A'_0\,.
\label{DPhoton:fields}
\end{equation}
The bare parameters in the Lagrangians of the two theories are related by
\begin{equation}
e_0 = \left(\zeta_\alpha^0\right)^{1/2} e'_0\,,\quad
a_0 = \zeta_A^0 a'_0\,.
\label{DPhoton:params}
\end{equation}
The \MS{} renormalized fields and parameters are related by
\begin{equation}
\begin{split}
&\psi(\mu) = \zeta_\psi^{1/2}(\mu) \psi'(\mu)\,,\quad
A(\mu) = \zeta_A^{1/2}(\mu) A'(\mu)\,,\\
&\alpha(\mu) = \zeta_\alpha(\mu) \alpha'(\mu)\,,\quad
a(\mu) = \zeta_A(\mu) a'(\mu)\,,
\end{split}
\label{DPhoton:renorm}
\end{equation}
where
\begin{equation}
\zeta_\psi(\mu) =
 \frac{Z'_\psi(\alpha'(\mu),a'(\mu))}{Z_\psi(\alpha(\mu),a(\mu))}
\zeta_\psi^0\,,\quad
\zeta_A(\mu) = \frac{Z'_A(\alpha'(\mu))}{Z_A(\alpha(\mu))} \zeta_A^0\,,\quad
\zeta_\alpha(\mu) = \frac{Z'_\alpha(\alpha'(\mu))}{Z_\alpha(\alpha(\mu))} \zeta_\alpha^0\,.
\label{DPhoton:renorm2}
\end{equation}

The gluon propagators in the two theories are related by
\begin{equation}
D_\bot(p^2) \left(g_{\mu\nu} - \frac{p_\mu p_\nu}{p^2}\right)
+ a_0 \frac{p_\mu p_\nu}{(p^2)^2}
= \zeta_A^0 \left[ D'_\bot(p^2) \left(g_{\mu\nu} - \frac{p_\mu p_\nu}{p^2}\right)
+ a'_0 \frac{p_\mu p_\nu}{(p^2)^2} \right]
+ \mathcal{O}\left(\frac{1}{M^2}\right)\,.
\label{DPhoton:D}
\end{equation}
This explains why the same decoupling constant $\zeta_A$
describes decoupling for both the photon field $A$
and the gauge-fixing parameter $a$.
It is most convenient to do matching at $p^2\to0$,
then the power-suppressed terms in~(\ref{DPhoton:D}) play no role.
The full-theory propagator near the mass shell is
\begin{equation}
D_\bot(p^2) = \frac{Z_A^{\text{os}}}{p^2}\,,\quad
Z_A^{\text{os}} = \frac{1}{1-\Pi(0)}\,.
\label{DPhoton:full}
\end{equation}
Only diagrams with muon loops contribute to $\Pi(0)$,
all the other diagrams contain no scale.
In the effective theory
\begin{equation}
D'_\bot(p^2) = \frac{Z_A^{\prime\text{os}}}{p^2}\,,\quad
Z_A^{\prime\text{os}} = \frac{1}{1-\Pi'(0)} = 1\,,
\label{DPhoton:eff}
\end{equation}
because all diagrams for $\Pi'(0)$ vanish.
Therefore,
\begin{equation}
\zeta_A^0 = \frac{Z_A^{\text{os}}}{Z_A^{\prime\text{os}}} = \frac{1}{1-\Pi(0)}\,.
\label{DPhoton:zeta0}
\end{equation}

At one loop, from~(\ref{PhotonZ:ZA}),
\begin{equation*}
Z_A(\alpha) = 1 - \frac{8}{3} \frac{\alpha}{4\pi\varepsilon}\,,\quad
Z'_A(\alpha) = 1 - \frac{4}{3} \frac{\alpha}{4\pi\varepsilon}
\end{equation*}
(because there are two lepton flavours in the full theory).
With this accuracy, we may put $\alpha'(\mu)=\alpha(\mu)$.
Re-expressing $\Pi(0)$~(\ref{OSPhoton:Pi}) via $\alpha(\mu)$
(and replacing $m_0$ by the on-shell muon mass $M$,
because the difference is beyond our accuracy),
we obtain
\begin{equation*}
\zeta_A(\mu) = \frac{Z'_A(\alpha(\mu))}{Z_A(\alpha(\mu))}\,\frac{1}{1-\Pi(0)}
= 1 - \frac{4}{3} \frac{\alpha(\mu)}{4\pi}
\left[ \left(\frac{\mu}{M}\right)^{2\varepsilon}
e^{\gamma\varepsilon} \Gamma(\varepsilon) - \frac{1}{\varepsilon} \right]\,,
\end{equation*}
and finally
\begin{equation}
\zeta_A(\mu) = 1 - \frac{8}{3} \frac{\alpha}{4\pi} \log\frac{\mu}{M}\,.
\label{DPhoton:zeta1}
\end{equation}
We can always find the $\mu$-dependence from the RG equation,
the initial condition is $\zeta_A(M)=1$ at one loop.

Let's find $\zeta_A(M)$ with two-loop accuracy.
We express $\zeta_A^0$~(\ref{DPhoton:zeta0}) via the renormalized quantities:
$\alpha(M)$ and the on-shell muon mass $M$.
Technically, this is nearly the same calculation as in Sect.~\ref{S:OSPhoton2}.
But now we express $e_0^2$ in the one-loop term via $\alpha(M)$,
and $Z_\alpha(\alpha)$ contains two lepton flavours:
\begin{equation}
\zeta_A^0 = 1 - \frac{4}{3} \frac{\alpha(M)}{4\pi\varepsilon}
- \left( \frac{16}{9} + 2 \varepsilon + 15 \varepsilon^2 \right)
\left(\frac{\alpha}{4\pi\varepsilon}\right)^2
\label{DPhoton:zeta0ren}
\end{equation}
(the extra term $16/9$ as compared to~(\ref{OSPhoton2:Z2})
comes from the doubled $\beta_0$).
We can neglect the difference between $\alpha(M)$ and $\alpha'(M)$
with our accuracy
(this difference is $\mathcal{O}(\alpha^3)$, see Sect.~\ref{S:DCharge}).
From~(\ref{PhotonZ2:ZAform}) and~(\ref{Charge:gammanf}) we have
\begin{equation*}
\frac{Z'_A(\alpha)}{Z_A(\alpha)} = 1 + \frac{4}{3} \frac{\alpha}{4\pi\varepsilon}
+ \left(\frac{32}{9} + 2 \varepsilon\right)
\left(\frac{\alpha}{4\pi\varepsilon}\right)^2\,.
\end{equation*}
All $1/\varepsilon$ poles cancel in the renormalized decoupling constant
$\zeta_A(M)$~(\ref{DPhoton:renorm2}), as they should, and we obtain
\begin{equation}
\zeta_A(M)
= 1 - 15 \left(\frac{\alpha(M)}{4\pi}\right)^2 + \cdots
\label{DPhoton:zeta2}
\end{equation}

The RG equation
\begin{equation}
\frac{d\log\zeta_A(\mu)}{d\log\mu}
+ \gamma_A(\alpha(\mu)) - \gamma'_A(\alpha'(\mu)) = 0
\label{DPhoton:RG}
\end{equation}
can be used to find $\zeta_A(\mu)$ for $\mu\neq M$.
For example, for $\mu=M(M)$,
the \MS{} muon mass~(\ref{OSm:mm1}) normalized at $\mu=M$,
the difference
\begin{equation}
M(M) - M = - 4 \frac{\alpha}{4\pi} M\,,
\label{DPhoton:MMM}
\end{equation}
so that $\zeta_A(M(M))$ is still $1+\mathcal{O}(\alpha^2)$:
\begin{equation}
\zeta_A(M(M))
= 1 - \frac{13}{3} \left(\frac{\alpha(M)}{4\pi}\right)^2 + \cdots
\label{DPhoton:zeta2M}
\end{equation}
This result can be easily obtained directly:
if we use $M_0=Z_m(\alpha(M))M(M)$ to renormalize the mass in the one-loop term
of $\zeta_A^0$ instead of $M_0= Z_m^{\text{os}}M$,
we get the formula similar to~(\ref{DPhoton:zeta0ren}),
but with $\alpha(M(M))$ and $13/3$ instead of $15$.

\subsection{Electron field}
\label{S:DElectron}

\begin{figure}[ht]
\begin{center}
\begin{picture}(52,20)
\put(26,10){\makebox(0,0){\includegraphics{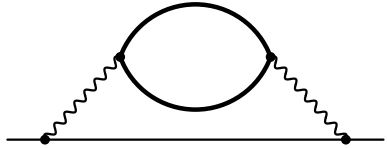}}}
\end{picture}
\end{center}
\caption{Decoupling of electron field}
\label{F:DElectron}
\end{figure}

The electron propagators in the full theory
and in the low-energy theory are related by
\begin{equation}
\rlap/p S(p) = \zeta_\psi^0\,\rlap/p S'(p)
+ \mathcal{O}\left(\frac{p^2}{M^2}\right)\,.
\label{DElectron:S}
\end{equation}
It is most convenient to do matching at $p\to0$,
where power corrections play no role.
The full-theory propagator near the mass shell is
\begin{equation}
S(p) = \frac{Z_\psi^{\text{os}}}{\rlap/p}\,,\quad
Z_\psi^{\text{os}} = \frac{1}{1-\Sigma_V(0)}\,.
\label{DElectron:full}
\end{equation}
Only diagrams with muon loops contribute to $\Sigma_V(0)$,
all the other diagrams contain no scale;
such diagrams first appear at two loops (Fig.~\ref{F:DElectron}).
In the effective theory
\begin{equation}
S'(p) = \frac{Z_\psi^{\prime\text{os}}}{\rlap/p}\,,\quad
Z_\psi^{\prime\text{os}} = \frac{1}{1-\Sigma_V'(0)} = 1\,,
\label{DElectron:eff}
\end{equation}
because all diagrams for $\Sigma_V'(0)$ vanish.
Therefore,
\begin{equation}
\zeta_\psi^0 = \frac{Z_\psi^{\text{os}}}{Z_\psi^{\prime\text{os}}}
= \frac{1}{1-\Sigma_V(0)}\,.
\label{DElectron:zeta0}
\end{equation}

At two loops (Fig.~\ref{F:DElectron}),
\begin{equation}
- i \rlap/p \Sigma_V(p^2) = \int \frac{d^d k}{(2\pi)^d}
i e_0 \gamma^\mu i \frac{\rlap/k+\rlap/p}{(k+p)^2} i e_0 \gamma^\nu
\left(\frac{-i}{k^2}\right)^2
i (k^2 g_{\mu\nu} - k_\mu k_\nu) \Pi(k^2)\,,
\label{DElectron:Sigma}
\end{equation}
where $i (k^2 g_{\mu\nu} - k_\mu k_\nu) \Pi(k^2)$
is the muon-loop contribution to the photon self-energy (Fig.~\ref{F:Photon1m}).
It is transverse; therefore, longitudinal parts of the photon propagators
($\sim\xi k_\alpha k_\beta$) do not contribute,
and the result is gauge invariant.
We only need the linear term in $p$ in both sides:
\begin{equation*}
\begin{split}
\rlap/p \Sigma_V(0) &{}= - i e_0^2 \int \frac{d^d k}{(2\pi)^d}
\gamma^\mu (k^2 \rlap/p - 2 p\cdot k\,\rlap/k) \gamma^\nu
(k^2 g_{\mu\nu} - k_\mu k_\nu) \frac{\Pi(k^2)}{(k^2)^4}\\
&{}= - i e_0^2 \int \frac{d^d k}{(2\pi)^d}
\left[ \gamma_\mu (k^2 \rlap/p - 2 p\cdot k\,\rlap/k) \gamma^\mu
- \rlap/k \rlap/p \rlap/k + 2 p\cdot k\,\rlap/k \right]
\frac{\Pi(k^2)}{(k^2)^3}\,.
\end{split}
\end{equation*}
Averaging over $k$ directions by $p\cdot k\,\rlap/k\Rightarrow(k^2/d)\rlap/p$,
we obtain
\begin{equation}
\Sigma_V(0) = - i e_0^2 \frac{(d-1)(d-4)}{d}
\int \frac{d^d k}{(2\pi)^d} \frac{\Pi(k^2)}{(k^2)^2}\,.
\label{DElectron:Sigma1}
\end{equation}

The muon-loop contribution to the photon self-energy (Fig.~\ref{F:Photon1m}) is
(we set $M=1$; the power of $M$ will be restored by dimensionality)
\begin{equation*}
i (p^2 g_{\mu\nu} - p_\mu p_\nu) \Pi(p^2)
= - \int \frac{d^d k}{(2\pi)^d}
\Tr i e_0 \gamma_\mu i \frac{\rlap/k+\rlap/p+1}{(k+p)^2-1}
i e_0 \gamma_\nu i \frac{\rlap/k+1}{k^2-1}\,.
\end{equation*}
Contracting in $\mu$ and $\nu$, we obtain
\begin{align*}
&\Pi(p^2) = - i \frac{e_0^2}{(d-1)(-p^2)}
\int \frac{d^d k}{(2\pi)^d} \frac{N}{D_1 D_2}\,,\\
&D_1 = 1 - (k+p)^2\,,\quad
D_2 = 1 - k^2\,,\quad
N = \Tr \gamma_\mu (\rlap/k+\rlap/p+1) \gamma^\mu (\rlap/k+1)\,.
\end{align*}

Let's calculate the integral
\begin{equation}
I = \int \frac{d^d p}{(2\pi)^d} \frac{\Pi(p^2)}{(-p^2)^2}\,,
\label{DElectron:Idef}
\end{equation}
which appears in~(\ref{DElectron:Sigma1}).
It is a two-loop massive vacuum diagram of Fig.~\ref{F:V2}:
\begin{equation*}
\begin{split}
&I = - i \frac{e_0^2}{d-1}
\int \frac{d^d k_1}{(2\pi)^d} \frac{d^d k_2}{(2\pi)^d}
\frac{N}{D_1 D_2 D_3^3}\,,\\
&D_1 = 1 - k_1^2\,,\quad
D_2 = 1 - k_2^2\,,\quad
D_3 = - (k_1-k_2)^2\,,\\
&N = \Tr \gamma_\mu (\rlap/k_1+1) \gamma^\mu (\rlap/k_2+1)\,.
\end{split}
\end{equation*}
All scalar products in the numerator can be expressed via the denominators:
\begin{equation*}
k_1^2 = 1 - D_1\,,\quad
k_2^2 = 1 - D_2\,,\quad
k_1 \cdot k_2 = \frac{1}{2} (2 + D_3 - D_1 - D_2)\,.
\end{equation*}
Calculating the trace and omitting terms with $D_1$ or $D_2$
(which produce vanishing integrals), we obtain
\begin{equation*}
N \Rightarrow 2 \left[4 - (d-2) D_3\right]\,.
\end{equation*}
Our integral $I$ becomes (see~(\ref{V2:def}))
\begin{equation*}
I = \frac{2i}{d-1} \frac{e_0^2}{(4\pi)^d}
\left[4 V(1,1,3) - (d-2) V(1,1,2)\right]\,.
\end{equation*}
Using~(\ref{V2:res}) and restoring the power of $M$,
we finally arrive at
\begin{equation}
I = \int \frac{d^d k}{(2\pi)^d} \frac{\Pi(k^2)}{(k^2)^2}
= - i \frac{e_0^2 M^{-4\varepsilon}}{(4\pi)^d} \Gamma^2(\varepsilon)
\frac{2(d-6)}{(d-2)(d-5)(d-7)}\,.
\label{DElectron:I}
\end{equation}

Therefore, we obtain from~(\ref{DElectron:zeta0})
\begin{equation}
\zeta_\psi^0 = 1 + \frac{e_0^4 M^{-4\varepsilon}}{(4\pi)^d} \Gamma^2(\varepsilon)
\frac{2(d-1)(d-4)(d-6)}{d(d-2)(d-5)(d-7)}\,.
\label{DElectron:zeta0res}
\end{equation}
The renormalized decoupling coefficient is
\begin{equation}
\zeta_\psi(\mu) = \zeta_\psi^0
\frac{Z_\psi'(\alpha'(\mu),a'(\mu))}{Z_\psi(\alpha(\mu),a(\mu))}\,.
\label{DElectron:zeraren}
\end{equation}
Its $\mu$-dependence can always be found
by solving the RG equation.
It is sufficient to obtain it at one point,
at some specific $\mu\sim M$,
to have the initial condition.
The most convenient point is $\mu=M$,
because $\alpha(M)=\alpha'(M)+\mathcal{O}(\alpha^3)$ (Sect.~\ref{S:DCharge})
and $a(M)=a'(M)+\mathcal{O}(\alpha^2)$ (Sect.~\ref{S:DPhoton}),
and the differences can be neglected with our accuracy.
The renormalization constant $Z_\psi$ is given by~(\ref{ElectronZ2:Z2}),
and $Z_\psi'$ --- by a similar formula with primed coefficients.
Their ratio is
\begin{equation}
\frac{Z_\psi(\alpha,a)}{Z_\psi'(\alpha,a)} = 1 + \frac{1}{4}
\left( \gamma_{\psi0} \Delta\beta_0
+ \frac{1}{2} \Delta\gamma_{A0} \gamma_{\psi0}'' a
- \Delta\gamma_{\psi1} \varepsilon \right)
\left(\frac{\alpha}{4\pi\varepsilon}\right)^2\,,
\label{DElectron:Zratio}
\end{equation}
where $\gamma_{\psi0}''$ is the coefficient of $a$ in $\gamma_{\psi0}$, and
\begin{equation*}
\Delta\beta_0 = -\frac{4}{3}\,,\quad
\Delta\gamma_{A0} = \frac{8}{3}\,,\quad
\Delta\gamma_{\psi1} = -4
\end{equation*}
are the single-flavour contributions to $\beta_0$, $\gamma_{A0}$, $\gamma_{\psi1}$.
We obtain
\begin{equation*}
\frac{Z_\psi(\alpha,a)}{Z_\psi'(\alpha,a)} = 1
+ \varepsilon \left(\frac{\alpha}{4\pi\varepsilon}\right)^2\,.
\end{equation*}
Re-expressing~(\ref{DElectron:zeta0res}) via the renormalized $\alpha(M)$,
\begin{equation*}
\zeta_\psi^0 = 1 + \varepsilon \left(1 - \frac{5}{6} \varepsilon + \cdots\right)
\left(\frac{\alpha}{4\pi\varepsilon}\right)^2\,,
\end{equation*}
we finally obtain
\begin{equation}
\zeta_\psi(M) = \frac{Z'_\psi(\alpha,a)}{Z_\psi(\alpha,a)} \zeta_\psi^0
= 1 - \frac{5}{6} \left(\frac{\alpha(M)}{4\pi}\right)^2 + \cdots
\label{DElectron:zetares}
\end{equation}

The RG equation
\begin{equation}
\frac{d\log\zeta_\psi(\mu)}{d\log\mu}
+ \gamma_\psi(\alpha(\mu),a(\mu)) - \gamma'_\psi(\alpha'(\mu),a'(\mu)) = 0
\label{DElectron:RG}
\end{equation}
can be used to find $\zeta_\psi(\mu)$ for $\mu\neq M$.
In contrast to the case of $\zeta_A(\mu)$~(\ref{DPhoton:RG}),
now $\gamma_\psi-\gamma_\psi'$ is of order $\alpha^2$,
so that changes of $\mu$ or order $\alpha$
(such as, e.g., (\ref{DPhoton:MMM}))
don't change the coefficient of $\alpha^2$ in~(\ref{DElectron:zetares}).

\subsection{Charge}
\label{S:DCharge}

The proper vertex $e_0\Gamma$ with the external propagators attached
(two electron propagators $S$ and one photon propagator $D$)
is the Green function of the fields $\bar{\psi}_0$, $\psi_0$, $A_0$
(i.e., the Fourier transform of the vacuum average
of the $T$-product of these three fields).
Therefore, the relation between this quantity in the full theory
and in the low-energy effective theory is
\begin{equation}
e_0 \Gamma S S D = \zeta_\psi^0 \left(\zeta_A^0\right)^{1/2}
e_0' \Gamma' S' S' D'\,,
\label{DCharge:Gamma1}
\end{equation}
or, taking into account $S=\zeta_\psi^o S'$, $D=\zeta_A^0 D'$,
\begin{equation}
e_0 \Gamma^\mu = \left(\zeta_\psi^0\right)^{-1} \left(\zeta_A^0\right)^{-1/2}
e_0' \Gamma^{\prime\mu}
\label{DCharge:Gamma2}
\end{equation}
In the full theory, the vertex at $p=p'$
on the mass shell ($p^2=0$) is
\begin{equation}
\Gamma^\mu = Z_\Gamma^{\text{os}} \gamma^\mu\,.
\label{DCharge:full}
\end{equation}
Only diagrams with muon loops contribute to $\Lambda^\mu(p,p)$,
all the other diagrams contain no scale;
such diagrams first appear at two loops
(see Fig.~\ref{F:muon}).
In the effective theory
\begin{equation}
\Gamma^{\prime\mu} = \gamma^\mu\,,
\label{DCharge:eff}
\end{equation}
because all diagrams for $\Lambda^{\prime\mu}(p,p)$ vanish.
Therefore,
\begin{equation}
\Gamma^\mu = \zeta_\Gamma^0 \Gamma^{\prime\mu}\,,\quad
\zeta_\Gamma^0 = \frac{Z_\Gamma^{\text{os}}}{Z_\Gamma^{\prime\text{os}}}
= Z_\Gamma^{\text{os}}\,,
\label{DCharge:zetaGamma0}
\end{equation}
and we obtain from~(\ref{DCharge:Gamma2})
\begin{equation}
\zeta_\alpha^0 = \left(\zeta_\Gamma^0 \zeta_\psi^0\right)^{-2}
\left(\zeta_A^0\right)^{-1}\,.
\label{DCharge:zeta0}
\end{equation}

The situation in QED is simpler,
due to the Ward identity (Sect.~\ref{S:OSe}).
In the full theory, we have~(\ref{OSe:Ward})
$Z_\psi^{\text{os}}Z_\Gamma^{\text{os}}=1$;
similarly, in the effective theory,
$Z_\psi^{\prime\text{os}}Z_\Gamma^{\prime\text{os}}=1$
(in fact, these two renormalization constants are equal to 1 separately).
Therefore,
\begin{equation}
\zeta_\Gamma^0 \zeta_\psi^0 = 1\,,\quad
\zeta_\alpha^0 = \left(\zeta_A^0\right)^{-1}\,.
\label{DCharge:Ward}
\end{equation}
Recalling also~(\ref{Vertex:ZalphaA}) $Z_\alpha=Z_A^{-1}$
and $Z'_\alpha=Z_A^{\prime-1}$,
we obtain
\begin{equation}
\zeta_\alpha = \zeta_\alpha^0 \frac{Z_\alpha'}{Z_\alpha}
= \left(\zeta_A^0 \frac{Z_A'}{Z_A}\right)^{-1}
= \zeta_A^{-1}\,.
\label{DCharge:Ward2}
\end{equation}
Finally, from~(\ref{DPhoton:zeta2}),
\begin{equation}
\zeta_\alpha(M) = 1 + 15 \left(\frac{\alpha(M)}{4\pi}\right)^2 + \cdots
\label{DCharge:zeta}
\end{equation}
This means that the running charges in the two theories are related by
\begin{equation*}
\alpha(M) = \zeta_\alpha(M) \alpha'(M)\,,
\end{equation*}
i.e., the running charge in the full QED (with both electrons and muons)
at $\mu=M$ is slightly larger than in the low-energy effective QED
(with only electrons).
If you prefer to do the matching at $\mu=M(M)$ instead of $\mu=M$,
you should use
\begin{equation}
\zeta_\alpha(M(M)) = 1 + \frac{13}{3} \left(\frac{\alpha(M)}{4\pi}\right)^2 + \cdots
\label{DCharge:zetaM}
\end{equation}
(see~(\ref{DPhoton:zeta2M})).

\subsection{Electron mass}
\label{S:DMass}

In the previous Sections, we considered QED with massless electrons
and heavy muons (with mass $M$).
Now let's take the electron mass into account as a small correction.
We shall expand everything up to linear terms in $m$.
The electron propagator in the full theory is given by~(\ref{Mass:S});
in the low-energy effective theory,
it involves $\Sigma'_{V,S}$ instead of $\Sigma_{V,S}$.
These two propagators are related by $\zeta^0_\psi$:
\begin{equation}
\frac{1}{1-\Sigma_V(p^2)} \frac{1}{\displaystyle
\rlap/p - \frac{1+\Sigma_S(p^2)}{1-\Sigma_V(p^2)} m_0}
= \zeta^0_\psi \frac{1}{1-\Sigma'_V(p^2)} \frac{1}{\displaystyle
\rlap/p - \frac{1+\Sigma'_S(p^2)}{1-\Sigma'_V(p^2)} m'_0}\,.
\label{DMass:S}
\end{equation}
Comparing the overall factors, we recover~(\ref{DElectron:zeta0}).
Comparing the denominators, we obtain
\begin{equation}
\frac{1+\Sigma_S(p^2)}{1-\Sigma_V(p^2)} m_0 =
\frac{1+\Sigma'_S(p^2)}{1-\Sigma'_V(p^2)} m'_0\,,
\label{DMass:mm}
\end{equation}
so that
\begin{equation}
\zeta_m^0 = \frac{m_0}{m'_0} = \frac{1-\Sigma_V(p^2)}{1-\Sigma'_V(p^2)}
\frac{1+\Sigma'_S(p^2)}{1+\Sigma_S(p^2)}
= \left(\zeta_\psi^0\right)^{-1} \frac{1}{1+\Sigma_S(0)}\,.
\label{DMass:zeta0}
\end{equation}
This is because $\Sigma'_S(0)=0$ in our approximation:
after we single out $m'_0$ in front of $\Sigma'_S(0)$
in the electron self-energy $\Sigma'(p^2)$ (similar to~(\ref{Mass:Sigma})),
we can set $m_0'=0$ in the calculation of $\Sigma'_S(0)$,
and it has no scale.
Only diagrams with muon loops contribute to $\Sigma_S(0)$;
such diagrams first appear at two loops (Fig.~\ref{F:DElectron}).
The renormalized mass decoupling constant is
\begin{equation}
\zeta_m(\mu) = \frac{m(\mu)}{m'(\mu)} = \zeta_m^0
\frac{Z'_m(\alpha'(\mu))}{Z_m(\alpha(\mu))}\,.
\label{DMass:zeta}
\end{equation}

At two loops (Fig.~\ref{F:DElectron}),
we have to take $m_0$ into account,
as compared to the massless case~(\ref{DElectron:Sigma}):
\begin{equation*}
- i \Sigma(p) = \int \frac{d^d k}{(2\pi)^d}
i e_0 \gamma^\mu i \frac{\rlap/k+\rlap/p+m_0}{(k+p)^2-m_0^2}
i e_0 \gamma^\nu
\left(\frac{-i}{k^2}\right)^2
i (k^2 g_{\mu\nu} - k_\mu k_\nu) \Pi(k^2)\,.
\end{equation*}
Again, this is gauge-invariant.
In order to extract $\Sigma_S$,
we have to retain $m_0$ in the numerator of the electron propagator;
after that, we can put $m_0=0$:
\begin{equation*}
\Sigma_S(0) = - i e_0^2 \int \frac{d^d k}{(2\pi)^d}
\frac{\gamma^\mu \gamma^\nu (k^2 g_{\mu\nu} - k_\mu k_\nu)}{(k^2)^3} \Pi(k^2)
= - i e_0^2 (d-1) \int \frac{d^d k}{(2\pi)^d} \frac{\Pi(k^2)}{(k^2)^2}\,.
\end{equation*}
Using the integral~(\ref{DElectron:I}), we obtain
\begin{equation}
\Sigma_S(0) = - \frac{e_0^4 M^{-4\varepsilon}}{(4\pi)^d} \Gamma^2(\varepsilon)
\frac{2(d-1)(d-6)}{(d-2)(d-5)(d-7)}\,.
\label{DMass:SigmaS}
\end{equation}
Therefore, from~(\ref{DMass:zeta0}) and~(\ref{DElectron:zeta0res}),
\begin{equation}
\zeta_m^0 = 1 + \frac{e_0^4 M^{-4\varepsilon}}{(4\pi)^d} \Gamma^2(\varepsilon)
\frac{8(d-1)(d-6)}{d(d-2)(d-5)(d-7)}\,.
\label{DMass:zeta0res}
\end{equation}

It is most easy to calculate $\zeta_m(M)$;
$\zeta_m(\mu)$ can then be found by solving the RG equation.
To our accuracy, we can take $\alpha(M)=\alpha'(M)$;
the ratio of the renormalization constants $Z_m(\alpha)/Z_m'(\alpha)$
is given by the formula similar to~(\ref{DElectron:Zratio}),
but $\gamma_m$ is gauge-invariant ($\gamma_{m0}''=0$),
and (see~(\ref{ElectronZ2:gammam})) $\gamma_{m0}=6$, $\Delta\gamma_{m1}=-20/3$:
\begin{equation*}
\frac{Z_m(\alpha)}{Z_m'(\alpha)} = 1
- \left(2 - \frac{5}{3} \varepsilon\right)
\left(\frac{\alpha}{4\pi\varepsilon}\right)^2\,.
\end{equation*}
Re-expressing~(\ref{DMass:zeta0res}) via the renormalized $\alpha(M)$,
\begin{equation*}
\zeta_m^0 = 1
- \left(2 - \frac{5}{3} \varepsilon + \frac{89}{18} \varepsilon^2 + \cdots\right)
\left(\frac{\alpha}{4\pi\varepsilon}\right)^2\,,
\end{equation*}
we finally obtain
\begin{equation}
\zeta_m(M) = 1 - \frac{89}{18} \left(\frac{\alpha(M)}{4\pi}\right)^2 + \cdots
\label{DMass:zetares}
\end{equation}
The RG equation
\begin{equation}
\frac{d\log\zeta_m(\mu)}{d\log\mu}
+ \gamma_m(\alpha(\mu)) - \gamma'_m(\alpha'(\mu)) = 0
\label{DMass:RG}
\end{equation}
can be used to find $\zeta_m(\mu)$ for $\mu\neq M$
(the difference $\gamma_m-\gamma_m'$ is of order $\alpha^2$).

\subsection{Decoupling in QCD}
\label{S:DQCD}

In QED, effects of decoupling of muon loops are tiny.
Also, pion pairs become important at about the same energies
as muon pairs, so that QED with electrons and muons
is a model with a narrow region of applicability.
Therefore, everything we discussed in the previous Sections
is not particularly important,
from the practical point of view.

In QCD, decoupling of heavy flavours is fundamental and omnipresent.
It would be a huge mistake to use the full 6-flavour QCD
at characteristic energies of a few GeV, or a few tens of GeV:
running of $\alpha_s(\mu)$ and other quantities would be grossly inadequate,
convergence of perturbative series would be awful because of huge logarithms.
In most cases, anybody working in QCD uses an effective low-energy QCD,
where a few heaviest flavours have been removed.
Therefore, it is important to understand decoupling in QCD.
And to this end the lessons of QED are very helpful.

Suppose we have a heavy flavour with on-shell mass $M$
and $n_l$ light flavours.
Then running of the full-theory coupling $\alpha_s^{(n_l+1)}(\mu)$
is governed by the $(n_l+1)$-flavour $\beta$-function;
running of the effective-theory coupling $\alpha_s^{(n_l)}(\mu)$
is governed by the $n_l$-flavour $\beta$-function;
their matching is given by
\begin{equation}
\alpha_s^{(n_l+1)}(\mu) = \zeta_\alpha(\mu) \alpha_s^{(n_l)}(\mu)\,,
\label{DQCD:alpha}
\end{equation}
with
\begin{equation}
\zeta_\alpha(M)
= 1 + \left( 15 C_F - \frac{32}{9} C_A \right) T_F
\left(\frac{\alpha_s(M)}{4\pi}\right)^2 + \cdots
\label{DQCD:zeta}
\end{equation}
Here the $C_F$ term can be obtained from the QED result~(\ref{DCharge:zeta})
by inserting the obvious colour factors;
the $C_A$ term is more difficult to obtain.
The RG equation
\begin{equation}
\frac{d\log\zeta_\alpha(\mu)}{d\log\mu}
+ 2 \beta^{(n_l+1)}(\alpha_s^{(n_l+1)}(\mu))
- 2 \beta^{(n_l)}(\alpha_s^{(n_l)}(\mu)) = 0
\label{DQCD:RG}
\end{equation}
can be used to find $\zeta_\alpha(\mu)$ for $\mu\neq M$.
The difference
\begin{equation*}
\beta^{(n_l+1)} - \beta^{(n_l)}
= - \frac{4}{3} T_F \frac{\alpha_s}{4\pi}
+ \mathcal{O}(\alpha_s^2)\,,
\end{equation*}
and therefore for the \MS{} mass $M(M)$~(\ref{OSQCD:mm1})
we have the well-known formula~\cite{LRV:95}
\begin{equation}
\zeta_\alpha(M(M))
= 1 + \left( \frac{13}{3} C_F - \frac{32}{9} C_A \right) T_F
\left(\frac{\alpha_s(M)}{4\pi}\right)^2 + \cdots
\label{DQCD:zetaM}
\end{equation}

\begin{figure}[ht]
\begin{center}
\begin{picture}(100,80)
\put(50,40){\makebox(0,0){\includegraphics{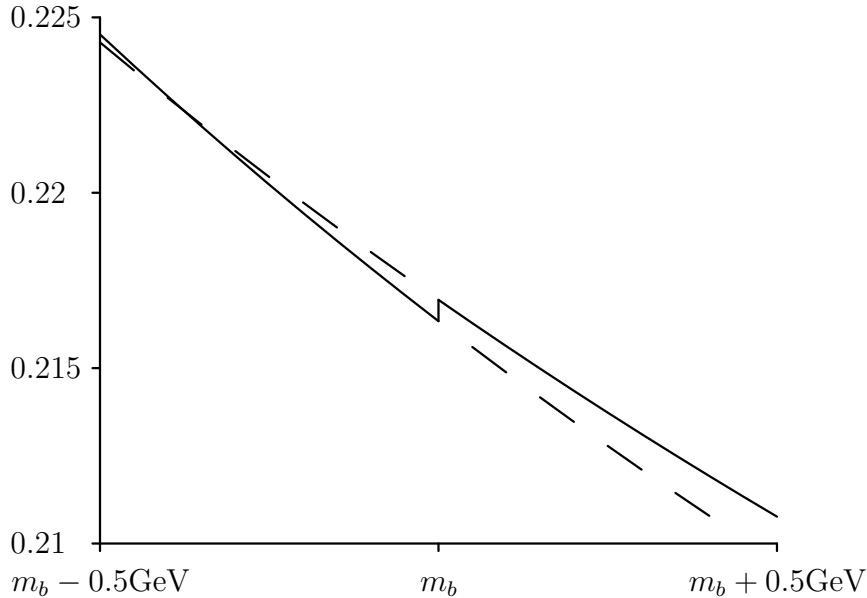}}}
\put(54,2){\makebox(0,0)[b]{$m_b$}}
\put(9,2){\makebox(0,0)[b]{$m_b-0.5\text{GeV}$}}
\put(99,2){\makebox(0,0)[b]{$m_b+0.5\text{GeV}$}}
\put(-3,9){\makebox(0,0)[l]{0.21}}
\put(-3,32.333333){\makebox(0,0)[l]{0.215}}
\put(-3,55.666667){\makebox(0,0)[l]{0.22}}
\put(-3,79){\makebox(0,0)[l]{0.225}}
\end{picture}
\end{center}
\caption{$\alpha_s^{(5)}(\mu)$ and $\alpha_s^{(4)}(\mu)$}
\label{F:as}
\end{figure}

The QCD running coupling $\alpha_s(\mu)$ not only runs when $\mu$ varies;
it also jumps when crossing heavy-flavour thresholds.
The behaviour of $\alpha_s(\mu)$ near $m_b$ is shown in Fig.~\ref{F:as}
(this figure has been obtained using the Mathematica package RunDec~\cite{CKS:00},
which takes into account 4-loop $\beta$-functions and 3-loop decoupling).
At $\mu>m_b$, the correct theory is the full 5-flavour QCD
($\alpha_s^{(5)}(\mu)$, the solid line);
at $\mu<m_b$, the correct theory is the effective low-energy 4-flavour QCD
($\alpha_s^{(4)}(\mu)$, the solid line);
the jump at $\mu=m_b$~(\ref{DQCD:zeta}) is shown.
Of course, both curves can be continued across $m_b$ (dashed lines),
and it is inessential at which particular $\mu\sim m_b$
we switch from one theory to the other one.
However, the on-shell mass $m_b$
(or any other mass which differs from it by $\mathcal{O}(\alpha_s)$,
such as, e.g., the \MS{} mass $m_b(m_b)$)
is most convenient, because the jump is small, $\mathcal{O}(\alpha_s^3)$.
For, say, $\mu=2m_b$ or $\mu=m_b/2$ it would be $\mathcal{O}(\alpha_s^2)$.

Light-quark masses $m_i(\mu)$ also rum with $\mu$,
and also jump when crossing a heavy-quark threshold.
The QCD result
\begin{equation}
m^{(n_l+1)}(M) = m^{(n_l)}(M)
\left[1 - \frac{89}{18} C_F T_F
\left(\frac{\alpha_s(M)}{4\pi}\right)^2 + \cdots \right]
\label{DQCD:m}
\end{equation}
can be obtained from the QED one~(\ref{DMass:zetares})
by inserting the obvious colour factors.

All QCD decoupling relations are currently known
with three-loop accuracy~\cite{CKS:98}.

\section{Conclusion: Effective field theories}
\label{S:EFT}

We already discussed effective field theories in Sect.~\ref{S:Dec}.
When using QCD, one rarely works works with all 6 flavours;
more often, one works in an effective low-energy QCD
with fewer flavours.

In fact, all our theories
(except The Theory of Everything, if such a thing exists)
are effective low-energy theories.
We don't know physics at arbitrarily small distances
(maybe, even our concept of space-time
becomes inapplicable at very small distances).
We want to describe phenomena at distances
larger than some boundary scale;
our ignorance about very small distances
is parametrized by local interactions of low-mass particles.
These observable particles and their interactions are thus described
by an effective field theory with all possible local operators
in its Lagrangian.
Coefficients of higher-dimensional operators
have negative dimensionalities,
and are proportional to negative powers
of the energy scale of a new physics
(the scale at which the effective low-energy theory breaks down).
At energies much lower than this scale,
these higher-dimensional terms in the Lagrangian are unimportant.
We may retain dimension-4 terms (renormalizable)
and, maybe, one or two power corrections.

The first historical example of an effective low-energy theory
is the Heisenberg--Euler effective theory in QED.
It is still the best example
illustrating typical features of such theories.

\begin{figure}[ht]
\begin{center}
\begin{picture}(80,35)
\put(17.5,20.5){\makebox(0,0){\includegraphics{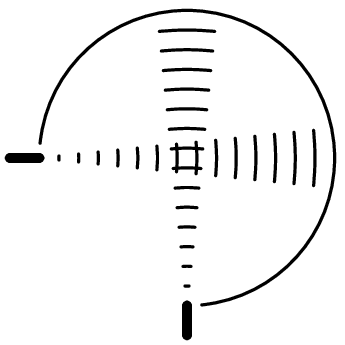}}}
\put(62.5,20.5){\makebox(0,0){\includegraphics{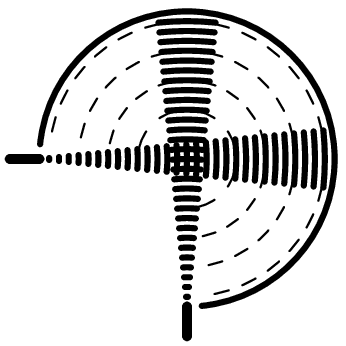}}}
\put(17.5,0){\makebox(0,0)[b]{a}}
\put(62.5,0){\makebox(0,0)[b]{b}}
\end{picture}
\end{center}
\caption{Scattering of low-energy photons}
\label{F:Photonia}
\end{figure}

In order to understand it better,
let's imagine a country, Photonia,
in which physicists have high-intensity sources
and excellent detectors of low-energy photons,
but they don't have electrons
and don't know that such a particle exists%
\footnote{We indignantly refuse to discuss the question
``Of what the experimentalists and their apparata are made?''
as irrelevant.}.
At first their experiments (Fig.~\ref{F:Photonia}a)
show that photons do not interact with each other.
They construct a theory, Quantum Photodynamics, with the Lagrangian
\begin{equation}
L = - \frac{1}{4} F_{\mu\nu} F^{\mu\nu}\,.
\label{EFT:L0}
\end{equation}
But later, after they increased the luminosity (and energy)
of their ``photon colliders'' and the sensitivity of their detectors,
they discover that photons do scatter,
though with a very small cross-section (Fig.~\ref{F:Photonia}b).
They need to add some interaction terms to this Lagrangian.
Lowest-dimensional operators having all the necessary symmetries
contain four factors $F_{\mu\nu}$.
There are two such terms:
\begin{equation}
L = - \frac{1}{4} F_{\mu\nu} F^{\mu\nu}
+ c_1 \left(F_{\mu\nu} F^{\mu\nu}\right)^2
+ c_2 F_{\mu\nu} F^{\nu\alpha} F_{\alpha\beta} F^{\beta\mu}\,.
\label{EFT:L1}
\end{equation}
They can extract the two parameters $c_{1,2}$ from two experimental results,
and predict results of infinitely many measurements.
So, this effective field theory has predictive power.

\begin{figure}[ht]
\begin{center}
\begin{picture}(22,22)
\put(11,11){\makebox(0,0){\includegraphics{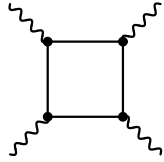}}}
\end{picture}
\end{center}
\caption{Photon--photon scattering in QED at one loop}
\label{F:gg1}
\end{figure}

We know the underlying more fundamental theory
for this effective low-energy theory,
namely QED,
and so we can help theoreticians from Photonia.
The amplitude of photon--photon scattering in QED at low energies
must be reproduced by the effective Lagrangian~(\ref{EFT:L1}).
At one loop, it is given by the diagram in Fig.~\ref{F:gg1}.
Expanding it in the photon momenta,
we can easily reduce it to the massive vacuum integrals~(\ref{V1:def}).
Due to the gauge invariance,
the leading term is linear in each of the four photon momenta.
Then we equate this full-theory amplitude
with the effective-theory one following from~(\ref{EFT:L1}),
and find the coefficients $c_{1,2}$
(this procedure is known as matching).
The result is
\begin{equation}
L = - \frac{1}{4} F_{\mu\nu} F^{\mu\nu}
+ \frac{\alpha^2}{180 m^4}
\left[ - 5 \left(F_{\mu\nu} F^{\mu\nu}\right)^2
+ 14 F_{\mu\nu} F^{\nu\alpha} F_{\alpha\beta} F^{\beta\mu} \right]\,.
\label{EFT:L2}
\end{equation}
It is not (very) difficult to calculate two-loop corrections
to this QED amplitude using the results of Sect.~\ref{S:V2},
and thus to obtain $\alpha^3$ terms in these coefficients.

\begin{figure}[ht]
\begin{center}
\begin{picture}(22,12)
\put(11,6){\makebox(0,0){\includegraphics{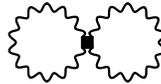}}}
\end{picture}
\end{center}
\caption{Radiative correction to the Stefan--Boltzmann law}
\label{F:StefanBoltzmann}
\end{figure}

There are many applications of the Lagrangian~(\ref{EFT:L2}).
For example, the energy density of the photon gas at temperature $T$
is $\sim T^4$ by dimensionality (Stefan--Boltzmann law).
What is the radiative correction to this law?
Calculating the vacuum diagram in Fig.~\ref{F:StefanBoltzmann}
at temperature $T$, one can obtain~\cite{KR:98}
a correction $\sim\alpha^2 T^8/m^4$.
Of course, this result is only valid at $T\ll m$.

The interaction terms in the Lagrangian~(\ref{EFT:L2})
contain the ``new physics'' energy scale,
namely the electron mass $m$, in the denominator.
If we want to reproduce more terms in the expansion
of QED amplitudes in the ratio $\omega/m$
($\omega$ is the characteristic energy),
we can include operators of higher dimensions in the effective Lagrangian;
their coefficients contain higher powers of $m$ in the denominator.
Such operators contain more $F_{\mu\nu}$ and/or its derivatives.
Heisenberg and Euler derived the effective Lagrangian
for constant field containing all powers of $F_{\mu\nu}$;
it is not sufficient for finding coefficients of operators
with derivatives of $F_{\mu\nu}$.
The expansion in $\omega/m$ breaks down when $\omega\sim m$.
At such energies the effective low-energy becomes useless,
and a more fundamental theory, QED, should be used;
in particular, real electron-positron pairs can be produced.

QED, the theory of electrons and photons,
is an effective low-energy theory too.
The leading non-renormalizable term in its Lagrangian
has dimensionality 5.
It is the Pauli interaction, the last term in~(\ref{DPhoton:L}).
It contains the scale of new physics,
in this case the muon mass $M$, in the denominator.
In reality, hadronic loops produce a comparable contribution
to the electron magnetic moment.
This hadronic contribution cannot be calculated theoretically,
because no one knows how to calculate low-energy hadronic processes
from the first principles of QCD%
\footnote{It can be calculated, with a good accuracy,
from experimental data about $e^+e^-$ annihilation into hadrons.}.
Therefore, the coefficient of the Pauli term in the QED Lagrangian
is a phenomenological parameter, to be extracted from experiment ---
from the measurement of the electron magnetic moment.

We were lucky that the scale of new physics in QED
is far away from the electron mass $m$.
Contributions of heavy-particle loops are also strongly suppressed
by powers of $\alpha$.
Therefore, the prediction for the electron magnetic moment
from the pure QED Lagrangian (without non-renormalizable corrections)
is in good agreement with experiment.
After this spectacular success of the simplest Dirac equation
(without the Pauli term) for electrons,
physicists expected that the same holds for the proton,
and its magnetic moment is $e/(2 m_p)$.
No luck here.
This shows that the picture of the proton
as a point-like structureless particle
is a very poor approximation at the energy scale $m_p$.

Another classical example of low-energy effective theories
is the four-fermion theory of weak interactions.
It was first proposed by Fermi as a fundamental theory of weak interactions;
now it is widely used (in a slightly modified form)
as an effective theory of weak interactions at low energies.

Let's consider, for example, $b$-quark decays.
In the Standard Model, they are described by diagrams with $W$ exchange.
However, the characteristic energy $E\sim m_b\ll m_W$,
and we can neglect the $W$ momentum as compared to $m_W$,
and replace its propagator by the constant $1/m_W^2$.
In other words, the $W$ propagation distance $\sim1/m_W$
is small compared to characteristic distances of the process $\sim1/E$.
We can replace the $W$ exchange by a local four-fermion interaction;
it is described by dimension-6 four-fermion operators,
their coefficients contain $1/m_W^2$.
There are 10 such operators.
We can calculate QCD corrections of any order
to diagrams containing such four-fermion vertex
without any problems;
UV divergences are eliminated by the matrix of renormalization constants
of these operators.
Their coefficients in the effective low-energy Lagrangian
obey a renormalization group equation.
The initial condition at $\mu=m_W$ is obtained by matching
the full-theory results with the effective-theory ones.

The ``new physics'' scale where this effective theory breaks down is $m_W$.
As is always the case in effective theories,
we have to fix the order of the expansion in $E^2/m_W^2$
we are interested in.
In this case, we work at the first order in $1/m_W^2$,
and there can be only one four-fermion vertex in any diagram.
What if we want to include the next correction in $E^2/m_W^2$?
In this case we have to include all the relevant dimension-8 operators
into the Lagrangian;
their coefficients contain $1/m_W^4$.
Their number is large but finite.
Elimination of UV divergences of diagrams with two four-fermion vertices
(i.e., renormalization of products of dimension-6 operators)
requires not only renormalization of these operators separately,
but also local overall counterterms
(i.e., dimension-8 operators).

We see that the fact that the theory with four-fermion interaction
is not renormalizable (as all effective field theories)
does not prevent us from calculating finite renormalized results
at any fixed order in the expansion in $E^2/m_W^2$.
The number of coefficients in the Lagrangian is finite
at each order of this expansion,
and the theory retains its predictive power.
In practice, this number quickly becomes large,
and one can only deal with few power corrections
in an effective theory.

In the past, only renormalizable theories were considered well-defined:
they contain a finite number of parameters,
which can be extracted from a finite number of experimental results
and used to predict an infinite number of other potential measurements.
Non-renormalizable theories were rejected
because their renormalization at all orders in non-renormalizable interactions
involve infinitely many parameters,
so that such a theory has no predictive power.
This principle is absolutely correct,
if we are impudent enough to pretend that our theory
describes the Nature up to arbitrarily high energies
(or arbitrarily small distances).

Our current point of view is more modest.
We accept the fact that our theories only describe the Nature
at sufficiently low energies (or sufficiently large distances).
They are effective low-energy theories.
Such theories contain all operators (allowed by the relevant symmetries)
in their Lagrangians.
They are necessarily non-renormalizable.
This does not prevent us from obtaining definite predictions
at any fixed order in the expansion in $E/M$,
where $E$ is the characteristic energy
and $M$ is the scale of new physics.
Only if we are lucky and $M$ is many orders of magnitude larger
than the energies we are interested in,
we can neglect higher-dimensional operators in the Lagrangian
and work with a renormalizable theory.

Practically all physicists believe that the Standard Model
is also a low-energy effective theory.
But we don't know what is a more fundamental theory
whose low-energy approximation is the Standard Model.
Maybe, it is some supersymmetric theory (with broken supersymmetry);
maybe, it is not a field theory, but a theory of extended objects
(superstrings, branes);
maybe, this more fundamental theory lives in a higher-dimensional space,
with some dimensions compactified;
or maybe it is something we cannot imagine at present.
The future will tell.

I am grateful to K.G.~Chetyrkin and V.A.~Smirnov for numerous discussions
of the topics discussed in these lectures;
to the organizers of the DIAS TH-2005 school in Dubna
for inviting me to give this course;
to S.V.~Mikhailov and A.A.~Vladimirov for useful remarks
on the contents of the lectures.

\newpage
\appendix
\section{Colour factors}
\label{S:Colour}

Here we shall discuss how to calculate colour diagrams
for $SU(N_c)$ colour group
(in the Nature, quarks have $N_c=3$ colours).
For more details, see~\cite{Cv}.

Elements of $SU(N_c)$ are complex $N_c\times N_c$ matrices $U$
which are unitary ($U^+U=1$) and have $\det U=1$.
Complex $N_c$-component column vectors $q^i$ transforming as
\begin{equation}
q \to U q
\quad\text{or}\quad
q^i \to U^i{}_j q^j
\label{Colour:Fund}
\end{equation}
form the space in which the fundamental representation operates.
Complex conjugate row vectors $q^+_i=(q^i)^*$ transform as
\begin{equation}
q^+ \to q^+ U^+
\quad\text{or}\quad
q^+_i \to q^+_j (U^+)^j{}_i
\quad\text{where}\quad
(U^+)^j{}_i = \left(U^i{}_j\right)^*\,;
\label{Colour:ConjFund}
\end{equation}
this is the conjugated fundamental representation.
The scalar product is invariant: $q^+ q'\to q^+ U^+ U q' = q^+ q'$.
In other words,
\begin{equation}
\delta^i_j \to \delta^k_l U^i{}_k (U^+)^l{}_j
= U^i{}_k (U^+)^k{}_j
= \delta^i_j
\label{Colour:delta}
\end{equation}
is an invariant tensor
(it is the colour structure of a meson).
The unit antisymmetric tensors $\varepsilon^{i_1\ldots i_{N_c}}$
and $\varepsilon_{i_1\ldots i_{N_c}}$ are also invariant:
\begin{equation}
\varepsilon^{i_1\ldots i_{N_c}} \to \varepsilon^{j_1\ldots j_{N_c}}\;
U^{i_1}{}_{j_1} \cdots U^{i_{N_c}}{}_{j_{N_c}}
= \det U \cdot \varepsilon^{i_1\ldots i_{N_c}}
= \varepsilon^{i_1\ldots i_{N_c}}
\label{Colour:eps}
\end{equation}
(they are the colour structures of baryons and antibaryons).

Infinitesimal transformations are given by
\begin{equation}
U = 1 + i \alpha^a t^a\,,
\label{Colour:ta}
\end{equation}
where $\alpha^a$ are infinitesimal real parameters,
and $t^a$ are called generators (of the fundamental representation).
They have the following properties:
\begin{equation}
\begin{array}{lll}
U^+ U = 1 + i \alpha^a \left( t^a - (t^a)^+ \right) = 1 & \Rightarrow & (t^a)^+ = t^a\,,\\[2mm]
\det U = 1 + i \alpha^a \Tr t^a = 1                     & \Rightarrow & \Tr t^a = 0\,,
\end{array}
\label{Colour:taprop}
\end{equation}
and are normalized by
\begin{equation}
\Tr t^a t^b = T_F \delta^{ab}\,;
\label{Colour:TF}
\end{equation}
usually, $T_F=1/2$ is used, but we shall not specialize it.
The space of unitary matrices is $N_c^2$-dimensional,
and that of traceless unitary matrices --- $(N_c^2-1)$-dimensional.
Therefore, there are $N_c^2-1$ generators $t^a$
which form a basis of this space.
Their commutators are $i$ times unitary traceless matrices,
therefore,
\begin{equation}
[t^a,t^b] = i f^{abc} t^c\,,
\label{Colour:fabc}
\end{equation}
where $f^{abc}$ are real constants.

The quantities
\begin{equation*}
A^a = q^+ t^a q'
\end{equation*}
transform as
\begin{equation}
A^a \to q^+ U^+ t^a U q' = U^{ab} A^b\,;
\label{Colour:Aa}
\end{equation}
this is the adjoint representation.
It is defined by
\begin{equation}
U^+ t^a U = U^{ab} t^b\,,
\label{Colour:adjdef}
\end{equation}
and hence
\begin{equation}
U^{ab} = \frac{1}{T_F} \Tr U^+ t^a U t^b\,.
\label{Colour:Uab}
\end{equation}
The components $(t^a)^i{}_j$ are some fixed numbers;
in other words, they form an invariant tensor
(see~(\ref{Colour:adjdef})):
\begin{equation}
(t^a)^i{}_j \to U^{ab} U^i{}_k (t^b)^k{}_l (U^+)^l{}_j = (t^a)^i{}_j\,.
\label{Colour:tinv}
\end{equation}
For an infinitesimal transformation,
\begin{equation*}
A^a \to q^+ (1 - i \alpha^c t^c) t^a (1 + i \alpha^c t^c) q'
= q^+ (t^a + i \alpha^c i f^{acb} t^b) q'\,,
\end{equation*}
so that
\begin{equation}
U^{ab} = \delta^{ab} + i \alpha^c (t^c)^{ab}\,,
\label{Colour:Uabinf}
\end{equation}
where the generators in the adjoint representation are
\begin{equation}
(t^c)^{ab} = i f^{acb}\,.
\label{Colour:fadj}
\end{equation}
As for any representation, these generators satisfy the commutation relation
\begin{equation}
(t^a)^{dc} (t^b)^{ce} - (t^b)^{dc} (t^a)^{ce} = i f^{abc} (t^c)^{de}\,;
\label{Colour:ffadj}
\end{equation}
it follows from the Jacobi identity
\begin{equation*}
\begin{split}
&[t^a,[t^b,t^d]] + [t^b,[t^d,t^a]] + [t^d,[t^a,t^b]] = 0\\
&{} = \left( i f^{bdc} i f^{ace} + i f^{dac} i f^{bce} + i f^{abc} i f^{dce} \right) t^e\,.
\end{split}
\end{equation*}

It is very convenient to do colour calculations in graphical form~\cite{Cv}.
Quark lines mean $\delta^i_j$, gluon lines mean $\delta^{ab}$,
and quark-gluon vertices mean $(t^a)^i{}_j$.
There is no need to invent names for indices;
it is much easier to see which indices are contracted ---
they are connected by a line.
Here are the properties of the generators $t^a$
which we already know:
\begin{equation}
\begin{array}{lll}
\Tr 1 = N_c &\quad\text{or}\quad& \raisebox{-5.25mm}{\includegraphics{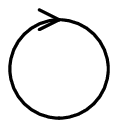}} = N_c\,,\\
\Tr t^a = 0 &\quad\text{or}\quad& \raisebox{-5.25mm}{\includegraphics{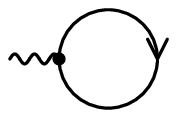}} = 0\,,\\
\Tr t^a t^b = T_F \delta^{ab} &\quad\text{or}\quad&
\raisebox{-5.25mm}{\includegraphics{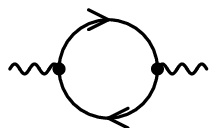}} =
T_F \raisebox{-0.25mm}{\includegraphics{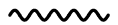}}\,.
\end{array}
\label{Colour:Cv0}
\end{equation}

There is a simple and systematic method for calculation of colour factors ---
Cvitanovi\'c algorithm~\cite{Cv}.
Now we are going to derive its main identity.
The tensor $(t^a)^i{}_j (t^a)^k{}_l$ is invariant, because $(t^a)^i{}_j$ is invariant.
It can be expressed via $\delta^i_j$,
the only independent invariant tensor with fundamental-representation indices
(it is clear that $\varepsilon^{i_1\ldots i_{N_c}}$ and $\varepsilon_{i_1\ldots i_{N_c}}$
cannot appear in this expression, except the case $N_c=2$;
in this case $\varepsilon^{ik}\varepsilon_{jl}$ can appear,
but it is expressible via $\delta^i_j$).
The general form of this expression is
\begin{equation}
(t^a)^i{}_j (t^a)^k{}_l
= a \left[ \delta^i_l \delta^k_j - b \delta^i_j \delta^k_l \right]\,,
\label{Colour:Cvt1}
\end{equation}
or graphically
\begin{equation}
\raisebox{-5mm}{\includegraphics{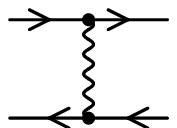}}
= a \left[ \raisebox{-5mm}{\includegraphics{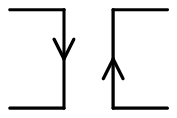}}
- b \raisebox{-5mm}{\includegraphics{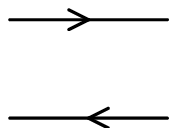}}
\right]\,,
\label{Colour:Cvg1}
\end{equation}
where $a$ and $b$ are some unknown coefficients.
If we multiply~(\ref{Colour:Cvt1}) by $\delta^j_i$,
\begin{equation*}
(t^a)^i{}_i (t^a)^k{}_l = 0
= a \left[ \delta^k_l - b N_c \delta^k_l \right]\,,
\end{equation*}
i.~e., close the upper line in~(\ref{Colour:Cvg1}),
\begin{equation*}
\raisebox{-5mm}{\includegraphics{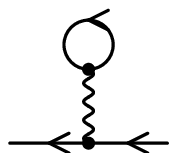}}
= a \left[ \raisebox{-5mm}{\includegraphics{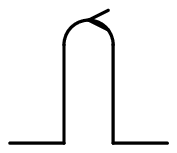}}
- b \raisebox{-5mm}{\includegraphics{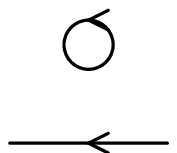}}
\right] = 0\,,
\end{equation*}
we obtain
\begin{equation*}
b = \frac{1}{N_c}\,.
\end{equation*}
If we multiply~(\ref{Colour:Cvt1}) by $(t^b)^j{}_i$,
\begin{equation*}
(t^b)^j{}_i (t^a)^i{}_j (t^a)^k{}_l = T_F (t^a)^k{}_l
= a \left[ (t^b)^k{}_l - \frac{1}{N_c} (t^b)^i{}_i \delta^k_l \right]\,,
\end{equation*}
i.~e., close the upper line in~(\ref{Colour:Cvg1}) onto a gluon,
\begin{equation*}
\raisebox{-5mm}{\includegraphics{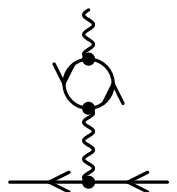}}
= a \left[ \raisebox{-5mm}{\includegraphics{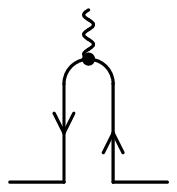}}
- \frac{1}{N_c} \raisebox{-5mm}{\includegraphics{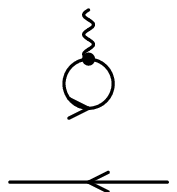}}
\right] = T_F \raisebox{-0mm}{\includegraphics{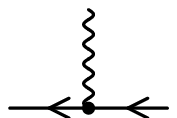}}\,,
\end{equation*}
we obtain
\begin{equation*}
a = T_F\,.
\end{equation*}
The final result is
\begin{equation}
(t^a)^i{}_j (t^a)^k{}_l
= T_F \left[ \delta^i_l \delta^k_j - \frac{1}{N_c} \delta^i_j \delta^k_l \right]\,,
\label{Colour:Cvt}
\end{equation}
or graphically
\begin{equation}
\raisebox{-5mm}{\includegraphics{cv7a.eps}}
= T_F \left[ \raisebox{-5mm}{\includegraphics{cv7b.eps}}
- \frac{1}{N_c} \raisebox{-5mm}{\includegraphics{cv7c.eps}}
\right]\,.
\label{Colour:Cvg}
\end{equation}
This identity allows one to eliminate a gluon exchange in a colour diagram:
such an exchange is replaced by the exchange of a quark--antiquark pair,
from which its colour-singlet part is subtracted.

It can be also rewritten as
\begin{equation}
\raisebox{-5mm}{\includegraphics{cv7b.eps}}
= \frac{1}{T_F} \left[ \raisebox{-5mm}{\includegraphics{cv7a.eps}}
+ \frac{1}{N_c} \raisebox{-5mm}{\includegraphics{cv7c.eps}} \right]\,,
\label{Colour:reducg}
\end{equation}
or
\begin{equation}
q^{\prime i} q^+_j
= \frac{1}{T_F} \left[ (q^+ t^a q') (t^a)^i{}_j
+ \frac{1}{N_c} (q^+ q') \delta^i_j \right]\,.
\label{Colour:reduct}
\end{equation}
This shows that the product of the fundamental representation
and its conjugate is reducible:
it reduces to the sum of two irreducible ones,
the adjoint representation and the trivial one.
In other words, the state of a quark--antiquark pair with some fixed colours
is a superposition of the colour-singlet and the colour-adjoint states.

The Cvitanovi\'c algorithm consists of elimination gluon exchanges~(\ref{Colour:Cvg})
and using simple rules~(\ref{Colour:Cv0}).
Let's consider a simple application:
counting gluon colours.
Their number is
\begin{equation*}
\begin{split}
&N_g = \raisebox{-8mm}{\includegraphics{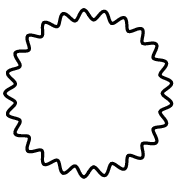}}
= \frac{1}{T_F} \raisebox{-8mm}{\includegraphics{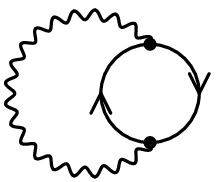}}
= \frac{1}{T_F} \raisebox{-8mm}{\includegraphics{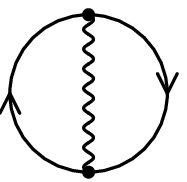}}\\
&{} = \raisebox{-8mm}{\includegraphics{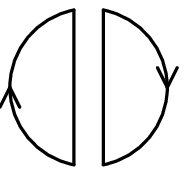}}
- \frac{1}{N_c} \raisebox{-8mm}{\includegraphics{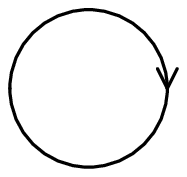}}
= N_c^2 - 1\,,
\end{split}
\end{equation*}
as we already know.

Now we consider a very important example:
\begin{equation*}
\begin{split}
&\raisebox{-0.25mm}{\includegraphics{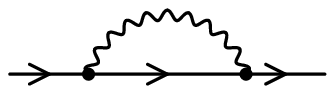}}
= T_F \left[ \raisebox{-0.25mm}{\includegraphics{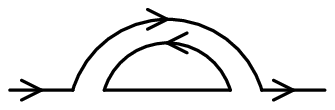}}
- \frac{1}{N_c} \raisebox{-0.25mm}{\includegraphics{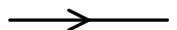}} \right]\\
&{} = T_F \left(N_c - \frac{1}{N_c}\right)
\raisebox{-0.25mm}{\includegraphics{cv9c.eps}}\,.
\end{split}
\end{equation*}
The result is
\begin{equation}
\raisebox{-0.25mm}{\includegraphics{cv9a.eps}}
= C_F \raisebox{-0.25mm}{\includegraphics{cv9c.eps}}
\quad\text{or}\quad
t^a t^a = C_F\,,
\label{Colour:CFdef}
\end{equation}
where the Casimir operator in the fundamental representation is
\begin{equation}
C_F = T_F \left(N_c - \frac{1}{N_c}\right)\,.
\label{Colour:CF}
\end{equation}

Colour diagrams can contain one more kind of elements:
3--gluon vertices $i f^{abc}$.
The definition~(\ref{Colour:fabc}) when written graphically is
\begin{equation}
\raisebox{-0.25mm}{\includegraphics{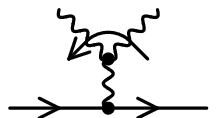}}
= \raisebox{-0.25mm}{\includegraphics{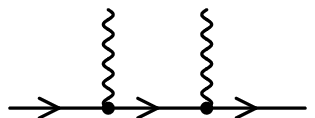}}
- \raisebox{-0.25mm}{\includegraphics{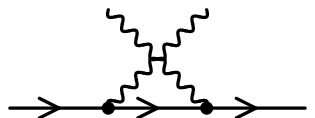}}\,.
\label{Colour:fabcg}
\end{equation}
Let's close the quark line onto a gluon:
\begin{equation*}
\raisebox{-20mm}{\includegraphics{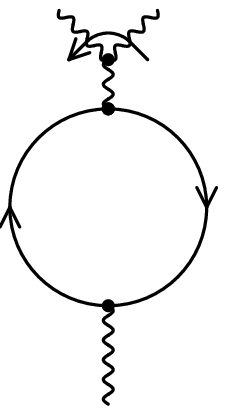}}
= \raisebox{-20mm}{\includegraphics{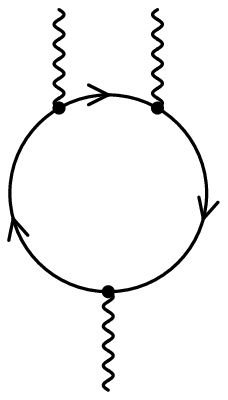}}
- \raisebox{-20mm}{\includegraphics{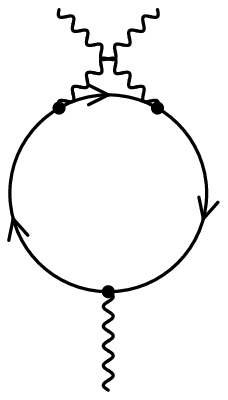}}\,.
\end{equation*}
Therefore,
\begin{equation}
\raisebox{-7.5mm}{\includegraphics{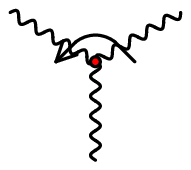}}
= \frac{1}{T_F} \left[ \raisebox{-7.5mm}{\includegraphics{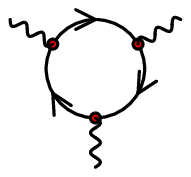}}
- \raisebox{-7.5mm}{\includegraphics{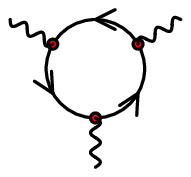}} \right]\,.
\label{Colour:Cvg3}
\end{equation}
This is the final rule of the Cvitanovi\'c algorithm:
elimination of 3--gluon vertices.

The commutation relation~(\ref{Colour:ffadj}) can be rewritten graphically,
similarly to~(\ref{Colour:fabcg}):
\begin{equation}
\raisebox{-11.25mm}{\includegraphics{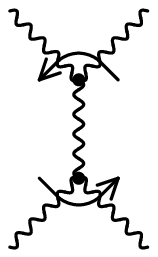}}
= \raisebox{-6.25mm}{\includegraphics{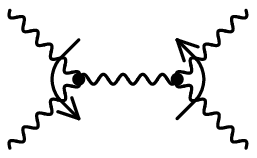}}
- \raisebox{-6.25mm}{\includegraphics{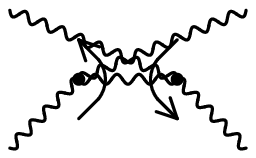}}\,.
\label{Colour:fabcadj}
\end{equation}
Sometimes  it is easier to use this relation
than to follow the Cvitanovi\'c algorithm faithfully.

Now we consider another very important example:
\begin{equation*}
\begin{split}
&\raisebox{-6.25mm}{\includegraphics{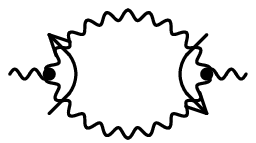}}\\
&{}= \frac{2}{T_F^2} \left[ \raisebox{-5.25mm}{\includegraphics{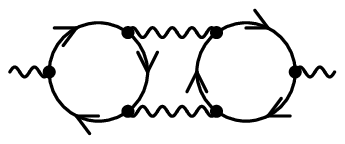}}
- \raisebox{-5.25mm}{\includegraphics{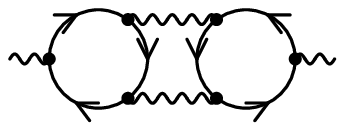}} \right]\\
&{} = \frac{2}{T_F} \left[ \raisebox{-6.25mm}{\includegraphics{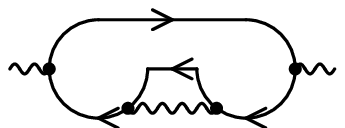}}
- \frac{1}{N_c} \raisebox{-3.25mm}{\includegraphics{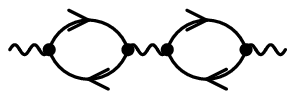}} \right.\\
&\hphantom{\displaystyle{}=\frac{2}{T_F}\Biggl[\Biggr.}
- \left. \raisebox{-6.25mm}{\includegraphics{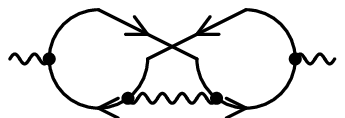}}
+ \frac{1}{N_c} \raisebox{-3.25mm}{\includegraphics{cv13e.eps}}
\right]\\
&{} = \frac{2}{T_F} \left[ \raisebox{-6.25mm}{\includegraphics{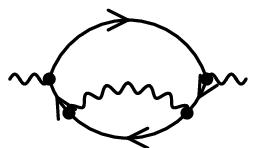}}
- \raisebox{-6.25mm}{\includegraphics{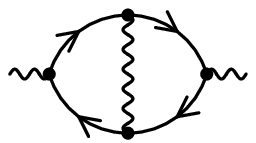}} \right]\\
&{} = 2 \left[ \raisebox{-13.25mm}{\includegraphics{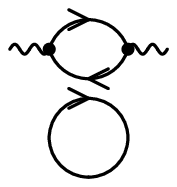}}
- \frac{1}{N_c} \raisebox{-3.25mm}{\includegraphics{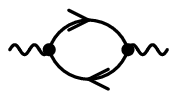}}
- \raisebox{-4.25mm}{\includegraphics{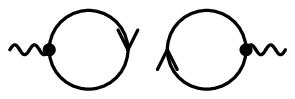}}
+ \frac{1}{N_c} \raisebox{-3.25mm}{\includegraphics{cv13h.eps}} \right]\\
&{} = 2 T_F N_c \raisebox{-0.25mm}{\includegraphics{gc.eps}}\,.
\end{split}
\end{equation*}
The result is
\begin{equation}
\raisebox{-6.25mm}{\includegraphics{cv13a.eps}}
= C_A \raisebox{-0.25mm}{\includegraphics{gc.eps}}
\quad\text{or}\quad
i f^{acd} i f^{bdc} = C_A \delta^{ab}\,,
\label{Colour:CAdef}
\end{equation}
where the Casimir operator in the adjoint representation is
\begin{equation}
C_A = 2 T_F N_c\,.
\label{Colour:CA}
\end{equation}

Another example:
\begin{equation}
\begin{split}
&\raisebox{-6.25mm}{\includegraphics{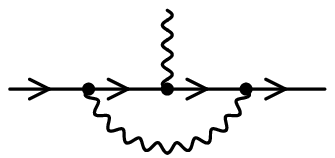}}
= T_F \left[ \raisebox{-7.45mm}{\includegraphics{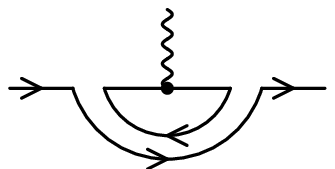}}
- \frac{1}{N_c} \raisebox{-0.25mm}{\includegraphics{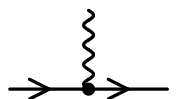}} \right]\\
&{} = - \frac{T_F}{N_c} \raisebox{-0.25mm}{\includegraphics{cv9f.eps}}\,,
\end{split}
\label{Colour:v1g}
\end{equation}
or
\begin{equation}
t^a t^b t^a = - \frac{T_F}{N_c} t^b\,,\quad
- \frac{T_F}{N_c} = C_F - \frac{C_A}{2}\,.
\label{Colour:v1t}
\end{equation}

One more example:
\begin{equation}
\begin{split}
&\raisebox{-10mm}{\includegraphics{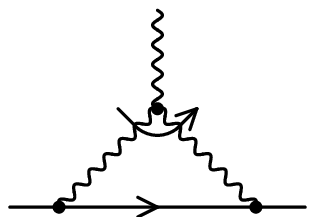}}
= \frac{1}{T_F} \left[ \raisebox{-10mm}{\includegraphics{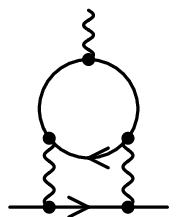}}
- \raisebox{-10mm}{\includegraphics{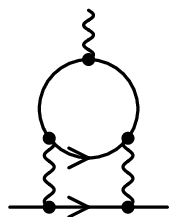}} \right]\\
&{} = \raisebox{-10mm}{\includegraphics{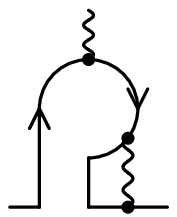}}
- \frac{1}{N_c} \raisebox{-10mm}{\includegraphics{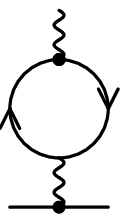}}
- \raisebox{-10mm}{\includegraphics{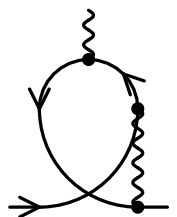}}
+ \frac{1}{N_c} \raisebox{-10mm}{\includegraphics{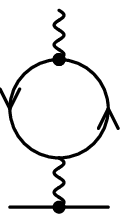}}\\
&{} = \raisebox{-7mm}{\includegraphics{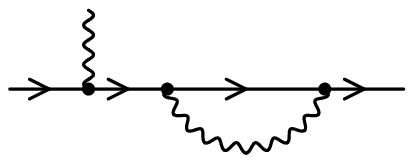}}
- \raisebox{-7mm}{\includegraphics{cv9d.eps}}\\
&{} = T_F \Biggl[ \raisebox{-5mm}{\includegraphics{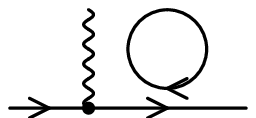}}
- \frac{1}{N_c} \raisebox{-5mm}{\includegraphics{cv9f.eps}}
- \raisebox{-5mm}{\includegraphics{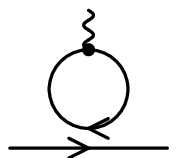}}
+ \frac{1}{N_c} \raisebox{-5mm}{\includegraphics{cv9f.eps}} \Biggr]\\
&{} = T_F N_c \raisebox{-4mm}{\includegraphics{cv9f.eps}}\,,
\end{split}
\label{Colour:v2g}
\end{equation}
or
\begin{equation}
i f^{abc} t^b t^a = \frac{C_A}{2} t^c\,.
\label{Colour:v2t}
\end{equation}
This can be derived in a shorter way.
Using antisymmetry of $f^{abc}$, (\ref{Colour:fabcg}) and~(\ref{Colour:CAdef}),
\begin{equation}
\raisebox{-10mm}{\includegraphics{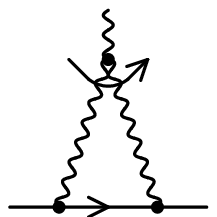}}
= \frac{1}{2} \left[ \raisebox{-10mm}{\includegraphics{cv16a.eps}}
- \raisebox{-10mm}{\includegraphics{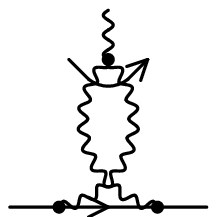}} \right]
= \frac{1}{2} \raisebox{-10mm}{\includegraphics{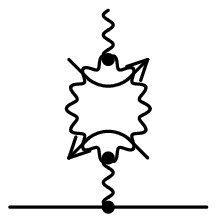}}
= \frac{C_A}{2} \raisebox{-4mm}{\includegraphics{cv9f.eps}}\,.
\label{Colour:alt1}
\end{equation}

We shall also need the result
\begin{equation}
\raisebox{-12mm}{\includegraphics{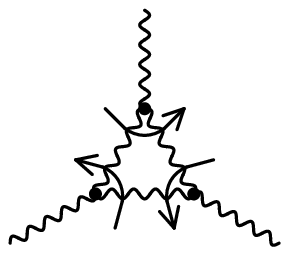}}
= \frac{C_A}{2} \raisebox{-7.5mm}{\includegraphics{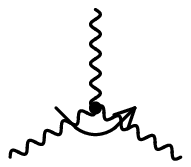}}
\quad\text{or}\quad
i f^{adf} i f^{bed} i f^{cfe} = \frac{C_A}{2} i f^{abc}\,.
\label{Colour:v3g}
\end{equation}
It can be, of course, obtained by using the Cvitanovi\'c algorithm.
A shorter derivation is similar to~(\ref{Colour:alt1}),
with the lower quark line replaced by the gluon one;
the commutation relation~(\ref{Colour:fabcadj}) in the adjoint representation
is used instead of~(\ref{Colour:fabcg}) in the fundamental one.

Sometimes the colour factor of a diagram vanishes,
and we don't need to undertake its (maybe, difficult) calculation.
For example,
\begin{equation}
\raisebox{-5.25mm}{\includegraphics{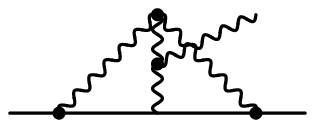}} = 0\,.
\label{Colour:zero}
\end{equation}
This can be obtained by using the Cvitanovi\'c algorithm.
A shorter derivation is following.
There is only one colour structure of the quark--gluon vertex:
\begin{equation*}
\raisebox{-5.25mm}{\includegraphics{p3.eps}}
= c \raisebox{-4mm}{\includegraphics{cv9f.eps}}\,.
\end{equation*}
Therefore, let's close this diagram:
\begin{equation*}
c \raisebox{-10.25mm}{\includegraphics{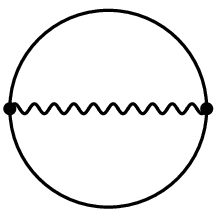}}
= \raisebox{-10.25mm}{\includegraphics{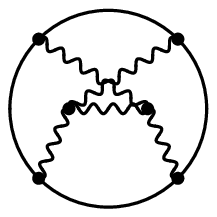}}
= \raisebox{-10.25mm}{\includegraphics{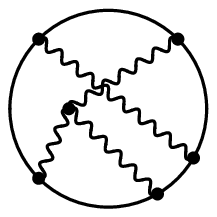}}
- \raisebox{-10.25mm}{\includegraphics{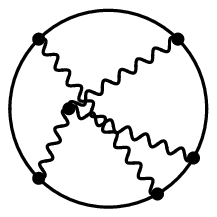}} = 0\,.
\end{equation*}

Colour factors of not too complicated diagrams
(e.g., propagator and vertex diagrams which have to be calculated
for obtaining QCD fields anomalous dimensions and $\beta$-function)
with $L$ loops (including $L_q$ quark loops) have the form
\begin{equation}
\left(T_F n_f\right)^{L_q} C_F^{n_F} C_A^{n_A}\,,\quad
n_F + n_A = L_g = L - L_q\,.
\label{Colour:Lg}
\end{equation}
This form is valid not only for $SU(N_c)$, but for any gauge group.
At $L_g\ge4$, new Casimir operators appear,
which are not reducible to $C_F$ and $C_A$.
Suppose we have calculated the colour factor for $SU(N_c)$
using the Cvitanovi\'c algorithm,
and we want to rewrite it in the form~(\ref{Colour:Lg})
valid for any group.
Our colour factor is $\left(T_F n_f\right)^{L_q}$ times a sum of terms of the form
$T_F^{L_g} N_c^n$ with various values of $n$.
Using~(\ref{Colour:CF}) and~(\ref{Colour:CA}),
we can rewrite each such term as
\begin{equation*}
\left(\frac{C_A}{2}\right)^x \left(\frac{C_A}{2} - C_F\right)^y
\quad\text{where}\quad
x + y = L_g\,,\quad
x - y = n\,.
\end{equation*}
Therefore, we may substitute
\begin{equation}
T_F^{L_g} N_c^n \rightarrow
\left[ \frac{C_A}{2} \sqrt{1 - \frac{2 C_F}{C_A}} \right]^{L_g}
\left[ \sqrt{1 - \frac{2 C_F}{C_A}} \right]^{-n}\,.
\label{Colour:Ctrick}
\end{equation}
Of course, if there are no errors, the result will contain no fractional powers,
just a polynomial in $C_F$ and $C_A$.%
\footnote{I am grateful to K.G.~Chetyrkin for explaining this trick to me.}

\end{document}